\documentclass[11pt]{article}
\usepackage{enumerate}
\usepackage{graphicx}
\RequirePackage{natbib}
\usepackage{amssymb}
\usepackage{algorithm}
\usepackage{amsmath}
\usepackage{setspace}
\usepackage{soul}
\usepackage{color}
\usepackage[left=2.54cm,top=2.54cm,right=2.54cm,bottom=2.54cm]{geometry}
\usepackage{enumerate}
\usepackage[utf8]{inputenc}
\usepackage{amsthm}
\newtheorem{theorem}{Theorem}

\newtheorem{lemma}{Lemma}
\theoremstyle{definition}
\newtheorem{definition}{Definition}
\usepackage{smile}   
\usepackage{cases}

\usepackage{booktabs}
\usepackage{multirow}
\usepackage{siunitx}
\usepackage{pgfplots}
\usepackage{tikz}
\usepackage{tabularx}
\usetikzlibrary{hobby}
\usepackage{bbm}
\usepackage[colorlinks,
linkcolor=red,
anchorcolor=blue,
citecolor=blue
]{hyperref}

\def\T{\mathrm{\scriptstyle T}}

\usepackage{fullpage}

\begin{document}
	\title{Model Linkage Selection for Cooperative Learning}
	\author{Jiaying Zhou, Jie Ding, Kean Ming Tan, Vahid Tarokh}
    \maketitle
 \bibliographystyle{plain} 

\begin{abstract}
We consider a distributed learning setting where each agent/learner holds a specific parametric model and data source. The goal is to integrate information across a set of learners to enhance the prediction accuracy of a given learner.  A natural way to integrate information is to build a joint model across a group of learners that shares common parameters of interest. However, the underlying parameter sharing patterns across a set of learners may not be a priori known. Misspecifying the parameter sharing patterns or the parametric model for each learner often yields a biased estimation and degrades the prediction accuracy.   We propose a general method to integrate information across a set of learners that is robust against misspecifications of both models and parameter sharing patterns. The main crux is to sequentially incorporate additional learners that can enhance the prediction accuracy of an existing joint model based on user-specified parameter sharing patterns across a set of learners. Theoretically, we show that the proposed method can data-adaptively select the most suitable way of parameter sharing and thus enhance the predictive performance of any particular learner of interest. Extensive numerical studies show the promising performance of the proposed method.

\end{abstract}


%%%%%%%%%%%%%%%%%%%%%%%%%%%%%%%%%%%%
%%%%%%%%%%%%%%%%%%%%%%%%%%%%%%%%%%%%
% Introduction
%%%%%%%%%%%%%%%%%%%%%%%%%%%%%%%%%%%%
%%%%%%%%%%%%%%%%%%%%%%%%%%%%%%%%%%%%
\section{Introduction}
In recent years, there has been a growing interest in statistical learning problems with a set of decentralized learners, where each learner encompasses a specific data modality and a statistical model built using domain-specific knowledge. The goal is to integrate information across decentralized learners to achieve higher statistical efficiency and predictive accuracy.  
Integrating information from different data sources is crucial in many scientific domains such as environmental science \citep{blangiardo2011bayesian,xingjian2015convolutional}, epidemiology \citep{yang2015accurate,guo2017monitoring}, statistical machine learning problems \citep{ngiam2011multimodal,kong2016land,cao2007hybrid,ye2020meta,xian2020assisted}, and computational biology \citep{simmonds2007covariate,liu2009modularization,liu2015multivariate,wen2014bayesian}.
For instance, in the context of epidemiology, a considerable amount of online search data from different platforms are integrated and used to form accurate predictions of influenza epidemics \citep{yang2015accurate,guo2017monitoring}.
The applications above raise a critical question: how to reliably integrate information from different data sources to enhance statistical efficiency in a robust manner?

We consider the setting where there are a set of learners, each consisting of a data set and a parametric statistical model.  The learners may or may not share common parameters among themselves. The goal is to develop a framework to enhance the statistical efficiency of any particular learner, say $\cL_1$, by integrating information from the other learners through parameter sharing.  
In general, learner $\cL_1$ can be assisted explicitly or implicitly by building a joint model with the other learners with potentially different statistical models and different data sources.
Explicit assistance could be achieved by joint modeling with a set of learners that share at least one parameter. 
On the other hand, implicit assistance could be achieved by joint modeling with learners whose parameters are not directly related to $\cL_1$ but are related to learners who could explicitly assist $\cL_1$. 
In principle, if the true underlying parameter sharing patterns among all learners are known a priori, and that the parametric statistical model for each learner is correctly specified, then one can build a joint model with constraints on the shared parameters based on a joint likelihood function.

Many existing modeling methods can be formulated as special instances of the above setting. For example, when multiple learners employ the same parametric model across different data sources, it is usually studied in the context of data integration~\citep{jensen2007bayesian,vonesh2006shared,liu2015multivariate,lee2017communication,jordan2019communication,danaher2014joint,ma2016joint,tang2016fused,li2018integrative,tang2019fusion,kidneycancer,shen2019fusion},
or in the context of distributed optimization~\citep{boyd2011distributed,shi2014linear,lee2017distributed,li2019communication}.
A recent topic named \textit{federated learning} is also related to the above setting, where a central server (interpreted as the main learner) sends the current global model to a set of clients (interpreted as other learners), each client updates the model parameters with local data source, and then returns them to the server~\citep{shokri2015privacy,konevcny2016federated,mcmahan2016communication}.

Though it is often helpful to establish a joint model from multiple learners, naively combining data sources and performing joint modeling can lead to severely degraded statistical performance due to four possible reasons: (i) misspecified statistical models for some learners; 
(ii) misspecified parameter sharing patterns among learners; 
(iii) heterogeneity from different data sources \citep{simmonds2007covariate,wen2014bayesian,liu2015multivariate}; and (iv) distinct learning objectives for different learners~\citep{zuech2015intrusion,sivarajah2017critical,xian2020assisted,wang2020information}.
Most existing data integration and distributed computing methods are based on the assumption that the statistical model for each learner is correctly specified and that the parameter sharing patterns are known a priori. 
A systematic statistical approach for decentralized learning that is robust against the four aspects above is relatively lacking.

In this paper, we propose a general approach to enhance the predictive performance of a specific learner $\cL_1$ by integrating information from the other learners.  
We consider the setting where there are $M$ learners and that the learners may have different statistical models and heterogeneous data sources.  
To characterize parameter sharing patterns among all learners, we introduce the notion of a model linkage graph. 
A model linkage graph $G=(V,E)$ consists of a set of $M$ vertices $V$ and a set of edges $E$, where each vertex represents a learner, and an edge between a pair of vertices encodes a unique parameter sharing pattern between the pair of learners. Also, a pair of learners do not share any common parameters if there is no edge between them. 
A joint model that enhances the predictive performance of $\cL_1$ can then be fit given a model linkage graph.
However, the ground truth or the most suitable model linkage graph is not known a priori. %and needs to be specified in practice.
Due to model misspecification within each learner and misspecified model linkages between pairs of learners, the prediction performance of $\cL_1$ may degrade after incorporating information from other learners based on an misspecified model linkage graph.

To enhance robustness against a misspecified model linkage graph, one could exhaustively build joint models for all possible sets of learners that are connected to $\cL_1$ within a given model linkage graph. Then, the set of learners that yields the largest conditional marginal likelihood of $\cL_1$ is selected.
However, such an approach is computationally infeasible since the number of possible sets of learners grows exponentially with the number of learners.  
We propose a greedy algorithm that is robust against a misspecified model linkage graph to address this challenge.
The proposed algorithm sequentially incorporates additional learners based on the user-specified model linkage graph, starting from learner $\cL_1$.  
In each iteration of our algorithm, we utilize the joint model built with a group of learners from the previous iteration and search for the next learner to improve the marginal likelihood of the current group of learners.
This process is continued until no more learners are included in the joint model.  
This approach approximately reduces the number of possible sets of model linkages from exponential to quadratic in the number of learners. 
Compared with a joint modeling approach, our proposed method is of distributed nature and does not require sharing data sources across learners.

To quantify the theoretical aspects of the proposed method, we introduce the notion of model linkage selection consistency and asymptotic prediction efficiency. They are different but conceptually parallel to asymptotic efficiency and selection consistency in the classical model selection literature (see, e.g., \citealp{DingOverview}).   
We show that the proposed algorithm achieves linkage selection consistency as long as the user-specified model linkage graph is a superset of the underlying model linkage graph. In other words, it will select data sources that are truly useful for enhancing the predictive performance of $\cL_1$ in a data-adaptive manner.  
In addition, we show that the proposed method achieves asymptotic prediction efficiency, meaning that the predictive performance of the selected model is asymptotically equivalent to that of the best joint model in hindsight.

The paper is outlined as follows. 
In Section~\ref{sec_form}, we provide the motivation, problem description, and definitions related to the model linkage graph.
In Section~\ref{sec_method}, we propose a general method for integrating information from different learners to enhance the predictive performance of $\cL_1$.
 The theoretical results for the proposed framework are provided in Section~\ref{general:theory}.
 In Section \ref{sec_simulation}, we perform numerical studies to evaluate the performance of the proposed method under different scenarios such as data contamination and model misspecification.
We close with a discussion in Section \ref{sec_con}, where we highlight some related literature on data integration and federated learning.
The technical proofs and the regularity conditions needed for the main results are included in the Appendix.

%%%%%%%%%%%%%%%%%%%%%%%%%%%%%%%%%%%%%%%%%%%
%%%%%%%%%%%%%%%%%%%%%%%%%%%%%%%%%%%%%%%%%%%
%-------------------------------    Section: Methodology   --------------------------------------------%
%%%%%%%%%%%%%%%%%%%%%%%%%%%%%%%%%%%%%%%%%%%
%%%%%%%%%%%%%%%%%%%%%%%%%%%%%%%%%%%%%%%%%%%
\section{Background and Motivation}
\label{sec_form}
Suppose that there are $M$ learners, $\cL_1, \cL_2, \ldots ,\cL_M$.  
Each learner is a pair $\cL_{\kappa}= (\Db^{(\kappa)},\mathcal P_\kappa)$, where $\Db^{(\kappa)}$ is a set of $n_{\kappa}$ observations from the  sample space $\mathcal D^{(\kappa)}$, and $\mathcal P_{\kappa}$ is a user-specified class of parametric model for modeling $\Db^{(\kappa)}$, parameterized by a $p_\kappa$-dimensional parameter $\btheta_{\kappa} \in \bTheta_{\kappa}$. 
Our goal is to develop a framework for enhancing the predictive performance of a particular learner, say $\cL_1$, by integrating information from other learners, $\cL_2,\ldots,\cL_M$.

We will focus on the regression setting where $\mathcal{D^{(\kappa)}}= (\mathcal{Y,X^{(\kappa)}})$, with $\cY =\RR$ and $\cX^{(\kappa)}=\RR^{k_\kappa}$. The covariates are allowed to have different dimensions across the $M$ learners due to different data sources.  
While we focus on the regression setting, the proposed approach can be applied more generally to data of different forms.
Let $\Db^{(\kappa)}=(\yb^{(\kappa)},\Xb^{(\kappa)}) \in  \cD^{{(\kappa)}}$, where $\yb^{(\kappa)}\in \RR^{n_\kappa}$ is an $n_\kappa$-dimensional vector of response and $\Xb^{(\kappa)}\in \RR^{n_\kappa\times k_{\kappa}}$ is an $n_\kappa\times k_\kappa$ matrix of covariates.  
Let $\mathcal P_\kappa = \left\{p_{\bm\theta_\kappa}^{(\kappa)}(\cdot| \btheta_k, \Xb^{(\kappa)}):\bm\theta_\kappa \in \bm\Theta_\kappa,\Xb^{(\kappa)} \in \mathcal X^{(\kappa)} \right\}$ be a class of user-specified parametric model for modeling $\yb^{(\kappa)}$ given  $\Xb^{(\kappa)}$.
For each learner, assume that the underlying response variable is generated independently according to the probability law $\cP^*_{\kappa}$ described by a conditional density $p_\kappa^*(\cdot |\xb^{(\kappa)})$, given the covariates $\xb^{(\kappa)} \in 
\RR^{k_{\kappa}}$.
We start with providing several definitions that will serve as a foundation of the proposed framework:~\emph{model misspecification}, \emph{model linkage}, and \emph{model linkage misspecification}.

A class of user-specified model $\cP_\kappa$ is said to be misspecified when it does not contain the underlying conditional density $p^\ast_\kappa(\cdot|\xb^{(\kappa)})$.
In practice, model misspecification often occurs due to an inappropriate functional form between the response and covariates, such as underfitting the model or neglecting dependent random noise (see, e.g.,  \citealp{misdepvar,ovb,uea3640}).  
We now provide a formal definition of model misspecification.

%%%%%%%%%%%%%%%%%%%%%%%%%%%%%%%%%%%%%%%%%%
%%%%%%%%%%%%%%%%%%%%%%%%%%%%%%%%%%%%%%%%%%
% Model Misspecification
%%%%%%%%%%%%%%%%%%%%%%%%%%%%%%%%%%%%%%%%%%
%%%%%%%%%%%%%%%%%%%%%%%%%%%%%%%%%%%%%%%%%%
\begin{definition}
\label{def:well-specified}
A model $\cP= \{p_{\bm\theta}:\bm\theta \in \bm\Theta  \}$ is \emph{well-specified} if there exists a $\bm\theta^\ast \in \bm \Theta$  such that $p_{\bm\theta^\ast} = p^\ast$ almost everywhere. 
%A model $\cP= \{p_{\bm\theta}:\bm\theta \in \bm\Theta  \}$ is \emph{misspecified} if $\sup\limits_{\bm\theta \in \bm\Theta}\mu\{y:p_{\bm\theta}(y|\bm\theta,\xb) = p^\ast (y|\xb)\}<1$ for any given covariates $\xb$, where $\mu$ is the Lebesgue measure. 
A model $\cP= \{p_{\bm\theta}:\bm\theta \in \bm\Theta  \}$ is \emph{misspecified} if $\sup_{\bm\theta \in \bm\Theta}\mathcal{P}^* \{y:p_{\bm\theta}(y|\bm\theta,\xb) = p^\ast (y|\xb)\}<1$ for some covariates $\xb$, where $\mathcal{P}^*$ denotes the probability measure corresponding to $p^\ast$. 

\end{definition}
  
\begin{figure}
    \centering
\includegraphics[scale=0.45]{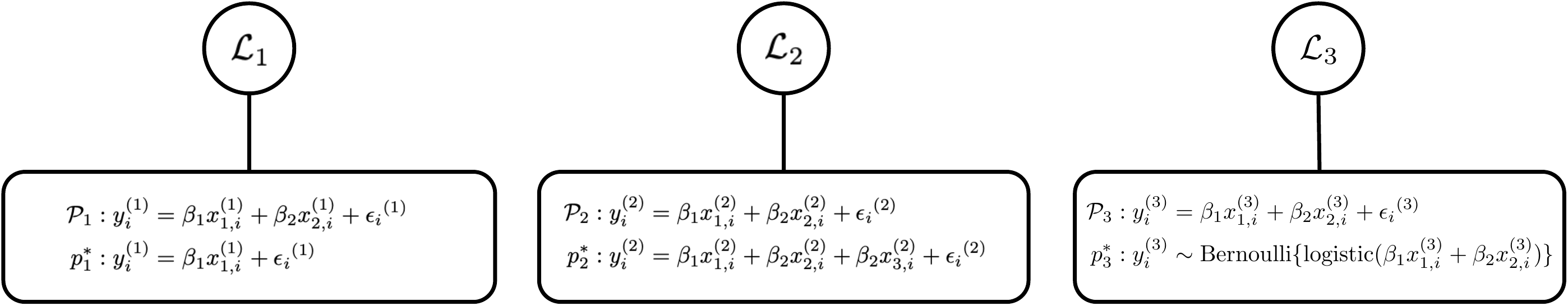}
\caption{Three examples of well-specified and misspecified models. Left panel: well-specified model. Middle panel: misspecified model. Right panel: misspecified model.}
\label{fig:wellmis}
\end{figure}

Figure~\ref{fig:wellmis} provides several examples of well-specified and misspecified models in the context of regression. 
The left panel of Figure~\ref{fig:wellmis} indicates that the learner $\cL_1$ is well-specified, since the underlying linear model $p^\ast_1 \in \cP_1$, where $\cP_1$ is a class of user-specified linear model with two covariates. 
On the other hand, the middle panel of Figure~\ref{fig:wellmis} presents a case when the learner $\cL_2$ is misspecified since $p^{\ast}_2 \notin \cP_2$. 
The model misspecification also occurs when the user-specified model class is different from the underlying data-generating process, as illustrated in the right panel of Figure~\ref{fig:wellmis}.

%%%%%%%%%%%%%%%%%%%%%%%%%%%
%%%%%%%%%%%%%%%%%%%%%%%%%%%
% Figure 1 on model misspecification
%%%%%%%%%%%%%%%%%%%%%%%%%%%
%%%%%%%%%%%%%%%%%%%%%%%%%%%

Next, we provide a new definition of model linkage. Suppose that there are two learners $\cL_i$ and $\cL_j$.  
A model linkage occurs between two learners $\cL_i$ and $\cL_j$ if they are restricted to share some common parameters.  A formal definition is given in Definition~\ref{def:model linkage}.

%%%%%%%%%%%%%%%%%%%%%%%%%%%%%%%%%%%%%%%%%%%
%%%%%%%%%%%%%%%%%%%%%%%%%%%%%%%%%%%%%%%%%%%
% Model Linkage
%%%%%%%%%%%%%%%%%%%%%%%%%%%%%%%%%%%%%%%%%%%
%%%%%%%%%%%%%%%%%%%%%%%%%%%%%%%%%%%%%%%%%%%
%---------------------                  Definition: Share Parameter                   ----------------------------%
\begin{definition}[Model linkage]
\label{def:model linkage}
Suppose that two learners $\cL_i$ and $\cL_j$ are well-specified.  Let $\btheta_{i,\cS_i}$ and $\btheta_{j,\cS_j}$ be subvectors of $ \btheta_i$ and $ \btheta_j$, indexed by the subsets $\cS_i\subseteq \{1,\ldots,p_i\}$ and $\cS_j\subseteq \{1,\ldots,p_j\}$, respectively.   
There exists a model linkage between $\cL_i$ and $\cL_j$ if 
 $\btheta_{i,\cS_i}=\btheta_{j,\cS_j}$, also denoted by $\btheta_{\cS_i,\cS_j}$ for notational convenience. 
We also refer to $\btheta_{\cS_i,\cS_j}$ as the \emph{shared common parameter} between $\cL_i$ and $\cL_j$.
\end{definition}
%-------------------------------------------------------------------------------------------------------------------------%

\begin{figure}[htp]
\centering
\includegraphics[scale = 0.5]{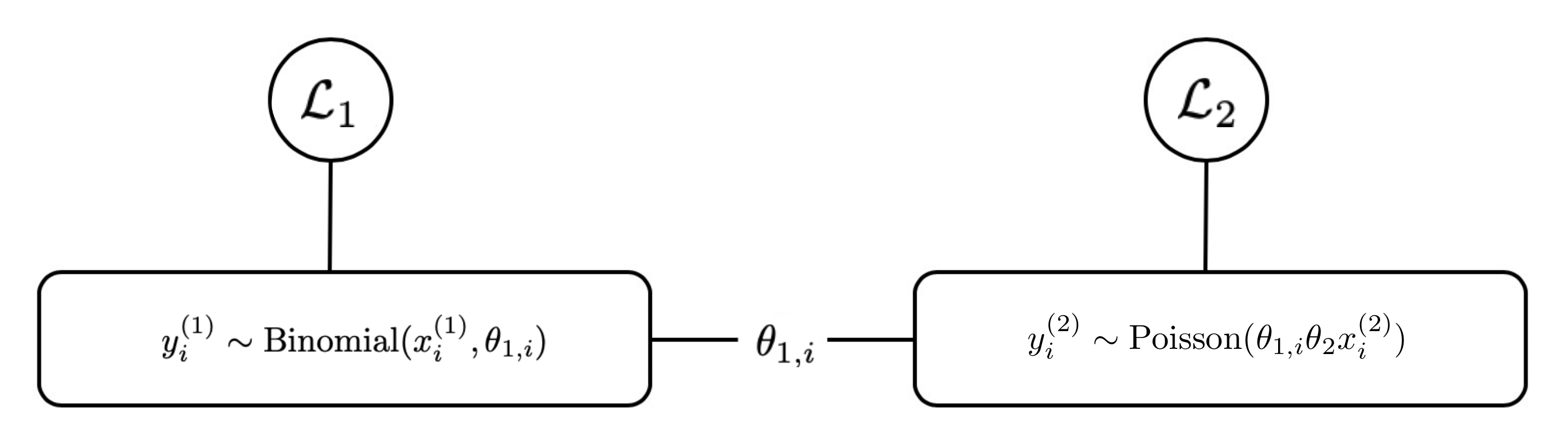}
\caption{Learners $\cL_1$ and $\cL_2$ share common parameters $\theta_{1,i}$, $i=1,\ldots,n$. }
\label{fig:share}
\end{figure}

To put the idea of model linkage into perspective, we consider an example in the context of an epidemiological study with two learners, illustrated in Figure~\ref{fig:share}. A similar example was considered in \citep{Plummer2015}. Suppose that both learners are well-specified.
Learner $\cL_1$ concerns estimating the human papilomavirus (HPV) prevalence with data $\Db^{(1)} = (\yb^{(1)},\xb^{(1)})$, where $\yb^{(1)}$ and $\xb^{(1)}$ are both $n$-dimensional vectors recording the number of women infected with high-risk HPV and the population size in different states, respectively.
A binomial model is specified for $y_i^{(1)}$ with a population size $x_i^{(1)}$ and HPV prevalence parameter $ \theta_{1,i}$, with $i=1,2,\ldots,n$.
Learner $\cL_2$ models the relationship between the HPV prevalence $\theta_{1,i}$ and cancer incidence in the form of Poisson regression.
In $\cL_2$, let $\Db_2 = (\yb^{(2)},\xb^{(2)})$, where $\yb^{(2)}$ is the number of cancer incidents and $\xb^{(2)}$ is the women-years of follow up.
Both $\cL_1$ and $\cL_2$ are restricted to share the same HPV prevalence parameters $\theta_{1,i},\ i=1,2,\ldots,n $, and thus there is a model linkage between $\cL_1$ and $\cL_2$. 

%------------------------------------------------------------------------------------------------------%
%%%%   End of Figure 2  %%%%%

The aforementioned definition and example focus mainly on whether there is a model linkage between two learners.  
Such an idea can be generalized to a set of model linkages among a group of learners, which we refer to as \emph{model linkage graph} in the following definition.

%%%%%%%%%%%%%%%%%%%%%%%%%%%%%%%%%%%
%%%%%%%%%%%%%%%%%%%%%%%%%%%%%%%%%%%
% Model Linkage Graph
%%%%%%%%%%%%%%%%%%%%%%%%%%%%%%%%%%%
%%%%%%%%%%%%%%%%%%%%%%%%%%%%%%%%%%%
\begin{definition}[Model linkage graph]
\label{def:model linkage graph}
Let $G=(V,E)$ be an undirected \emph{model linkage graph}, where $V$ is a set of $M$ vertices representing the $M$ learners $\cL_1,\ldots,\cL_M$, and $E$ is an edge set encoding model linkages between pairs of learners.  
\end{definition}
%One can think of a model linkage graph as a generalization of enforcing a prior distribution on some parameter of interest to the setting of enforcing a prior on the collaboration among a group of learners.
In practice, the model linkage graph and model linkages between pairs of learners are pre-specified by the user, usually based on domain-specific knowledge, before model fitting.
Therefore, the prediction performance of a learner may not be improved after incorporating information from the model linkage graph due to the potential misspecification of models and model linkages.  
A concept in parallel with model misspecification in the context of model linkage misspecification is provided in Definition~\ref{def_linkageMis}.

%-----------------------    Definition:Collaboration Misspecification --------------------------------%
\begin{definition}[Model linkage misspecification]
\label{def_linkageMis}
Suppose that there is a model linkage between two learners $\cL_i$ and $\cL_j$. In other words, a pair of subsets of parameters in two learners are restricted to be the same, say $\btheta_{i,\cS_i}=\btheta_{j,\cS_j}$.  A model linkage is misspecified if either $\cL_i$ or $\cL_j$ is misspecified, or that $\btheta_{i,\cS_i}^*\ne \btheta_{j,\cS_j}^*$.
More generally, a model linkage graph $G=(V,E)$ is misspecified if there exists a misspecified model linkage between a pair of learners.
\end{definition}
%------------------------------------------------------------------------------------------------------------------%

\begin{figure}[htp]
\centering
\includegraphics[scale = 0.5]{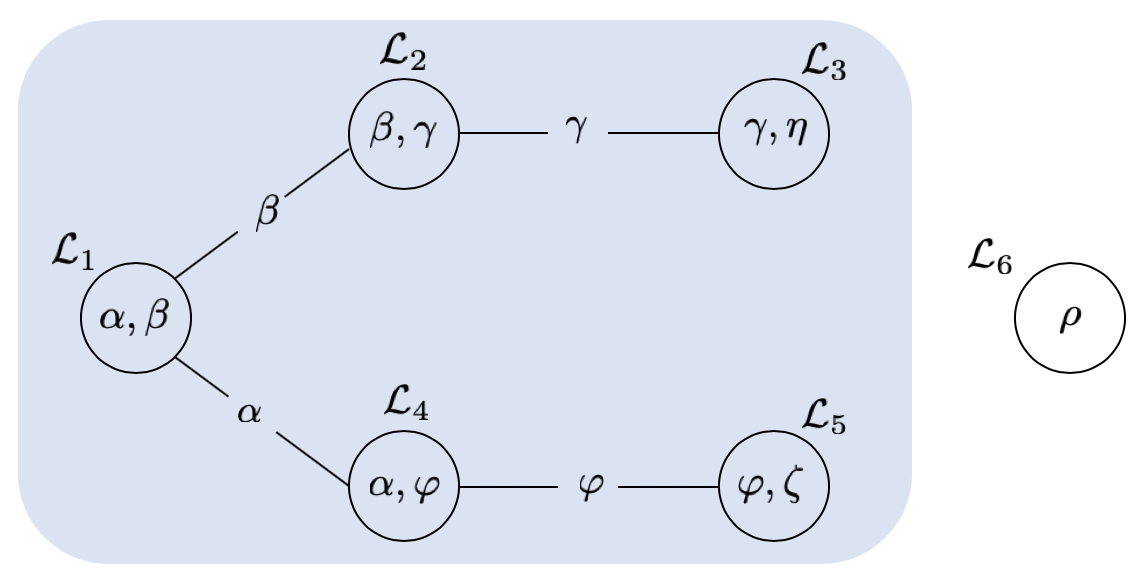}
\caption{A model linkage graph with six learners.   Learner $\cL_1$ shares common parameters with $\cL_2$ and $\cL_4$, and is implicitly connected to $\cL_3$ and $\cL_5$ through $\cL_2$ and $\cL_4$. }
\label{fig1:information flow}
\end{figure}

Recall that we are interested in enhancing the predictive performance of $\cL_1$ by integrating information from other learners $\cL_2,\ldots,\cL_M$.
Thus far, it is clear that a model linkage should exist between $\cL_1$ and $\cL_{\kappa}$ if the pair of learners shares common parameters.  
We now introduce the notion of \emph{information flow} in which learners that do not share common parameters directly with $\cL_1$ can also enhance its predictive performance by sharing common parameters with learners that can, in turn, assist $\cL_1$.
Consider an example with six learners as illustrated in Figure~\ref{fig1:information flow}.
For simplicity, assume that the statistical models for all learners are all well-specified and that the true model linkage graph is known.  
Learners $\cL_2$ and $\cL_4$ share common parameters with $\cL_1$, and thus there exist model linkages between $\cL_1$ and $\cL_2$, and $\cL_1$ and $\cL_4$. 
In addition, learner $\cL_3$ has a shared common parameter with $\cL_2$, and thus in principle, $\cL_3$ can help enhance the predictive performance of $\cL_1$ implicitly.  
This process of positive feedback transmission is an \emph{information flow} that enables implicit assistance to $\cL_1$. 
In Figure~\ref{fig1:information flow}, there are two such information flows, namely $\cL_3 \to \cL_2 \to \cL_1$ and $\cL_5\to \cL_4\to \cL_1$.  
Thus, in the model linkage graph, there exist paths from $\cL_3$ and $\cL_5$ to $\cL_1$.
Learner $\cL_6$ does not share common parameters with learners that are related to $\cL_1$, and thus there is no model linkage between $\cL_6$ and the others.

However, in practice, the true model linkage graph is not known a priori and needs to be specified. 
A misspecified model linkage graph may hamper the predictive performance of $\cL_1$.      
In the following section, we propose a data-adaptive framework to identify an appropriate model linkage graph for prediction. 
The same idea can be used to improve parameter estimation accuracy.

%%%%%%%%%%%%%%%%%%%%%%%%%%%%%%%%%%%%%%%%%
%%%%%%%%%%%%%%%%%%%%%%%%%%%%%%%%%%%%%%%%%
% Linkage Selection
%%%%%%%%%%%%%%%%%%%%%%%%%%%%%%%%%%%%%%%%%
%%%%%%%%%%%%%%%%%%%%%%%%%%%%%%%%%%%%%%%%%
\section{General Bayesian Framework for Model Linkage Selection}
\label{sec_method}
\subsection{Proposed Method} 
\label{sec_selection}
We propose a Bayesian framework to enhance the predictive performance of the learner $\cL_1$ by leveraging data from the other learners $\cL_2,\ldots,\cL_M$ according to a \emph{user-specified model linkage graph}, $G=(V,E)$.  
Suppose that there exists a model linkage between two learners, say learners $\cL_i$ and $\cL_j$. Some of the elements in $\btheta_i$ and $\btheta_j$ are restricted to be the same, say $\btheta_{i,\cS_i}=\btheta_{j,\cS_j}=\btheta_{\cS_i,\cS_j}$.
From the Bayesian perspective, a natural way to integrate information is to compute the posterior distribution of the parameters $\btheta=(\btheta_{i,-\cS_i}^\T, \btheta_{\cS_i,\cS_j}^\T,\btheta_{j,-\cS_j}^\T)^\T$, where $\btheta_{i,-\cS_i}$ and $\btheta_{j,-\cS_j}$ are obtained by removing the elements from $\btheta_i$ and $\btheta_j$, indexed by the sets $\cS_i$ and $\cS_j$, respectively.  %Let $\tilde{\btheta_i}=(\btheta_{i,-\cS_i}^\T,\btheta_{\cS_i,\cS_j}^\T )^\T$ and $\tilde{\btheta_j}=(\btheta_{j,-\cS_j}^\T,\btheta_{\cS_i,\cS_j}^\T )^\T$. 
Then, the posterior distribution of $\btheta$ can be computed as 
\[
\pi(\btheta\mid \Db^{(i)}, \Db^{(j)}) =  \frac{p_{{\btheta}_i}^{(i)}(\yb^{(i)} | {\btheta}_i,\Xb^{(i)})p_{{\btheta}_j}^{(j)}(\yb^{(j)} | {\btheta}_j,\Xb^{(j)}) \pi (\btheta)  }{p(\yb^{(i)},\yb^{(j)} | \Xb^{(i)},\Xb^{(j)})}
\]
where $\pi(\btheta)$ is a prior distribution on $\btheta$, and $p(\yb^{(i)},\yb^{(j)} | \Xb^{(i)},\Xb^{(j)})$ is the marginal likelihood obtained by integrating the numerator of the above equation with respect to $\btheta$.

More generally, one can compute the posterior distribution of the parameters according to the user-specified model linkage graph. 
Let $\cC(G)$ be a set of indices recording the vertices that form a connected component with learner $\cL_1$ in the user-specified graph $G$, including learner $\cL_1$.
That is, $\cC(G)$ is a set containing all indices of learners that have at least a path to $\cL_1$ as well as learner $\cL_1$. 
For brevity, throughout the paper, we let $\btheta$ denote a vector obtained by concatenating the entries of $\btheta_\kappa$ for all $\kappa \in \cC(G)$ without duplication.  That is, the shared parameters between pairs of learners appear only once in $\btheta$. 
%Moreover, for two learners $\cL_\kappa$ and $\cL_{\kappa'}$ that are joint with a model linkage, let $\tilde{\btheta}_\kappa= (\btheta_{\kappa,-\cS_\kappa}^\T,\btheta^\T_{\cS_\kappa,\cS_{\kappa'}})^\T$ and $\tilde{\btheta}_{\kappa'}= (\btheta_{{\kappa'},-\cS_{\kappa'}}^\T,\btheta^\T_{\cS_\kappa,\cS_{\kappa'}})^\T$ be their corresponding parameters, where $\btheta_{\cS_\kappa,\cS_{\kappa'}}$ are the shared parameters between learners   $\cL_\kappa$ and $\cL_{\kappa'}$. 
%When there are multiple model linkages to a learner, its corresponding parameter vector is defined similarly as the aforementioned.
The posterior distribution for $\btheta$ under a specific model linkage can then be computed by
\[
\pi(\btheta\mid \cup_{\kappa \in \cC(G)}\Db^{(\kappa)}) =  \frac{ \pi (\btheta)\prod_{\kappa\in \cC(G)}p_{{\btheta}_\kappa}^{(\kappa)}(\yb^{(\kappa)} | {\btheta}_\kappa,\Xb^{(\kappa)})  }{p(\cup_{\kappa\in \cC(G)} \yb^{(\kappa)} | \cup_{\kappa\in \cC(G)}\Xb^{(\kappa)})}
\]
Then, the posterior predictive distribution for a new observation in learner $\cL_1$ can be computed as
\begin{equation}
\label{eq:joint:predictive}
p(\tilde{y} |  \cup_{\kappa \in \cC(G)}\Db^{(\kappa)},\tilde{\xb}) = \int_{\bm \Theta} p_{{\btheta}_1}^{(1)}(\tilde{y}|{\btheta}_1,\tilde{\xb}) \pi(\btheta\mid  \cup_{\kappa\in \cC(G)}\Db^{(\kappa)})  d\btheta,
\end{equation}
where $\tilde{\xb}$ denotes the future observed covariates and $\bm \Theta$ is the parameter space.  
Note that the posterior predictive distribution has integrated information from learners in $\cC(G)$ through the parameters in $\btheta$. 

In practice, the true model linkage graph is not known a priori, and the user-specified model linkage graph $G$ may be misspecified, as defined in Definition~\ref{def_linkageMis}.  
%In fact, it is often the case that the learners themselves are misspecified or that the model linkages between pairs of learners are misspecified.
A joint model as in \eqref{eq:joint:predictive} with a misspecified model linkage graph $G$ can lead to severely biased parameter estimation, which affects the predictive quality of $\cL_1$.
Let $|\cC(G)|$ be the cardinality of the set $\cC(G)$.      
One way to solve the above problem is to exhaustively search all possible sets of model linkages over a graph with $|\cC(G)|$ learners, and pick the set of model linkages that yields the largest marginal likelihood for $\cL_1$ conditioned on other learners.   
However, the number of possible sets of model linkages grows exponentially with $|\cC(G)|$, and it is computationally infeasible to evaluate all possible subgraphs of $G$.

To address the above challenge, we propose a greedy algorithm that is computationally feasible with theoretical guarantees. The proposed algorithm reduces the possible sets of model linkages from exponential in $|\cC(G)|$ to quadratic in $|\cC(G)|$. 
The main idea is to successively search for the next learner that will improve the conditional marginal likelihood of the current group of learners, starting from a singleton set $\cL_1$. 
%Specifically, in each iteration of the proposed algorithm, we integrate information from a learner that is not already part of the current group of learners that increases the marginal likelihood of the current group of learners. 
The algorithm will terminate and output an estimated model linkage graph $\hat{G}=(V,\hat{E})$ when adding any further learner no longer increases the marginal likelihood of the maintained learners. 
The greedy algorithm is outlined in Algorithm \ref{algo_greedy}.

%
%We exemplify the algorithm with the example presented in Figure~\ref{fig1:information flow}. In that context, we have $\cC(G) = \{1,2,3,4,5\}$ since learner $\cL_1,\ldots,\cL_5$ form a connected component containing $\cL_1$.  
%Suppose that $G$ happens to represent the underlying data generating process. The algorithm is expected to proceed as follows.
%Learner $\cL_1$ will be linked with $\cL_2$ and $\cL_4$ during the first two iterations of Algorithm~\ref{algo_greedy}, and subsequently, $\cL_3$ and $\cL_5$ will be merged into the connected component formed by $\cL_1,\cL_2,$ and $\cL_4$ in the following iterations.  
%The algorithm will then be terminated since there is no model linkage between $\cL_6$ and learners in $\cC(G)$ from the user-specified graph.

\begin{algorithm}[tb]
  \centering
  \caption{Greedy Algorithm for Model Linkage Selection.}\label{algo_greedy}
  \footnotesize
  \begin{algorithmic}[1]
    \renewcommand{\algorithmicrequire}{\textbf{Input:}}
    \renewcommand{\algorithmicensure}{\textbf{Output:}}
\REQUIRE 
User-specified graph $G$, data $\Db^{(\kappa)}$,  parameter vector $ \btheta_\kappa$, parametric distribution $p_{\btheta_\kappa}^{(\kappa)}(\cdot\mid \btheta_{\kappa},\Xb^{(\kappa)})$, prior distribution $\pi_\kappa(\cdot)$ on  parameters $\btheta_\kappa$, for $\kappa =1,\ldots,M$. 
\STATE Initialize the index $\ell=1$, linkage set $\zeta^{(1)}=\left\{1\right\}$
\FOR{$\ell = 2,\ldots, M$}
\STATE Let $N_G(\zeta^{(\ell-1)})$ denote the neighboring set of $\zeta^{(\ell-1)}$ within $G$, namely the set of learners in $G \backslash \zeta^{(\ell-1)}$ that have a model linkage with at least one learner in $\zeta^{(\ell-1)}$

\STATE Calculate $j_{\text{opt}}= \text{argmax}_{j\in N_G(\zeta^{(\ell-1)})} p(\cup_{\kappa \in \zeta^{(\ell-1)}} \yb^{(\kappa)} \mid \cup_{\kappa \in \zeta^{(\ell-1)}} \Xb^{(\kappa)} , \Db^{(j)})$
  
\STATE \textbf{if} $p(\cup_{\kappa \in \zeta^{(\ell-1)}}\yb^{(\kappa)}| \cup_{\kappa \in \zeta^{(\ell-1)}}\Xb^{(\kappa)}) \geq  p(\cup_{\kappa \in \zeta^{(\ell-1)}} \yb^{(\kappa)} \mid \cup_{\kappa \in \zeta^{(\ell-1)}} \Xb^{(\kappa)} , \Db^{(j_{\text{opt}})}) $ \quad\textbf{break}
 
\STATE Let $\zeta^{(\ell)}=\left\{j_{\text{opt}}\right \}\cup \zeta^{(\ell-1)}$
\STATE \textbf{if} ${\zeta^{(\ell)}} = \cC(G)$\quad\textbf{break}\\

\ENDFOR
\STATE  For a new predictor vector, let 
$\hat p^{(\ell)}=p(\cdot| \cup_{\kappa \in \zeta^{(\ell)}} \Db^{(\kappa)},\tilde{\xb})$ and 
$\hat \pi^{(\ell)}=\pi(\cdot| \cup_{\kappa \in \zeta^{(\ell)}} \Db^{(\kappa)})$.
\ENSURE
	Predictive distribution $\hat p=\hat p^{(\ell)}$,  posterior distribution $\hat\pi=\hat\pi^{(\ell)}$, and model linkage graph $\hat{G}$.
\end{algorithmic}
\label{algo:greedy}
\end{algorithm}
When the algorithm terminates at the $\ell$th iteration, $\cL_1$ will integrate information from learners in $ \zeta^{(\ell)}$, which leads to the following posterior distribution of $\btheta$
\[
\pi( \bm\theta | \cup_{\kappa \in \zeta^{(\ell)}} \Db^{(\kappa)})  = \frac{ \pi(\btheta)     \prod_{\kappa\in\zeta^{(\ell)}} p_{\btheta_k}^{(\kappa)}(\yb^{(\kappa)}| \btheta_k,\Xb^{(\kappa)})}{ p(\cup_{\kappa \in \zeta^{(\ell)}} \yb^{(\kappa)} |\cup_{\kappa \in \zeta^{(\ell)}} \Xb^{(\kappa)})}.
\]
Recall that the above $\btheta$ is the vector obtained by concatenating free parameters of $\btheta_\kappa$ for all $\kappa \in \zeta^{(\ell)}$ without duplication.
The posterior predictive distribution for a new observation $\tilde{\xb}$ in learner $\cL_1$ obtained from Algorithm~\ref{algo_greedy} is computed as 
\[
p(\tilde{y} | \cup_{\kappa \in \zeta^{(\ell)}} \Db^{(\kappa)},\tilde{\xb}) = \int_{\bm \Theta} p_{\btheta_1}^{(1)}(\tilde{y}|\btheta_1,\tilde{\xb}) \pi(\bm\theta| \cup_{\kappa \in \zeta^{(\ell)}} \Db^{(\kappa)}) d\bm\theta.
\]

\begin{figure}[tb]
\centering
\includegraphics[width=0.9\linewidth]{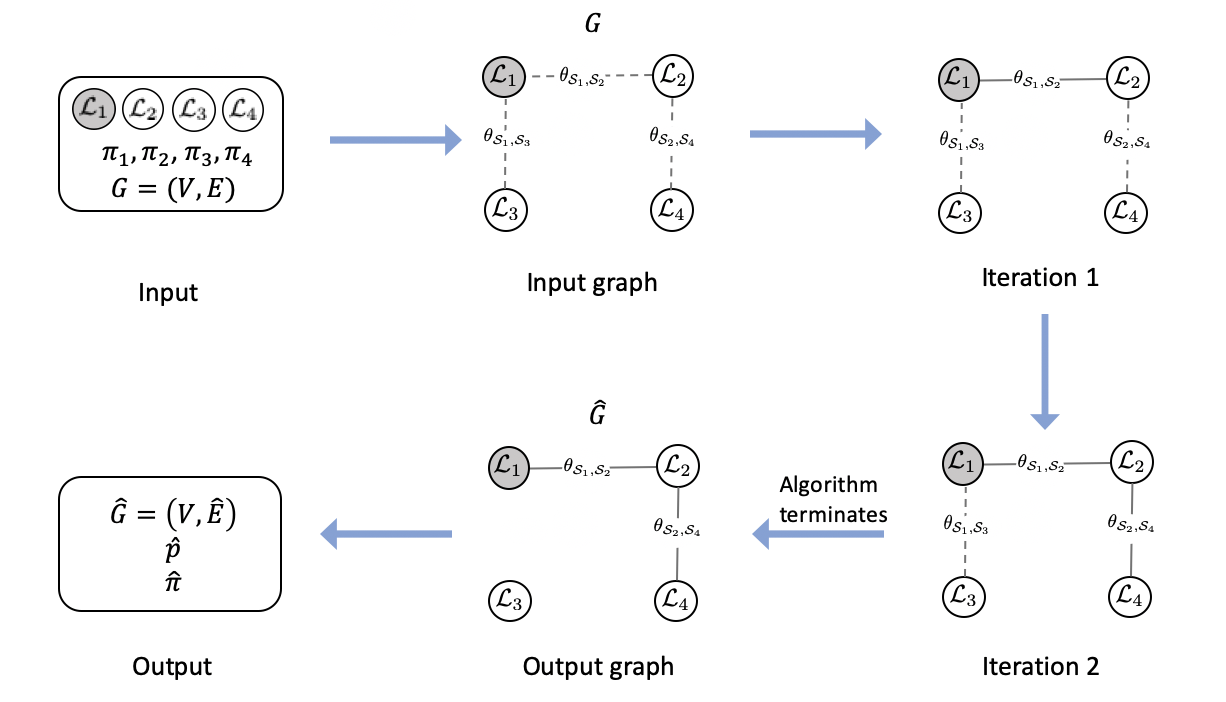}
\caption{A schematic diagram illustrating Algorithm~\ref{algo_greedy} with four learners. Each learner $\cL_\kappa = (\Db^{(\kappa)},\cP_\kappa)$ consists of the data and a user-specified class of parametric model. The algorithm starts with a user-specified graph that may or may not  be misspecified.
At each iteration, the algorithm selects a learner that maximizes the marginal likelihood of the current group of learners. The algorithm terminates when no such learners can be found. Finally, the algorithm outputs an estimated linkage graph $\hat G$, predictive distribution $\hat p$, and the posterior distribution $\hat \pi$ for $\btheta$. }
\label{algo:example}
\end{figure}

We now illustrate Algorithm~\ref{algo_greedy} with a toy example in Figure~\ref{algo:example}. 
In this example, there are four learners, and the user-specified model linkage forms a connected component among all learners, i.e., there is a path from one learner to another learner. 
Suppose that the goal is to assist learner $\cL_1$. It can be seen from the user-specified model linkage graph $G$ that $\cL_1$ is directly connected to $\cL_2$ and $\cL_3$, and implicitly connected with $\cL_4$ through $\cL_2$. 
At the first iteration of Algorithm~\ref{algo_greedy}, $\cL_1$ computes its marginal likelihood $p(\Db^{(1)})$, and conditional marginal likelihoods $p(\Db^{(1)} \mid \Db^{(2)})$ and $p(\Db^{(1)} \mid \Db^{(3)})$. 
If none of the conditional marginal likelihoods is larger than the marginal likelihood, Algorithm~\ref{algo_greedy} will be terminated; otherwise, it includes the learner that produces the larger conditional marginal likelihood into the model linkage set.  
In this case, learner $\cL_2$ is included in the model linkage set.  
%The algorithm is terminated automatically when including additional learner does not yield a larger conditioner marginal likelihood.  
Note that  $\cL_1$ does not need to get access to raw data of the other learners.  It only needs the posterior distribution (in the form of, e.g., Monte Carlo samples) locally calculated by other learners.

At the second iteration, the linkage set $\{\cL_1,\cL_2\}$ is treated as one learner since they are linked together during the first iteration.  On the other hand, the candidate learners to establish a linkage are $\cL_3$ and $\cL_4$. In Figure~\ref{algo:example}, $\cL_4$ is included in the linkage set following a similar argument as in the first iteration.  
Finally, $\cL_3$  is not included in the linkage set based on our criterion and the algorithm terminates. Consequently, $\cL_1$ will obtain a parameter estimation and predictive model that are trained from the union of $\cL_1$, $\cL_2$, and $\cL_4$.

%%%%%%%%%%%%%%%%%%%%%%%%%%%%%%%%%%%
%%%%%%%%%%%%%%%%%%%%%%%%%%%%%%%%%%%
% Subsection: Efficiency and Consistency
%%%%%%%%%%%%%%%%%%%%%%%%%%%%%%%%%%%
%%%%%%%%%%%%%%%%%%%%%%%%%%%%%%%%%%%
\subsection{Theoretical Results}
We provide definitions on \emph{asymptotic prediction efficiency} and \emph{model linkage selection consistency} in Section~\ref{subsec:theory1}.  Theoretical results for the proposed framework are presented in Section~\ref{general:theory}.
Throughout the section, let $G=(V,E)$ and $\hat{G}=(V,\hat{E})$ be the user-specified and estimated model linkage graphs, respectively.  
The user-specified $G=(V,E)$ is usually specified based on prior scientific knowledge of practitioners, which may or may not be well-specified.
Let $G^* = (V,E^*)$ be the largest subgraph of $G$, whose underlying model linkages are all well-specified after the statistical model for each learner is specified.  
That is, for any pair of learners $\cL_i$ and $\cL_j$ connected on $G$, their linkage is well-specified if and only if there exists an edge between them in $G^*$. 
Recall that more linkages imply a fewer number of free parameters in the joint model from $G$. 
Thus, intuitively speaking, $G^*$ represents the most parsimonious parameterization of the underlying data-generating process. 
%In other words, $G^*$ encodes the underlying parameter sharing patterns for $\cL_1,\ldots,\cL_M$ after the parametric statistical models and the user-specified model linkage graph are specified for all learners.
Ideally, the proposed algorithm can data-adaptively select $\hat{G}=G^*$, thus identifying the correct model linkages and filtering out misspecified ones within $G$.

\subsubsection{Definitions}
\label{subsec:theory1}
Recall that the goal of the proposed framework is to enhance the predictive performance of $\cL_1$ by borrowing information from other learners, $\cL_2,\ldots, \cL_M$.
To evaluate the predictive performance of $\cL_1$, we consider a general class of proper scoring rules \citep{gneiting2007strictly,parry2012proper,shao2019bayesian}.  
Examples of proper scoring functions are the logarithmic score $s(p,y,\xb)= -\log p(y|\xb)$ and the Brier score $s(p,y,\xb) = -p(y|\xb)+0.5 \int_{\mathbb{R}} p(\tilde{y} | \xb)^2 d\tilde{y}$~\citep{parry2016linear}.
%We refer the reader to \citet{parry2012proper} for more details on the definition of proper scoring rules.

%%%%%%%%%%%%%%%%%%%%%%%%%%%%%%%%%%%
%%%%%%%%%%%%%%%%%%%%%%%%%%%%%%%%%%%
% Definition: proper scoring function
%%%%%%%%%%%%%%%%%%%%%%%%%%%%%%%%%%%
%%%%%%%%%%%%%%%%%%%%%%%%%%%%%%%%%%%
\begin{definition}[Proper scoring function]
\label{def:scoringrule}
Let $p^\ast$ be the true data-generating density function. A scoring function $s: (p,y,\xb) \mapsto s(p,y,\xb)$ is proper if for any conditional density function $p$, we have $\int_{\cY} s(p,y,\xb) p^\ast(y|\xb) dy \geq \int_{\cY} s(p^\ast,y,\xb) p^\ast(y|\xb) dy$ almost surely. 
\end{definition}

Let $\mathbb E$ be the expectation with respect to the data-generating distribution of $y$ conditional on $\xb$, denoted by $p^*$. Let $(\tilde{y}, \tilde\xb)$ be a new observation.
The value $\mathbb E[s(p^*,\tilde{y},\tilde\xb) \mid \tilde\xb]$ is referred to as the \emph{oracle score}, and $\mathbb E[s(\hat{p},\tilde{y},\tilde\xb ) - s(p^*,\tilde{y},\tilde\xb) \mid \tilde\xb]$ is a non-negative expected prediction loss since $s(\cdot)$ is a proper scoring rule. 
It can be seen that the non-negativity of the Kullback-Leibler divergence from $p^\ast(\cdot|\xb)$ to $\hat p(\cdot|\xb)$, defined as
$D_{\textsc{kl}}\{p^\ast(\cdot|\xb) \rVert \hat p(\cdot|\xb)\} = \int_{\mathcal Y}p^\ast(y|\xb) [\log \{p^\ast(y|\xb)\} - \log \{\hat p(y|\xb)\}]dy$,
implies that the logarithmic score is proper.

Recall that $\cC(G)$ is a set of indices recording a set of learners that forms a connected component with $\cL_1$ in a model linkage graph $G$. Next, we define linkage selection consistency.    
\begin{definition}[Linkage selection consistency]
\label{def_consistency}
Given a pre-specified model linkage graph $G $, suppose that $\psi: \{\Db^{(\kappa)}: \kappa\in\cC(G)\} \mapsto \hat G$ is a linkage selection criterion in order to assist $\cL_1$. 
Then, the linkage selection criterion $\psi$ achieves linkage selection consistency if $\Pr(\hat E = E^*)\rightarrow 1$ as $n \rightarrow \infty$. 
\end{definition}
Here, the probability is defined over the observed data. In other words, a consistent linkage selection criterion $\psi$ selects all the correct model linkages present in the user-specified linkage graphs.     
%Note that if $E^*\subseteq E$, then a linkage selection consistent criterion $\psi$ recovers $E^*$ as the estimated model linkage edge set.  
%On the other hand, if $E^* \not \subset E$, then $\psi$ is selection consistent if it data-adaptively selects model linkages that are present only 
%In other words, a data integrating criterion $\psi$ achieves linkage selection consistency if it is robust to a misspecified model linkage graph.  
Next, we introduce the notion of asymptotic prediction efficiency.
Suppose that $C$ denotes a generic set of learners other than $\cL_1$. Let $\hat{p}_{C}$ denote the predictive distribution of $\cL_1$ conditional all the learners in $C$.

\begin{definition}[Asymptotic prediction efficiency]
\label{def_efficiency}
Let $\hat p$ be a constructed marginal predictive distribution for $\cL_1$, and let $s(\cdot)$ be a proper scoring function. Then, $\hat p$ is asymptotically prediction efficient if 
\begin{equation}
\label{efficiency}
\frac{ \mathbb E[s(\hat{p}_{\cC(G^*)},\tilde {y},\tilde\xb) - s(p^*,\tilde {y},\tilde\xb)  ]}{\mathbb E[s(\hat{p},\tilde{y},\tilde\xb) - s(p^*,\tilde {y},\tilde\xb)  ]}
\end{equation}
converges in probability to one
as the number of observations $n_{\kappa} \rightarrow \infty$ for $\kappa=1,\ldots,M$. 
Here, the expectation is taken over a new observation $(\tilde {y},\tilde\xb)$. 
If $\hat p$ is the posterior predictive distribution for $\cL_1$ under a certain model linkage ${\hat G}$, then ${\hat G}$ is also referred to as an asymptotically efficient model linkage graph.
\end{definition}
The ratio \eqref{efficiency} contrasts the expected prediction loss of the constructed predictive density function $\hat{p}$ and that of the predictive density function induced by $G^*$. %obtained by enumerating all possible model linkages with the criterion of minimizing the expected prediction loss.

%------------------------------      Fast Greedy Algorithm     ------------------------------%

%%%%%%%%%%%%%%%%%%%%%%%%%%%%%%%%%%%%%%%%%
%%%%%%%%%%%%%%%%%%%%%%%%%%%%%%%%%%%%%%%%%
\subsubsection{Theoretical Properties of Algorithm~\ref{algo_greedy}}
\label{general:theory}
We now proceed to study the theoretical properties of Algorithm~\ref{algo_greedy}. 
For technical convenience, we assume that learner $\cL_1$  is well-specified.  
Note that more generally, learner $\cL_1$ may or may not be well-specified. In either case, the predictive performance can be evaluated by a proper scoring rule, e.g., the logarithmic rule.
Throughout the theoretical studies, we consider the regime in which the number of learners $M$ is fixed and the number of observations $n_{\kappa}$ satifies $n_{\kappa}/n \rightarrow c_{\kappa} $, where $n=\sum_{\kappa=1}^M n_{\kappa}$ and $c_{\kappa}$ is a positive constant. 
Recall that $G= (V,E)$ is a user-specified graph. Moreover, recall that $G^*=(V,E^*)$ is the true model linkage graph, defined as the largest subgraph of $G$ with correct model linkages between pairs of learners (given that each learner's model is specified).
We denote the estimated model linkage graph from Algorithm~\ref{algo_greedy} as $\hat{G}=(V,\hat{E})$.  

%-------------------------------------theorem:   Prediction Efficiency  --------------------------------------%
\begin{theorem}
\label{thm:consistency}
Under some regularity conditions in Appendix~\ref{appendix:regularity} and given a user-specified graph $G$, the estimated model linkage edge set $\hat{E}$ from Algorithm~\ref{algo_greedy}  achieves linkage selection consistency, namely $\Pr(\hat E = E^*) \rightarrow 1$ as $n\rightarrow \infty$. 
\end{theorem}

%------------------------------------------------------------------------------------------------------------------%
Theorem~\ref{thm:consistency} indicates that the estimated model linkage graph $\hat{E}$ from Algorithm~\ref{algo_greedy} is consistent in linkage selection, only selecting all the well-specified model linkages from $G$.  
Thus, the proposed approach is asymptotically robust against model linkage misspecification.
%When the user-specified model linkage edge set $E$ is a superset of the true underlying model linkage edge set $E^*$, the proposed method recovers $E^*$ with probability one asymptotically, namely $\mathrm{Pr}(\hat{E}=E^*) \rightarrow 1$. 
We will experimentally verify the finite sample performance of Algorithm~\ref{algo_greedy} in the next section, considering various settings such as model misspecification, model linkage misspecification, and data contamination. 
The following theorem guarantees that the predictive distribution constructed from Algorithm~\ref{algo_greedy} is asymptotically prediction efficient.  

\begin{theorem}
\label{thm:efficiency}
Let $\hat p$ be the constructed predictive distribution for $\cL_1$ via Algorithm~\ref{algo_greedy} based on its selected model linkage graph $\hat{G}$. Let $s(\cdot)$ be a proper scoring function. Under the same conditions as in Theorem~\ref{thm:consistency}, $\hat p$ is asymptotically prediction efficient. %, namely 
%\begin{equation}
%\label{theory:efficient}
%\frac{ \mathbb E[s(\hat{p}_{\cC(G^*)},\tilde {y},\tilde\xb) - s(p^*,\tilde {y},\tilde\xb)]}{\mathbb E[s(\hat{p},\tilde{y},\tilde\xb) - s(p^*,\tilde {y},\tilde\xb)]}
%\rightarrow 1.
%\end{equation}
\end{theorem}
Note that Theorem~\ref{thm:efficiency} holds even when the statistical models for certain learners are misspecified due to the definition of $G^*$.  In other words, the proposed method yields a predictive distribution that is robust to model misspecification and model linkage graph misspecification.

%-----------------------------------------------     Section   ------------------------------------------------------%
%------------------------------                 Simulation               -----------------------------------------------%
\section{Numerical Studies}
\label{sec_simulation}
\subsection{Linear Regression Example}
\begin{figure}[!htp]
\centering
\includegraphics[scale=0.4]{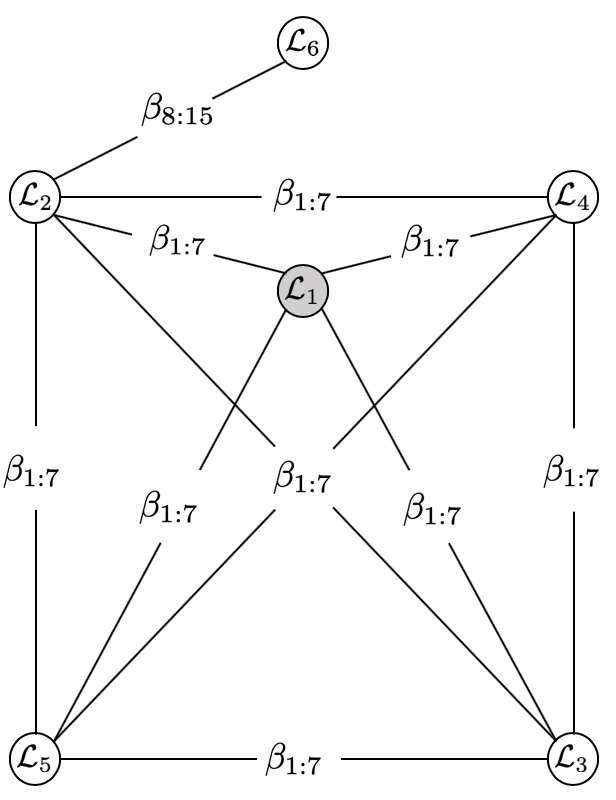}\qquad\qquad\qquad
\includegraphics[scale=0.4]{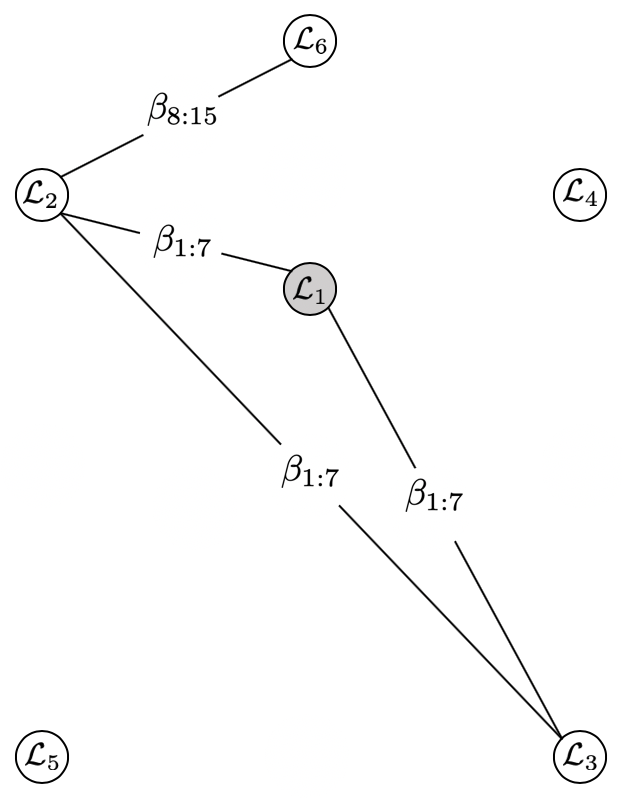}
\caption{\label{fig:linearprior}   The user-specified model linkage graph $G$ and the largest well-specified model linkage graph $G^*$ within $G$ are shown on the left and right panels, respectively. }
\end{figure}

We consider a regression setting with six learners $\cL_1,\ldots,\cL_6$.  
The goal is to enhance the predictive performance of $\cL_1$ by incorporating information from other learners.  
The data for the six learners are generated in the following way.
\begin{equation*}
\begin{split}
y_i^{(\kappa)}= \begin{cases} \sum_{j=1}^7 \beta_{j}^{(\kappa)} x_{ij}^{(\kappa)}+\epsilon_{i}^{(\kappa)} & \mathrm{for~} \kappa=1,3,4, \\ 
\sum_{j=1}^{15} \beta_{j}^{(\kappa)} x_{ij}^{(\kappa)}+\epsilon_{i}^{(\kappa)} & \mathrm{for~} \kappa=2,\\ 
\sum_{j=1}^7 \beta_{j}^{(\kappa)} (x_{ij}^{(\kappa)}+5)^2+\epsilon_{i}^{(\kappa)}& \mathrm{for~} \kappa=5,\\
\sum_{j=8}^{15} \beta_{j}^{(\kappa)} x_{ij}^{(\kappa)}+\epsilon_{i}^{(\kappa)} & \mathrm{for~} \kappa=6,\\
\end{cases}
\end{split}
\end{equation*}  
where all regression coefficients are set to equal 0.3 except that $\beta_j^{(4)}=0.6$ for $j=1,\ldots,7$.
Each covariate $x_{ij}^{(\kappa)}$ is generated from a standard normal distribution for all the learners.
Additionally, the random noise are generated from a standard normal distribution.
Since learners $\cL_1,~\cL_2$, and $\cL_3$ share common parameters, and $\cL_2$ shares common parameters with $\cL_6$, there are model linkages among learners $\cL_1,\cL_2,\cL_3,$ and $\cL_6$.  The true underlying model linkage graph $G^*$ is illustrated on the right panel of Figure~\ref{fig:linearprior}. 
Note that there is a model linkage between $\cL_2$ and $\cL_3$ since both share the same common parameters with $\cL_1$.
For simplicity, we set the sample size for each learner to be $n$.  
%%%%%%%%%%%%%%%%%%%%%%%%%%%%%%%%%%%%%%%%%
%%%%%%%%%%%%%%%%%%%%%%%%%%%%%%%%%%%%%%%%%

%%%%%%%%%%%%%%%%%%%%%%%%%%%%%%%%%%%%%%%%%%%

 In practice, the user needs to specify a model linkage graph for the six learners and a statistical model for each learner.  
For the numerical studies, we specify correct statistical models for $\kappa=\{1,2,3,4,6\}$, and we misspecify $\cL_5$ by assuming that the covariates are linearly related to the response $y$. 
We further restrict $\beta_j^{(\kappa)}$ to be the same for $\kappa=1,\ldots,5$ and $j=1,\ldots,7$. 
Moreover, we restrict $\beta_{j}^{(2)}=\beta_{j}^{(6)}$ to be the same for $j=8,\ldots,15$.
Hence, the model linkage graph is misspecified in the sense that we assume that there exist model linkages between $\cL_4$ and $\{\cL_1,~\cL_2,~\cL_3,~\cL_5\}$, and between $\cL_5$ and $\{\cL_1,\cL_2,\cL_3\}$.
The user-specified graph $G$ is illustrated on the left panel of Figure~\ref{fig:linearprior}.
We apply Algorithm~\ref{algo_greedy} with the aforementioned user-specified graph and impose a multivariate normal distribution, $N_p(\mathbf{0},4\Ib_p)$, as the prior distribution for the regression coefficients for all learners.

We will compare the proposed Algorithm~\ref{algo_greedy} to fitting the model using data only from $\cL_1$, and using the combined data from $\cL_1$ and $\cL_4$.   Recall that the true regression coefficients in $\cL_4$ are different from that of $\cL_1$, and thus combining data in $\cL_1$ and $\cL_4$ can lead to severely biased estimates of regression coefficients.

\begin{figure}[!htp]
\centering
\includegraphics[scale=0.45]{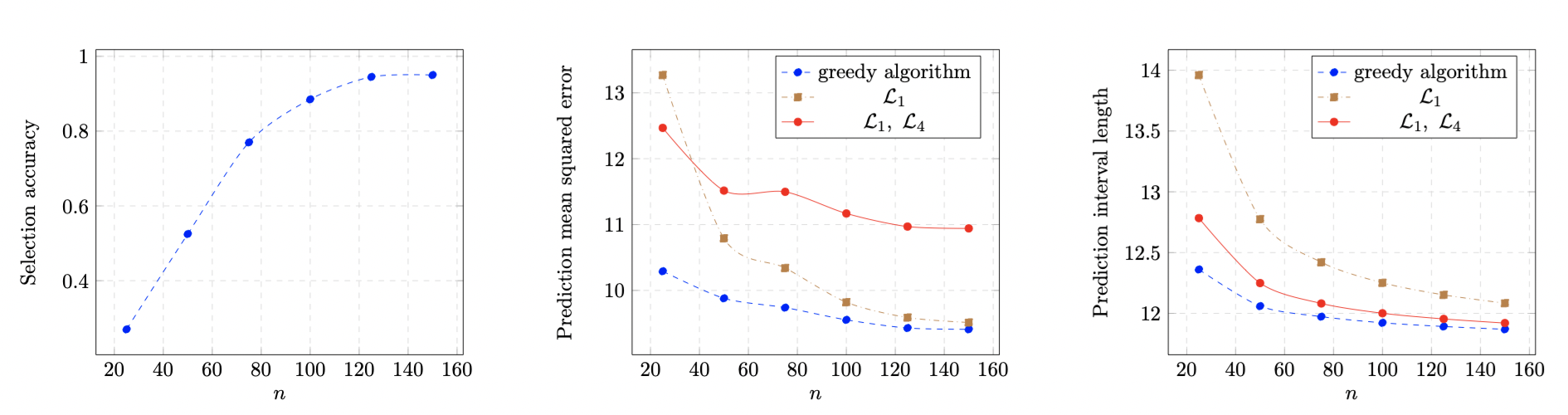}
\caption{The results for modeling based on the proposed algorithm (``greedy algorithm''), modeling of the single agent $\cL_1$ (``$\cL_1$''), and combined modeling of $\cL_1$ and $\cL_4$ (``$\cL_1,\cL_4$'') in the regression experiment. The evaluation uses the selection accuracy (left), prediction mean squared error (middle), and length of the 95\% prediction interval (right), averaged over $200$ replications.  }
\label{linear case1}

\end{figure}

To assess the model linkage selection accuracy, we calculate the selection accuracy as the proportion of times when the estimated model linkage graph $\hat{G}$ from Algorithm~\ref{algo_greedy} is equal to $G^*$. 
To evaluate the performance across different models, we generate $50$ test data for $\cL_1$, and calculate the mean squared error between the predicted response and the true response in the test data. Note that the mean squared error is a surrogate of the predictive logarithmic score under the normal noise assumption. The predicted response is obtained by taking the mean of the posterior predictive distribution for each model.
In addition, we calculate the length of the 95\% prediction interval obtained from the posterior predictive distribution. The results for a range of sample sizes $n \in \{50,75,\ldots,150\}$, averaged over $200$ replications, are shown in Figure~\ref{linear case1}.

From the left panel of Figure~\ref{linear case1}, we see that the selection accuracy from  Algorithm~\ref{algo_greedy} increases as the sample size increases. When $n\approx 120$, the estimated model linkage graph $\hat{G}=G^*$ with probability approximately one, supporting the selection consistency results in Theorem~\ref{thm:consistency}.
Notably, Algorithm~\ref{algo_greedy} yields a consistent model linkage graph even though the user-specified model linkage graph $G$ is misspecified as shown in Figure~\ref{fig:linearprior}.
From the middle panel of Figure~\ref{linear case1}, we see that the proposed greedy algorithm yields the lowest prediction mean squared error across a range of sample sizes $n$.
The results indicate that combining data from $\cL_1$ and $\cL_4$ imprudently can lead to higher prediction mean squared errors. Meanwhile, properly integrating data can improve the predictive performance of a single learner $\cL_1$.   
Finally, the average length of the 95\% prediction interval for the different models is presented in the right panel of Figure \ref{linear case1}.
We see that the proposed algorithm yields the narrowest prediction interval across the range of $n$.

%%%%%%%%%%%%%%%%%%%%%%%%%%%%%%%%%%%%%%%%%%%%%%%%
%%%%%%%%%%%%%%%%%%%%%%%%%%%%%%%%%%%%%%%%%%%%%%%%

%------------------------------                    Logistic regression.                       ------------------------------%
\subsection{Logistic Regression Example}

In this section, we illustrate that the proposed framework can be employed for classification problems.
We perform a numerical study with five learners $\cL_1,\ldots,\cL_5$, with the goal of enhancing the prediction accuracy of $\cL_1$.
For each learner, we generate the covariates independently from a uniform distribution on the closed interval $[-1,1]$. Then, the response variable is generated as the following:
\begin{equation*}
\begin{split}
\mathrm{logit}(\Pr(y_i^{(\kappa)} \mid \xb_i^{(\kappa)}) )= \begin{cases}  (\xb_{i}^{(\kappa)})^\T\bbeta^* & \mathrm{for~} \kappa=1,2,3,4, \\ 
0& \mathrm{for~} \kappa=5,\\
\end{cases}
\end{split}
\end{equation*}  
where $\bbeta^\ast= \{-0.8,-0.5,-0.2,0.1,0.4,0.7,1.0,1.3,1.6\}^\T$.
Learners $\cL_1,~\cL_2$, $\cL_3$, and $\cL_4$ share common parameters $\bbeta^*$, and there are model linkages among $\cL_1$, $\cL_2,\cL_3$, and $\cL_4$.  
Learner $\cL_5$ indicates that $y_i^{(5)}$ follows a Bernoulli distribution with probability 0.5, and is independent of the covariates.
Thus, there are no model linkages between $\cL_5$ and the other learners.   
We again set the sample sizes for all learners to the same $n$ for simplicity.  

\begin{figure}
\includegraphics[scale = 0.5]{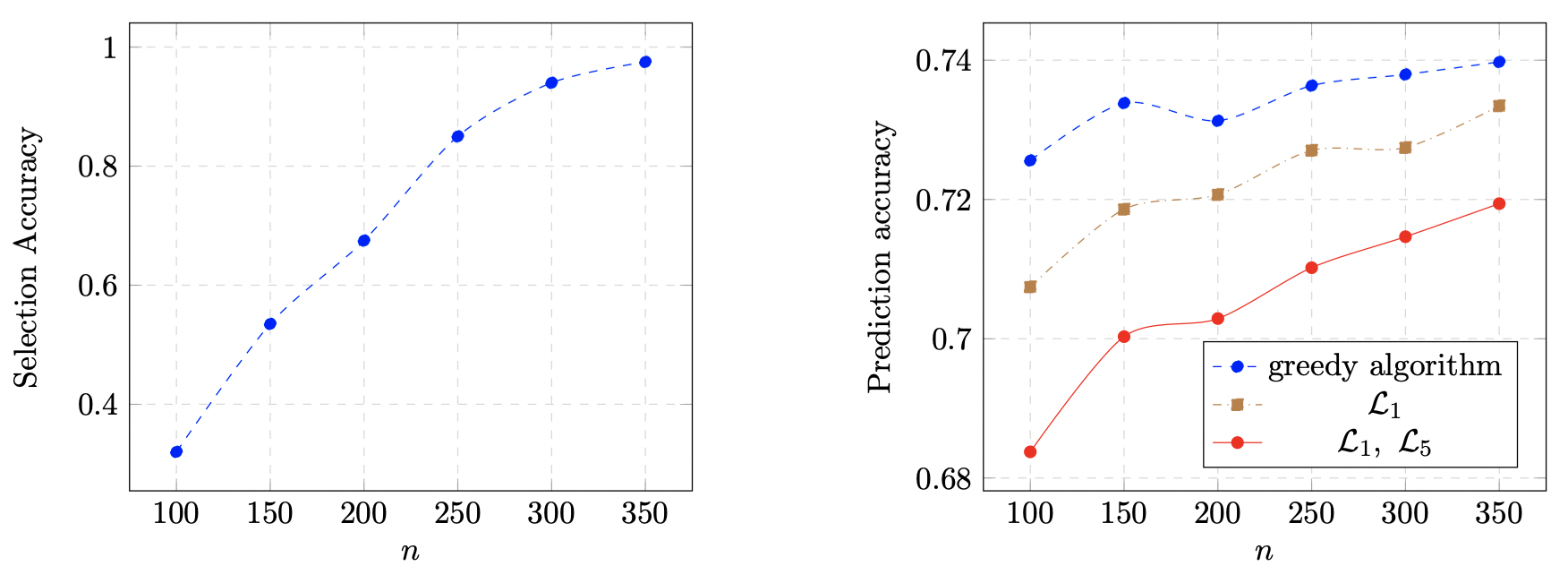}	
\caption{The results for modeling based on the proposed algorithm (``greedy algorithm''), modeling of the single agent $\cL_1$ (``$\cL_1$''), and combined modeling of $\cL_1$ and $\cL_5$ (``$\cL_1,\cL_5$'') in the logistic regression experiment. The evaluation uses selection accuracy and prediction mean squared error, averaged over $200$ replications. }
\label{logistic case1}

\end{figure}

We compare Algorithm~\ref{algo_greedy} to the modeling based on the data from $\cL_1$ and that based on the combined data from $\cL_1$ and $\cL_5$. 
To evaluate the performance across different methods, we calculate the selection accuracy and prediction error. 
For Algorithm~\ref{algo_greedy}, we specify an incorrect model linkage graph where all learners are connected and fit the same logistic regression model for all learners.    
We set a multivariate normal distribution, $\cN (\mathbf 0,4\Ib)$, as the prior distribution for the regression coefficients for all learners.
The results for a range of sample sizes $n=\{ 100,150,\ldots,350\}$, averaged over $200$ replications, are shown in Figure~\ref{logistic case1}.

 From the left panel of Figure~\ref{logistic case1}, we see that the selection accuracy converges to one as we increase the sample size $n$ for each learner.  
 That is, the greedy algorithm chooses not to include information from $\cL_5$, even when the user-specified model linkage graph contains model linkages between $\cL_5$ and the other learners. 
 In addition, the proposed greedy algorithm yields the highest prediction error, whereas the modeling based on the joint data from $\cL_1$ and $\cL_5$ yields the lowest prediction error.

%%%%%%%%%%%%%%%%%%%%%%%%%%%%%%%%%%%%%%
%%%%%%%%%%%%%%%%%%%%%%%%%%%%%%%%%%%%%%
% Figure
%%%%%%%%%%%%%%%%%%%%%%%%%%%%%%%%%%%%%%
%%%%%%%%%%%%%%%%%%%%%%%%%%%%%%%%%%%%%%
%%%%%   Logistic regression greedy      %%%%%%%%

\subsection{Logistic Regression with Data Contamination using Breast Cancer Data}
\label{sec:spam}
Data contamination is an important issue when one decides whether to incorporate information from other learners. 
In practice, when several data sources are collected such that they have the same covariates, users tend to analyze the combined dataset to leverage more information.
However, if certain data sources are corrupted or contaminated, it is crucial to discriminate against them and avoid incorporating information from the contaminated learners.
In this section, we illustrate that Algorithm~\ref{algo_greedy} is robust against data contamination on some data sources.

We consider the Wisconsin Breast Cancer database \citep{doi:10.1287/opre.43.4.570}.  The data consist of a response variable recording whether a cancer tissue is benign or malignant with 9 covariates from a total of  699 subjects. 
We randomly choose 100 samples as the test data for evaluating prediction accuracy.  Then, the data are randomly divided into 10 learners, in which each learner has $n$ samples.    
Since the data from the 10 learners $\cL_1,\ \cL_2,\ldots,\ \cL_{10}$ are subsamples of the original data set, we assume that the regression coefficients are the same across all learners.
We then contaminate the data in $\cL_{10}$ such that the binary response is flipped.

We apply Algorithm~\ref{algo_greedy} with a misspecified model linkage graph by assuming that all learners are linked among each other.   For each learner, we assume a logistic regression model with an intercept and 9 covariates.   For simplicity, we impose the prior distribution $\cN(0,4^2)$ on all of the regression coefficients.  
The prediction accuracy for the proposed method, the method that uses $\cL_1$, and the method that uses the combined data $\cL_1$ and $\cL_{10}$, averaged over $200$ replications, are reported in Table~\ref{tab:wpbc}.

From Table~\ref{tab:wpbc}, we see that naively combining data from $\cL_1$ and $\cL_{10}$ will lead to a much lower prediction accuracy than the model using only data from $\cL_1$. 
Our proposed method, on the other hand, chooses not to incorporate information from $\cL_{10}$. 
Also, by combining data sources adaptively, our proposed method yields a prediction accuracy that is much higher than the model fit using data from $\cL_1$ alone, for both cases of $n=\{25,50\}$.

%----------------   WPBC result   -----------------%
\begin{table}[h!]
  \begin{center}
    \caption{Performance results of the proposed approach, model of $\cL_1$ alone, and joint model of $\cL_1$ and $\cL_{10}$, as evaluated by the prediction accuracy. The results are averaged over $200 $ replications, with $n=\{25,50\}$. }
    \label{tab:wpbc}
    \begin{tabular}{c|c|c|c} % <-- Changed to S here.
      \textbf{number of samples} & \textbf{proposed method} &\textbf{$\cL_1$} & \textbf{$\cL_1$ and $\cL_{10}$}\\
     
      \hline
      $n=25$ & 0.928 & 0.844 & 0.500\\
      $n=50$ & 0.952 & 0.894 & 0.498\\
      
    \end{tabular}
  \end{center}
\end{table}
%-------------------------------------------------%

%%%%    Real Data: KIDNEY PROTEIN DATA      %%%%
\subsection{Integrating Information on Kidney Cancer Data}
In this section, we analyze the kidney cancer data considered in \cite{kidneycancer}.
The kidney cancer data consist of $33$ different types of tumors, with a total number of $n=8108$ samples and up to $p=198$ proteins for the different types of tumors.
Each tumor type may have different number of proteins.
To study the association between patients' survival time and proteins, \citet{kidneycancer} fit an accelerated failure time model with a log-normal assumption, from which they identified eight proteins that are most related to  patients'  survival time.  

We now illustrate that integrating information from related cancer tumors using the proposed method can improve the prediction accuracy of patients' survival time.  
For simplicity, we consider only patients that are not alive at the observed survival time, and fit a linear regression on the log-transformed survival time.   
We consider three types of tumors: (i) kidney renal clear cell carcinoma (KIRC), $\cL_1$, (ii) kidney renal papillary cell carcinoma (KIRP), $\cL_2$, and (iii) uterine corpus endometrial carcinoma (UCEC), $\cL_3$, each of which has $146$, $24$, and $34$ samples, respectively. 
Moreover, we pick up three proteins ``PCADHERIN'', ``GAB2'', and ``HER3\_pY1298'' as the covariates, following~\citep{kidneycancer} .
The three proteins have been well studied and are all well-known for kidney tumor growth and invasion \citep{PCADHERIN,GAB2,HER3pY1298}. 
In particular, the PCADHERIN has been considered as one of the most important proteins for kidney cancer~\citep{kidneycancer}.
Both KIRC and KIRP originate from cells in the proximal convoluted tubules of the nephron~\citep{chen2016multilevel}, and thus it is reasonable to assume that PCADHERIN has a similar effect on the log-transformed survival time of patients with KIRC and KIRP.
On the other hand, PCADHERIN is expected to have a different effect on UCEC, since UCEC is a type of uterine cancer.
Inspired by the above domain knowledge, we set up a model linkage graph in the following way. We assume that there are linkages among KIRP, KIRC, and UCEC by sharing the effect of PCADHERIN on survival time across three tumor types.

We fit a linear regression model with a log-transformed response, namely
$\log(y_i^{(\kappa)}) = \sum_{j=1}^3 x_{ij}^{(\kappa)} \beta_{j}^{(\kappa)} + \epsilon_{i}^{(\kappa)}$, where the indices $i$, $j$, and $\kappa$ denote the $i$th subject, the $j$th protein, and the $\kappa$th tumor type.
For simplicity, we assume that the random noise is normally distributed with different variances to account for heterogeneity across different tumor types, namely $\epsilon_{i}^{(\kappa)}\sim \cN (0,\sigma_\kappa^2)$. Let $\beta_{1}^{(1)}, \beta_{1}^{(2)}, \beta_{1}^{(3)}$ be the regression coefficient for PCADHERIN across the three cancer types, which are linked on the specified linkage graph.

We compare the prediction accuracy of the proposed greedy algorithm with the model using only data from $\cL_1$ (namely KIRC).  To that end, we sample $20$ data points from $\cL_1$ such that the sample sizes across three tumor types are approximately the same. We treat the remaining data points for test purposes.   
The prior distributions of the regression coefficients are assumed to be standard normal, and the prior distributions for the three intercepts are assumed to follow a normal distribution with a mean of $10$ and variance one.  
Moreover, we assume that $\sigma_\kappa^2\sim \text{InvGamma}(2,1)$ for $\kappa=1,2,3$, where $\text{InvGamma}$ denotes the inverse gamma distribution. 
We replicate the experiments above $100$ times by sub-sampling $20$ data observations from $\cL_1$ for training the models.  

The results indicate that Algorithm~\ref{algo_greedy} connects $\cL_2$ to $\cL_1$ in 65\% of the replications. Meanwhile, it connects $\cL_2$ to $\cL_3$ 20\% times. %it is very likely that the two kidney cancer data share the same regression coefficient for PCADHERIN. Among all experiments, our method chooses $\cL_3$ 20\% of the time, indicating that there is less chance that UCEC also shares the same regression coefficient for PCADHERIN than that of $\cL_2$.
Here, Algorithm~\ref{algo_greedy} selects different linkages in multiple replications due to the randomness of finite samples. 
The average prediction mean squared error (with its standard error) of the proposed method is $1.69 (0.027)$. In comparison, the values are $1.74 (0.031)$ if using data only from $\cL_1$, and $1.66 (0.024)$ if using the joint data from $\cL_1,\ \cL_2$ throughout the $100$ replications.
The results imply that collaborating with $\cL_2$ increases the predictive performance of $\cL_1$, and Algorithm~\ref{algo_greedy} tends to favor such a collaboration.

\subsection{Epidemiological Data Study}

In this experiment, we revisit the example illustrated in Figure~\ref{fig:share} of Section~\ref{sec_form}. 
Recall that in this example, $\mathcal{L}_1$ employs $n$ Binomial models $y_i^{(1)}\sim \text{Binomial}(x_i^{(1)}, \theta_{1,i})$, and learner $\mathcal{L}_2$ has poisson models $y_i^{(2)}\sim \text{Poisson}(\theta_{1,i} \theta_2 x_i^{(2)})$, with $i=1,2,\ldots,n$. 
The datasets in two learners are denoted by $\Db^{(1)} = (\yb^{(1)}, \xb^{(1)})$ and $\Db^{(2)} = (\yb^{(2)}, \xb^{(2)})$.
Note that the two learners' models are heterogeneous, and the parameters of $\cL_2$ are not identifiable since $\theta_{1,i} \theta_2 = c \theta_{1,i} \cdot c^{-1} \theta_2$ for any positive constant $c$.  
With a joint modeling of $\cL_1$ and $\cL_2$, the parameters become identifiable. 
We first use the real data from~\citep{maucort2008international}, where each learner has $n=13$ data. For $\mathcal{L}_1$, the population size $x_i^{(1)}$ ranges from $37$ to $700$ and the empirical infection rate $y_i^{(1)}/x_i^{(1)}$ ranges from $0$ to $0.2$. For $\mathcal{L}_2$, the women-years follow up $x_i^{(2)}$ ranges from $20000$ to $550000$, and the number of cancer incidences $y_i^{(2)}$ ranges from $10$ to $700$. In this real data experiment, we divided each $x_i^{(2)}$ by $1000$ for computation easiness. Note that such a scaling does not make an essential differences for parameter estimation. We used $\text{Uniform}(0,1)$ as the prior distribution for each $\theta_{1,i}$ and $\text{Gamma}(5,1)$ for $\theta_{2}$. 
The result of Algorithm~\ref{algo:greedy} is that no linkage is established between $\cL_1$ and $\cL_2$. 

To develop more insights into the nature of the above models and data size, we perform four cases of simulated data experiments. We use $n=13$ and the identical prior distributions as in the real-data experiment. 
In Case 1, we generated simulated data with $\mathbf x^{(1)} = (200, \overbrace{ 1000,...,1000 }^{12})$ and $\mathbf x^{(2)} = (\overbrace{ 1000,...,1000 }^{13})$, $\theta_2=10$, and $\theta_{1,i}=0.1$ for $i=1,2,\ldots,n$. It serves as a case with large data, which tends to produce an accurate parameter estimation. 
Case 2 is similar to Case 1, except that $\mathbf x^{(1)} = (20, \overbrace{ 100,...,100 }^{12})$ and $\mathbf x^{(2)} = (\overbrace{ 100,...,100 }^{13})$. 
The latter two experimental cases are based on parameters estimated from the real data. 
Case 3 simulates the setting where there exists no underlying linkage between two learners. In particular, we simulate ${y}_i^{(1)}$ from the Binomial model with the original population size $x_i^{(1)}$ and the empirical rate ${\theta}_{1,i}$ that is calculated from the original data observations. We simulate $y_i^{(2)}$ from the Poisson model with the expectation that equals the original count observation. %The data generating process for all models indicates that there is no model linkage between two learners. 
In contrast, Case 4 simulates the setting where there exists an underlying linkage between two learners. 
%Suppose the real data in two learners are $\breve{\yb}^{(1)}$, $\breve{\xb}^{(1)}$, $\breve{\yb}^{(2)}$, and $\breve{\xb}^{(2)}$. 
%The data sets are therefore denoted as $\breve{\Db}^{(1)} = (\breve{\yb}^{(1)},\breve{\xb}^{(1)})$ and $\breve{\Db}^{(2)} = (\breve{\yb}^{(2)},\breve{\xb}^{(2)})$.
In particular, we estimate a joint model of the two learners, and use the posterior of $\theta_{1,i}$'s and $\theta_2$ to generate $\mathbf y^{(1)}$ in Binomial models and $\mathbf {y}^{(2)}$ in Poisson models, with the original population sizes $\mathbf x^{(1)}$ and $\mathbf x^{(2)}$.

\begin{table}[tb]
\centering
\caption{Predictive performance of using a joint modeling of $\mathcal{L}_1$ and $\mathcal{L}_2$ (``Joint''), the proposed method (``Proposed''), and a single-agent modeling (``$\mathcal{L}_1$ or $\mathcal{L}_2$ alone''), evaluated for each learner.  The evaluation is based on the expected log-predictive distribution, numerically computed from $1000$ out-sample data and $100$ replications. Standard errors are reported in the parentheses.}
\label{table:hpv_ex}
\resizebox{1\columnwidth}{!}{%
\begin{tabular}{ccccccc}
\toprule
\multirow{2}{*}{}              & \multicolumn{3}{c}{$\mathcal{L}_1$ (Binomial)}      & \multicolumn{3}{c}{$\mathcal{L}_2$ (Poisson)}         \\
                               & Joint & Proposed       & $\mathcal{L}_1$ alone      & Joint   & Proposed        & $\mathcal{L}_2$  alone       \\ \midrule
Case 1      & -3.76(0.006) & -3.86(0.016) & -4.05(0.009) & -3.76(0.002)  & -3.76(0.002)  & -3.76(0.002)  \\ \hline
Case 2       & -4.29(0.033) & -4.38(0.041) & -4.87(0.019) & -4.03(0.011)  & -4.03(0.011)  & -4.03(0.010)  \\ \hline
Case 3     & -8.37(0.069) & -4.58(0.135) & -4.04(0.016) & -3.62(0.009) & -3.63(0.009) & -3.63(0.009) \\ \hline
Case 4 & -3.56(0.033) & -3.56(0.033) & -4.05(0.016) & -3.26(0.008)  & -3.26(0.008)  & -3.31(0.008)  \\ \bottomrule
\end{tabular}
}
\end{table}

To calculate the out-sample test performance of $\cL_1$ or $\cL_2$, whomever is being assisted, we generate $1000$ test data based on the underlying data distributions. In particular, we numerically calculate $\mathbb{E} [\log p(\bm y^f \mid \bm x^f)]$ where $p$ is the predictive distribution and $(\bm y^f, \bm x^f)$ denotes the test data.  
%For the first learner, each binomial model fix the population size 1000, and generate 1000 test data. For the second learner, if there exists model linkage between two learners, the $i$-th poisson model generates the data with mean $\theta_{1,i}^* \theta_2^* \tilde{x}_{i,j}$, where the future data $\tilde{x}_{i,j}$ is fixed as 100 for all $i$ and $j$. If there is no model linkage, the $i$-th poisson model generates the data with mean $\tilde{\theta}_{1,i}^* \theta_2^* x_{i,j}$. Note that without linkage, the value of $\tilde{\theta}_{1,i}^*$ and $\theta_2^*$ is not identifiable, but the product $\tilde{\theta}_{1,i}^* \theta_2^*$ is identifiable.
The results are shown in Table~\ref{table:hpv_ex}. Recall that in Cases 1, 2, and 4, there exists a model linkage between $\mathcal{L}_1$ and $\mathcal{L}_2$. In Case 3, there exists no underlying model linkage.
From the results, the test performance of the proposed method can data-adaptively approach the better performance of two options, namely with and without a linkage through the parameters $\theta_{1,i}$. 
Also, though $\cL_2$ alone is not identifiable, it can improve $\cL_1$'s performance in collaborative cases. 
%using $\mathcal{L}_1$ alone, and close to the results from $\mathcal{L}_1$, $\mathcal{L}_2$ when there is a model linkage between two learners.   

\section{Conclusion and Further Remarks}
\label{sec_con}

With the rapid growth of low-cost data collection devices and decentralized learners, data analysts are faced with an important challenge to integrate information across a set of learners with diverse data sources. 
However, prudently combining data sources and fitting a joint model on all the data sources can lead to biased estimates with low prediction accuracy due to misspecified models or model linkages.  
We proposed a general approach to enhance the predictive performance of learner $\cL_1$ by robustly integrating information from other learners.
As the information is integrated through parameter linkages by sharing parameters, the data sources do not need to be transmitted across learners. As such, the method is naturally compatible with decentralized learning that involves {internet-of-things} requiring low-energy consumption~\citep{da2014internet}, {smart sensors} with limited hardware capacities~\citep{zhou2016learning}, and {decentralized networks} with limited communication bandwidths~\citep{xiao2005universal}. 
We showed that the proposed method can be linkage selection-consistent and asymptotically prediction-efficient.  
The theoretical properties are established under the regime where the number of learners $M$ and the parameter dimensions are fixed.  
An interesting future work is to study the theoretical properties of the proposed framework under the regime in which the number of learners or the dimension of parameters is allowed to diverge with the sample size.

In the following, we briefly describe the connection between the proposed framework and existing methods on data integration and distributed learning.

\textbf{Data Integration}.
Integrating information from different data sources has been studied in the context of data integration (see, for instance, \citealp{tang2016fused,li2018integrative}, and the references therein).
When there is a unified model across multiple data sources, it is possible to improve statistical efficiency through parameter sharing or by fitting a model using the combined data.
For instance, \citet{tang2016fused} employed a fused lasso approach to encourage the regression coefficients for different data sources to be similar. 
\citet{li2018integrative} developed an integrative linear discriminant analysis method by combining different data sources, and showed that the classification accuracy can be improved compared with using a single data source. 
To address multiple parametric models, some earlier work pre-specified certain constraints on latent variables to utilize heterogenous data sources. For example, in the study of  gene regulatory networks, \citet{jensen2007bayesian} proposed a Bayesian hierarchical model to integrate  gene expression data, ChIP binding data, and promoter sequence data to infer statistical relationships between  transcription factors and genes. 
The uniqueness of our work compared with the existing methods is that our proposed method allows for a set of learners with diverse learning objectives and distinct statistical models.  The set of learners share information only through linked parameters of interest.
Therefore, the method in Section~\ref{sec_method} can be used to help any learner to efficiently identify cooperative learners when prior information is lacking.

\citet{lunn2000winbugs} and \citet{Plummer2015} also studied data integrations in the context of cut distributions, which can be seen as a probabilistic version of a two-step estimator. 
The main idea is to cut the propagation from uncertain models to precise models during joint learning to reduce biases propagated from incorrectly specified models~\citep{lunn2009combining,ogle2013feedback}.
\citet{jacob2017better} proposed a predictive score principle for choosing the most appropriate joint modeling approach among the cut, full posterior, prior, and two-step approaches over a set of learners. 
It is possible to incorporate cut distribution into our proposed method to possibly improve the performance in the finite-sample regime.
% The number of possible predictive scores exponentially increases with the number of learners, and thus the enumeration of all possible joint models could be computationally infeasible for many learners. %Moreover, they focused mainly on mean estimation problems.
Nevertheless, the number of possible candidates exponentially increases with the number of learners, and thus the search space of each greedy selection step can be computationally prohibitive. 
%The above toy example only concerns one cut distribution because of the specific way parameters are defined (namely $\bm \theta_1$ for $\cL_1$ and $\theta_2$ for $\cL_2$). In general, there can be many possible directions of `cutting' the information flow, causing a large search space and infeasible computation costs.
Also, a systematic theoretical study of the cut distribution remains a challenging problem.

\textbf{Distributed Learning}.
Data privacy has gained much attention in recent years, especially in distributed learning where data curators do not wish to share the original data. 
It motivates some recent advancements in distributed learning method such as the {federated learning}, where a central server sends the current global statistical model to a set of selected clients, and then each client updates the model parameter with local data and returns the updates to the central server~\citep{shokri2015privacy,konevcny2016federated,diao2020heterofl}.
The objective function for federated learning is typically formulated as
\begin{equation}
\label{eq:FL}
\min_{\btheta} \ F(\bm\theta):=\sum_{\kappa=1}^M F_\kappa(\bm\theta), \ \ \ \text{where} \ \ \ F_\kappa(\bm\theta) =  \sum_{(\xb,y) \in \Db^{(\kappa)}} f(\bm\theta ; \xb,y),
\end{equation}
where $f$ denotes a global loss function, $\bm\theta$ parameterizes a global model to learn, and $\Db^{(\kappa)}$ is a labeled dataset of the $\kappa$th client. %and $n$ is the overall sample size. 
To optimize over $\bm\theta$, each client locally takes several iterations of (stochastic) gradient descent on the current parameter using its local data, and then the server takes a weighted average of the resulting parameters.
Within a similar context, \citet{jordan2019communication} proposed a communication-efficient surrogate likelihood framework  for solving distributed statistical estimation problems, which provably improves upon simple averaging schemes.

In the context of our proposed framework, consider a set of learners each holding a data source $\Db^{(\kappa)}$ and personal objective function $f_{\kappa}$, and a set of optimization constraints $\cC$. 
A frequentist counterpart of our Bayesian approach is to minimize a proper scoring function~\citep{dawid2015bayesian}, e.g., the negative log-likelihood function, added with some form of regularization. 
The unknown parameters can be estimated by solving the following optimization problem
\begin{align}
	\min_{\btheta_1,\ldots,\btheta_M} \ F(\bm\theta)&:=
	\sum_{\kappa=1}^M F_{\kappa}(\btheta_\kappa) + R(\btheta_1,\ldots,\btheta_M), \quad \textrm{ subject to } \cC, \label{eq_stat} \\
	&\textrm{ where } F_{\kappa}(\btheta) =  \sum_{(\xb,y) \in \Db^{(\kappa)}} f_{\kappa}(\btheta; \xb,y),
	\nonumber
\end{align}
where $R$ is a suitably chosen regularization function. 
The federated learning falls into the above formulation, when the models are restricted to be the same among different learners, namely $f_\kappa=f, \btheta_\kappa=\btheta$ for all $\kappa$.
Without the constraint $\cC$ and regularization $R$,~\eqref{eq_stat} is equivalent to optimizing $M$ individual objectives separately.  

Both the developed Bayesian formulation or the frequentist counterpart in (\ref{eq_stat}) can also be regarded as forms of personalized federated learning, where decentralized and heterogeneous agents participate in a joint learning with peer agents to boost local learning performance.
A key characteristic of personalization is that each agent has a specific local task, and consequently agent-specific loss, model, and data. As such, the order of establishing linkages and the selected set of collaborative learners may depend on whom to assist.
This is reflected through the asymmetric nature of the proposed Algorithm~\ref{algo_greedy} in finite-sample regimes. To illustrate this point, we provide a toy example that involves three learners, each with a Gaussian model $y^{(\kappa)} \sim \mathcal{N}(\mu_{\kappa}, 1)$, $\kappa=1,2,3$. It can be regarded as a regression with $x^{(\kappa)} =1$ and unknown parameters $\mu_{\kappa}$'s. Suppose that $G$ is a fully connected graph, and the prior of  $\mu_{\kappa}$ is $\mathcal{N}(0,10^2)$. The observed data are $y^{(1)} =2$, $y^{(2)}=-0.3$, $y^{(3)}=-2$, respectively. It can be verified that if $\cL_1$ is the one to be assisted, Algorithm~\ref{algo_greedy} will link it with $\cL_2$ at the first step; Then, $\cL_1$ and $\cL_2$ are not linked with $\cL_3$, terminating the algorithm. On the other hand, if $\cL_2$ will be assisted, the algorithm first links it with $\cL_3$ and then stops. Consequently, $\cL_1$ and $\cL_2$  will not establish a linkage in that case. The above indicates the asymmetric nature of collaboration: $\cL_1$ being assisted by $\cL_2$ doest not mean that $\cL_2$ can assist $\cL_1$ in the presence of other learners. Nevertheless, it can be verified that there is no such asymmetry in the case of two learners, meaning that $\cL_1$ selecting $\cL_2$ is equivalent to $\cL_2$ selecting $\cL_1$ in Algorithm~\ref{algo_greedy}.

Even in the vanilla federated learning context, considerations of user misspecification or adversarial attacks are relatively new~\citep{bhagoji2019analyzing,jere2020taxonomy}, %\citep{yin2018byzantine,alistarh2018byzantine,cao2019distributed}, 
and the proposed notion of prediction efficiency and selection consistency are readily applicable.
Moreover, in a general statistical learning where parameters may lose interpretability, it is still possible to build linkages among learners to reduce the overall model complexity and generalization errors. 
For example, \citet{DingRRNN,DingController} recently showed that appropriately restricting deep neural network parameters can significantly improve the performance of multi-modal image generation, compared with state-of-the-art methods that train an image generator separately from each data modality. 
From a theoretical perspective, the risk bound can be reduced by restricting the size of the function spaces through $\cC$. When each learner's predictive performance is not severely biased by other learners compared with its reduced variance, it is worth establishing a joint optimization in the form of (\ref{eq_stat}).

\section*{Acknowledgement}
The authors thank the action editor and two anonymous referees for their helpful comments.  
The authors thank Dr.~Veera Baladandayuthapani for sharing the kidney cancer data in \cite{kidneycancer}.
Jiaying Zhou was supported by the the Army Research Laboratory and the Army Research Office under grant number W911NF-20-1-0222, and National Science Foundation under grant number ECCS-2038603.
Jie Ding and Vahid Tarokh were supported by the Office of Naval Research under grant number N00014-18-1-2244. 
Kean Ming Tan was supported by National Science Foundation under grant numbers DMS-1949730 and DMS-2113346, and National Institutes of Health under grant number RF1-MH122833.

%----------------------------------------------    APPENDIX     -----------------------------------------------%

%\newpage
\appendix
\section{Notation and Regularity Conditions}
\label{appendix:regularity}
We start with introducing some regularity conditions needed for the theoretical development. 
These regularity conditions are generalizations of those in \citet{walker1969asymptotic} from scalar to multidimensional vector.
Suppose that each observation $(y_i,\xb_i),\ 1\leq i \leq n,$ is modeled via a joint distribution with a density function $p(y,\xb\mid \bm\theta)$ with respect to a $\sigma$-finite measure $\mu$.  Moreover, $\xb \in \mathbb{R}^k$  is modeled using the density function $h(\xb)$ that is independent of the parameter $\bm \theta= (\theta_1,\theta_2,\ldots,\theta_p)^\T \in \mathbb{R}^p$.
The joint density can thus be written as $p(y,\xb\mid \bm\theta) = p(y\mid \xb,\bm\theta) h(\xb)$.  Let the true conditional density function of $y$ given $\xb$ be $p^*(y\mid \xb)$, and thus the true joint density of $(y, \xb)$ can be written as $p^*(y,\xb)= p^*(y\mid \xb)h^*(\xb)$. Note that $h(\xb)$ and $h^*(\xb)$ are not necessarily the same due to potential model misspecification when modeling $\xb$.

Let $\btheta^*$ be an interior value in the parameter space $\bTheta$ defined as the minimizer of the Kullback-Leibler divergence between $p(y\mid \xb,\btheta^*)$ and $p^*(y\mid \xb)$:

 \begin{align}
 \btheta^* &= \underset{\btheta\in \RR^p}{\argmax} \int_{\{\cY,\cX\}} \log \frac{p(y,\xb\mid \btheta)}{p^*(y,\xb)}p^*(y,\xb)d\mu \nonumber\\
& =\underset{\btheta\in \RR^p}{\argmax} \int_{\{\cY,\cX\}} [\log p(y\mid \xb,\btheta)+\log\{h(\xb)\}]p^*(y\mid \xb)h^*(\xb)d\xb dy \nonumber\\
& = \underset{\btheta\in \RR^p}{\argmax} \int_{\{\cX\}}h^*(\xb)\int_{\{\cY\}} \log p(y\mid \xb,\btheta)p^*(y\mid \xb) dy d\xb\nonumber\\
& = \underset{\btheta\in \RR^p}{\argmax} \int_{\{\cY\}} \log p(y\mid \xb,\btheta)p^*(y\mid \xb) dy.
 \end{align}
Note that when the parametric model $p(y\mid \xb,\btheta)$ is well-specified, $p(y\mid \xb,\btheta^*)= p^*(y\mid \xb)$.
 Let $\ell(\btheta)=\sum_{i=1}^n \log p(y_i,\xb_i\mid \btheta)$ be the log-likelihood function for the $n$ observations and let $\hat\btheta=\underset{\btheta\in \RR^p}{\argmax}\ \ell(\btheta)$ be the maximum likelihood estimator (MLE) of $\btheta$.
%Let $p_\ast$ denote the underlying data generating distribution and $\mathbb E$ its corresponding expectation.  
Let $\bm I_n(\bm\theta)$ be the observed Fisher information matrix with $\left\{ \bm I_n(\bm\theta)\right \} _{i,j}=-\frac{\partial^{2}\ell(\bm\theta)}{\partial \theta_i \partial \theta_j }$, and let $\bm I(\bm\theta)$ be the expected Fisher information matrix for a single observation $(y,\xb)$ with $\left\{ \bm I(\bm\theta)\right \} _{i,j}=-\mathbb{E} \left\{ \frac{\partial^{2}\log p(y,\xb\mid  \bm\theta)}{\partial \theta_i \partial \theta_j }\right \} $.
Let $\{\bm J(\bm\theta)\}_{i,j} = \mathbb E\{ \frac{\partial\log p(y,\xb\mid  \bm\theta)}{\partial \theta_i }\frac{\partial\log p(y,\xb\mid  \bm\theta)}{\partial \theta_j}\}$.
We define $T(n) = \Theta\{f(n)\}$, namely when $n$ is large, there exist some fixed constants $c_1$ and $c_2$ such that $c_1 f(n) \leq |T(n)| \leq c_2 f(n)$.

Some regularity conditions that are needed in the theoretical development are listed in the following. 

\begin{itemize}
\item [(\uppercase\expandafter{\romannumeral 1})] The parameter space $\bm\Theta \subseteq \mathbb R^p$ is compact. % is a closed set of vectors in $\mathbb R^p$.

\item [(\uppercase\expandafter{\romannumeral 2})]The set of points $\{\mathcal Y, \mathcal X\}=\left\{ (y,\xb):p(y,\xb\mid  {\bm\theta})>0\right \} $ is independent of ${\bm\theta}$.

\item [(\uppercase\expandafter{\romannumeral 3})]If ${\bm\theta}_{\alpha}\neq {\bm\theta}_{\beta}$, then $\mu\left\{ (y,\xb):p(y,\xb \mid  {\bm\theta}_{\alpha})\neq p(y,\xb\mid  {\bm\theta}_{\beta})\right \} >0$.

\item[(\uppercase\expandafter{\romannumeral 4})]For all $(y,\xb) \in \{\mathcal Y,\mathcal X\}$ and $\delta>0$, we have $\left |\log p(y,\xb| {\bm\theta})-\log p(y,\xb| {\bm\theta}^\prime)\right | < H_\delta(y,\xb,{\bm\theta}^\prime)$ as long as $\|{\bm\theta}-{\bm\theta}^\prime \|_2<\delta$. Here, function $H_\delta$ has the property that $\underset{\delta \to 0}{\operatorname{\lim}}\ H_\delta(y,\xb,{\bm\theta}^\prime)=0$ and that 
\[
\underset{\delta \to 0}{\operatorname{\lim}}\int_{\{\mathcal Y,\mathcal X\}} H_\delta(y,\xb,{\bm\theta}^\prime)p^*(y,\xb)d\mu=0. \quad
%\forall \tilde{\bm\theta} \in \bm\Theta.
\]
\item[(\uppercase\expandafter{\romannumeral 5})]If $\bm\Theta$ is not bounded, then for any ${\bm\theta}_M \in \bm\Theta,$ and sufficiently large $\Delta,$ we have
\[
\log p(y,\xb|{\bm\theta}) - \log p(y,\xb|{\bm\theta}_M) < K_\Delta(y,\xb,{\bm\theta}_M),
\]
where $\|{\bm\theta}\|_2 > \Delta$ and $K_\Delta$ has the property that
\[
\underset{\Delta \to \infty}{\operatorname{lim}}\int_{\{\mathcal Y, \mathcal X\}} K_\Delta(y,\xb,{\bm\theta}_M)p^*(y,\xb)d\mu < 0.
\]
%Note that this limit needs not to be finite.
\item [(\uppercase\expandafter{\romannumeral 6})]
The maximum likelihood estimator (MLE), denoted by $\hat{\bm\theta}$, exists, and the matrix $\bm I_n(\hat{\bm\theta})$ is positive definite almost surely.
%\end{itemize}I_n

%The next two set of conditions will use the interior value $\btheta^*$.
%\begin{itemize}
\item[(\uppercase\expandafter{\romannumeral 7})]The log-likelihood function $\log p(y,\xb| {\bm\theta})$ is twice continuously differentiable with respect to ${\bm\theta}$ in some neighborhood of $\btheta^*$.  %that is to say, the twice differentiation of $\log p(y,\xb| {\bm\theta})$ is continuous at $\btheta^*$.

\item[(\uppercase\expandafter{\romannumeral 8})] %The derivatives and integrations are exchangeable in our proofs. %
The first and second derivatives with respect to $\btheta$, and the integral of $\log p(y,\xb|\btheta)$, are exchangeable. 
 
\item[(\uppercase\expandafter{\romannumeral 9})]There exists a $\delta>0$ such that 
\[
\left|\frac{\partial^2 \log p(y,\xb\mid \btheta)}{\partial \theta_i\partial \theta_j} -\frac{\partial^2 \log p(y,\xb\mid \btheta^*)}{\partial \theta_i\partial \theta_j}         \right| < M_\delta(y,\xb,\btheta^*)
\]
 for any pair $(i,j)$ and $\|{\bm\theta}-\btheta^*\|_2<\delta$, where the function $M_\delta$ satisfies
\[
\underset{\delta \to 0}{\operatorname{lim}}\int_{\{\mathcal Y,\mathcal X\}} M_\delta(y,\xb,\btheta^*)p^*(y,\xb) d\mu = 0.
\]

\item[(\uppercase\expandafter{\romannumeral 10})]The prior density function is continuous at ${\bm\theta} = \btheta^*$ and $\pi(\btheta^*) > 0.$

\item[(\uppercase\expandafter{\romannumeral 11})]If $\cL_1$ has a well-specified model $p_{\btheta_1}$, and $\cL_2$ has misspecified model $p_{\btheta_2}$, the model linkage (defined in Definition \ref{def:model linkage}) between $\cL_1$ and $\cL_2$ is misspecified (Definition~\ref{def_linkageMis}).
\end{itemize}

Conditions (\uppercase\expandafter{\romannumeral 1})--(\uppercase\expandafter{\romannumeral 5}) ensure that when $\btheta^*$ is an interior value of $\bm \Theta$, $\ell(\bm\theta) - \ell(\btheta^*)$ is sufficiently small for values of $\bm\theta$ that are not in the vicinity of $\btheta^*$, with probability tending to one as $n \to \infty$. 
We note that Condition (\uppercase\expandafter{\romannumeral 1}) is a stronger than necessary assumption made to simplify the technical arguments. Alternatively, one may remove this assumption and show that the parameter estimate falls into a compact set with probability increasing to one as the sample size increases. 
Conditions (\uppercase\expandafter{\romannumeral 6})--(\uppercase\expandafter{\romannumeral 9}) ensure that when $\bm\theta = \btheta^*,\ n^{1/2}(\hat{\bm\theta} - \btheta^*)$ has a limiting distribution $\mathcal{N}\big(\bm 0, \bm I(\btheta^*)^{-1}\bm J(\btheta^*)\bm I(\btheta^*)^{-1}\big)$. 
Conditions (\uppercase\expandafter{\romannumeral 7})--(\uppercase\expandafter{\romannumeral 9}) assume that  $\bm{I}_n(\bm\theta)$ is smooth in the vicinity of $\btheta^*$.  % and the asymptotic normality is used  to show that $\hat{\bm\theta}-\btheta^*=\Theta(n^{-\frac 1 2})$. 
Condition~(\uppercase\expandafter{\romannumeral 11}) is needed to guarantee that learners with misspecified models are not included into the joint model to enhance the statistical performance of $\cL_1$.

%%%%%%%%%%%%%%%%%%%%%%%%%%%%%
%%%%%%%%%%%%%%%%%%%%%%%%%%%%%
% Scoring Rule
%%%%%%%%%%%%%%%%%%%%%%%%%%%%%
%%%%%%%%%%%%%%%%%%%%%%%%%%%%%
\section{Assumptions on the Scoring Rule}
\label{app:assumption}
Let $\yb = \{y_1,y_2,
	\ldots,y_n\}^\T$ be an $n$-dimensional vector of the response and $\Xb = \{\xb_1,\xb_2,\ldots,\xb_n\}^\T$ $\in \RR^{n\times k}$ be the design matrix of covariates.  
In the following, we state some assumptions on the score function $s$ defined in Definition~\ref{def:scoringrule}. 
\begin{itemize}
	\item[(A1)] $\underset{p, y, \xb}{\sup}\,\mathbb E[s(p,y,\xb)]\in (0,\infty)$.
	\item[(A2)] Assume that the model $p(y|\xb,\btheta)$ is well-specified. 
	Let $p(\tilde y|\tilde \xb,\yb,\Xb)$ be the Bayesian predictive distribution of $\tilde y$ given the new predictor $\tilde \xb$ and the data $\yb$, $\Xb $. As $n\to \infty$, 
	\begin{equation}
		\mathbb E \biggl[ s\{p(\tilde{y}|\tilde{\xb},\yb,\Xb),\tilde y,\tilde \xb\}-s\{p(\tilde{y}|\tilde{\xb},\btheta^*),\tilde y,\tilde \xb\} \biggr] = \Theta(n^{-\frac 1 2}).
	\end{equation}
	\end{itemize}
The first assumption indicates that the expected loss is bounded by some positive constant for any given density function and prediction point. The second assumption indicates that the difference between the prediction score and oracle score is at the order of $n^{-1/2}$. 
Many commonly used scoring rules such as the Kullback-Leibler divergence and cross-entropy satisfy the aforementioned assumptions.  

%\textcolor{red}{Include justifications of A2 here}.

%------------------------------        Some lemmas before the proof       ------------------------------%
%------------------------------            Lemma 1                        -------------------------------------%
\section{Proof of Theorems~\ref{thm:consistency}--\ref{thm:efficiency}}
\label{proof:thm1}
We start with some technical lemmas that will be helpful for the proof of Theorems~\ref{thm:consistency}--\ref{thm:efficiency}.   
Let $L(\btheta|\yb,\Xb) = p(\yb,\Xb | {\bm\theta})=\prod_{i=1}^n p(y_i,\xb_{i}| {\bm\theta})$ be the likelihood function for the $n$ observations and let $\ell(\btheta)$ be the log-likelihood function. Let
$p(\yb,\Xb) = \int p(\yb,\Xb | {\bm\theta})\pi(\bm\theta) d\bm\theta$ be the marginal likelihood of $(\yb,\Xb)$.
The following lemma is a multidimensional counterpart of \citet{walker1969asymptotic} that provides limiting properties for the maximum likelihood estimator and marginal likelihood.

 %%%%%%%%%%%%%%%%%%%%%%%%%%%%%%%%
%%%%%%%%%%%%%%%%%%%%%%%%%%%%%%%%
% Lemma Regularity
%%%%%%%%%%%%%%%%%%%%%%%%%%%%%%%%
%%%%%%%%%%%%%%%%%%%%%%%%%%%%%%%%
\begin{lemma}
\label{lemma:regularity}
Assume that the regularity conditions in Appendix \ref{appendix:regularity} hold. 
Let $\btheta^* \in \RR^p$ be an interior value in the parameter space $\bm\Theta$, and assume that the data pair $(y_i, \xb_{i}) \in \mathbb R^{k+1}$ has density $p(y_i,\xb_i|\btheta)$ for $i=1,2,\ldots,n$. 
Let $\hat{\btheta}$ be the maximum likelihood estimator of $\btheta$. 
As $n\to \infty$, the following results hold:
\begin{enumerate}[(i)]
\item \label{lemma:regularity:(a)} Let $N(\delta)=\left\{ {\bm\theta}:\|{\bm\theta}-\btheta^*\|_2<\delta\right \} $ be a neighborhood of $\btheta^*$ contained in ${\bm\Theta}$. For any positive $\delta$, 
there exists a positive number $k(\delta)$ depending on $\delta$ such that
\[
\underset{n \to \infty}{\operatorname{lim}}\ \Pr\left [\underset{{\bm\theta} \in \bm\Theta \setminus N({\delta})}{\operatorname{sup}}n^{-1}\left\{ \ell({\bm\theta})-\ell(\btheta^*)\right\}<-k(\delta)\right ]=1.
\]

\item \label{lemma:regularity:(b)} 
Let ${({\hat \xi}^2)}^{-1}= { \mathrm{det}} |\bm I_n(\hat{\btheta})|$, we have 
\begin{math}
\underset{n \to \infty}{\operatorname{lim}}\ n^{-p}\left\{  {{({\hat \xi}^2)}^{-1}}\right\}= { \rm{det}}|\bm I(\btheta^*)|.
\end{math}
\item \label{lemma:regularity:(c)} \begin{math}\ell(\btheta^*)-\ell(\hat{\bm\theta})=\Theta(1).\end{math}
\item \label{lemma:regularity:(d)} 
\begin{math}\underset{n \to \infty}{\operatorname{lim}}\ \left\{ p(\yb,\Xb | \hat{\btheta})\hat \xi\right\}^{-1}p(\yb,\Xb)=(2\pi)^{\frac p 2}\pi(\btheta^*).\end{math}\\
\end{enumerate}

\end{lemma}
Lemma~\ref{lemma:regularity}(\ref{lemma:regularity:(a)}) indicates that the difference between ${\ell({\bm\theta})}$ and $\ell({\btheta^*})$ will be large when $\bm\theta$ is not in the $\delta$-neighborhood of $\btheta^*$. 
Lemma~\ref{lemma:regularity}(\ref{lemma:regularity:(b)}) establishes that the determinant of the observed Fisher information matrix converges to the determinant of the expected Fisher information matrix. Lemma~\ref{lemma:regularity}(\ref{lemma:regularity:(c)}) shows that the log-likelihood function evaluated at $\btheta^*$ and $\hat\btheta$ are at the same order.
The proof of Lemma~\ref{lemma:regularity} is provided in Appendix~\ref{app:lemma:regularity}.

%------------------------------------------------------------------------------------------------------------------------%

Next, we present a key lemma that provides similar results as those of Lemma~\ref{lemma:regularity},  
but under the setting with non-identically distributed data that arise from two different data sources from two different learners. 
To this end, we define some notation.
Without loss of generality, we consider two learners $\cL_1$ and $\cL_2$ with sample size $n_1$ and $n_2$, respectively.   
In particular, for each learner $\cL_{\kappa}$ with $\kappa=1,2$, the user-specified density function of the data pair
 $(y_i^{(\kappa)},\xb_{i}^{(\kappa)}) \in \mathbb R^{k_\kappa+1}$ 
is denoted as $p_{\btheta_\kappa}^{(\kappa)}(y_i^{(\kappa)},\xb_i^{(\kappa)}|\bm\theta_\kappa)$.
Let $p_{\kappa}^*(y_i^{(\kappa)},\xb_i^{(\kappa)})$ be the true underlying density function. 
Similar to Lemma~\ref{lemma:regularity}, let $\bm\theta_\kappa^* \in \RR^{p_\kappa}$ be an interior value in the parameter space $\bm\Theta_\kappa$.

As defined in Definition \ref{def:model linkage}, let $\btheta_{1,\cS_1}=\btheta_{2,\cS_2}=\btheta_{\cS_1,\cS_2}\in \mathbb R^{p_s}$ be the shared parameter between $\cL_1$ and $\cL_2$.
Moreover, let $\btheta_{\cC} = (\btheta_{1,-\cS_1}^\T, \btheta_{\cS_1,\cS_2}^\T,\btheta_{2,-\cS_2}^\T)^\T \in \mathbb R^{p_\cC}$ be the vector obtained by concatenating  entries of $\btheta_1$ and $\btheta_2$ without duplication, which has a dimension of $p_\cC = p_1+p_2-p_s$. Let $\btheta_{\cC}^* \in \mathbb R^{p_\cC}$ be its corresponding interior value in $\bTheta_\cC$.
 We denote $\tilde \btheta_1 = (\btheta_{1,-\cS_1}^\T,\btheta_{\cS_1,\cS_2}^\T)^\T$ and $\tilde \btheta_2 = (\btheta_{2,-\cS_2}^\T,\btheta_{\cS_1,\cS_2}^\T)^\T$ as the parameters for $\cL_1$ and $\cL_2$ after incorporating information from the model linkage between the two learners.
 We note that $\tilde{\btheta}_1$ and $\tilde{\btheta}_2$ are equivalent to $\btheta_1$ and $\btheta_2$, respectively, after some reordering of the elements.  The notation are simply defined to facilitate the proof of the theoretical results.
  Let $\yb^{(\kappa)} = (y_1,y_2,\ldots,y_{n_\kappa})^\T$ and $\Xb^{(\kappa)} = (\xb_{1}^{(\kappa)},\xb_{2}^{(\kappa)},\ldots,\xb_{n_\kappa}^{(\kappa)})^\T$. 
Denote $L(\tilde\btheta_1|\yb^{(1)},\Xb^{(1)}),\ L(\tilde\btheta_2|\yb^{(2)},\Xb^{(2)})$, and $ L(\btheta_\cC|\yb^{(1)},\yb^{(2)},\Xb^{(1)},\Xb^{(2)})$ as the likelihood functions  of $\cL_1$, $\cL_2$, and $(\cL_1,\ \cL_2)$, where  $L(\tilde\btheta_1|\yb^{(1)},\Xb^{(1)})= p_{\tilde\btheta_1}^{(1)}(\yb^{(1)},\Xb^{(1)}|\tilde\btheta_1)$, $L(\tilde\btheta_2|\yb^{(2)},\Xb^{(2)}) = p_{\tilde\btheta_2}^{(2)}(\yb^{(2)},\Xb^{(2)}|\tilde\btheta_2)$, and
\begin{align*}
L(\btheta_\cC|\yb^{(1)},\yb^{(2)},\Xb^{(1)},\Xb^{(2)})
&=p_{\btheta_\cC}^{(\cC)}(\yb^{(1)},\yb^{(2)},\Xb^{(1)},\Xb^{(2)}|\btheta_\cC) \\
&= p_{\tilde\btheta_1}^{(1)}(\yb^{(1)},\Xb^{(1)}|\tilde\btheta_1)p_{\tilde\btheta_2}^{(2)}(\yb^{(2)},\Xb^{(2)}|\tilde\btheta_2).
\end{align*} 
Let $\ell_1,\ \ell_2$, and $\ell_{\cC}$ be their corresponding log-likelihood functions and let $\hat{\bm\theta}_1,\hat{\bm\theta}_2,$ and $\hat{\bm\theta}_\cC$ be the MLEs obtained from maximizing $\ell_1,\ \ell_2$, and $\ell_\cC$, respectively. Let $\bI_{n_\kappa}^{(\kappa)}(\tilde\btheta_\kappa)$ and $\bI^{(\kappa)}(\tilde\btheta_\kappa)$ be the observed Fisher information matrix and expected Fisher information matrix on single observation, respectively, for $\kappa = 1,2$.
Let $\bI_n(\btheta_{\cC})$ be the matrix with element $\{\bI_n(\btheta_{\cC}) \}_{i,j} = -\frac{\partial^2 \ell_{\cC}(\btheta_{\cC})}{\partial \theta_{\cC,i}\partial \theta_{\cC,j}} $, where $\theta_{\cC,i}$ is the $i$th element of $\btheta_{\cC}$.
Next, let $\pi(\tilde\btheta_1),\ \pi(\tilde\btheta_2)$, and $\pi(\btheta_\cC)$ be the prior density of $\tilde\btheta_1,\ \tilde\btheta_2$, and $ \btheta_\cC$, respectively.
Let $p(\yb^{(\kappa)},\Xb^{(\kappa)}) = \int p_{\tilde\btheta_\kappa}^{(\kappa)}(\yb^{(\kappa)},\Xb^{(\kappa)} | {\tilde\btheta_\kappa})\pi(\tilde\btheta_\kappa) d\tilde\btheta_\kappa$ be the marginal likelihood of $(\yb^{(\kappa)},\Xb^{(\kappa)}),$ for $\kappa=1,2$. 
Moreover, the marginal likelihood of $(\yb^{(1)},\yb^{(2)},\Xb^{(1)},\Xb^{(2)})$ is expressed as 
$$p(\yb^{(1)},\yb^{(2)},\Xb^{(1)},\Xb^{(2)}) = \int p_{\tilde\btheta_1}^{(1)}(\yb^{(1)},\Xb^{(1)} | {\tilde\btheta_1})p_{\tilde\btheta_2}^{(2)}(\yb^{(2)},\Xb^{(2)} | {\tilde\btheta_2})\pi(\bm\theta_\cC) d\bm\theta_\cC.$$
We now present the results in the following lemma.

%%%%%%%%%%%%%%%%%%%%%%%%%%%
%%%%%%%%%%%%%%%%%%%%%%%%%%%
% Lemma 2
%%%%%%%%%%%%%%%%%%%%%%%%%%%
%%%%%%%%%%%%%%%%%%%%%%%%%%%
\begin{lemma}
	\label{lemma:multiple}
	Assume that the regularity conditions in Appendix \ref{appendix:regularity} hold. 
	Suppose that $n=n_1+n_2$, and $n_1/n \to c_1$, $n_2/n \to c_2$, as $n\rightarrow \infty $, where $c_1,c_2\in (0,1)$.   % $n_1/n_2 = \Theta(1)$, 
Then, the following results hold:
\begin{enumerate}[(i)]
\item \label{lemma2:regularity:(a)} Let $N_{\cC}(\delta)=\left\{ {\bm\theta}:\|{\bm\theta}-\btheta_{\cC}^*\|_2<\delta\right \} $ be a neighborhood of $\btheta_{\cC}^*$ contained in ${\bm\Theta_\cC}.$ 
For any $\delta>0$, 
there exists a positive number $k(\delta)$ depending on $\delta$ such that
\[
\underset{n \to \infty}{\operatorname{lim}}\ \Pr\left [\underset{{\bm\theta} \in \bm\Theta_\cC \setminus N_{\cC}({\delta})}{\operatorname{sup}}n^{-1}\left\{ \ell_{\cC}({\bm\theta})-\ell_{\cC}(\btheta_{\cC}^*)\right\}<-k(\delta)\right ]=1.
\]

\item \label{lemma2:regularity:(b)} 
Let ${({\hat\xi_\cC}^2)}^{-1} = \det|\bm I_n(\hat{\bm\theta}_{\cC})|$,
we have 
\begin{math}
\underset{n \to \infty}{\operatorname{lim}}\ n^{-p_{\cC}}\left\{  {{({\hat \xi_\cC}^2)}^{-1}}\right\}= \Theta(1).
\end{math}
\item \label{lemma2:regularity:(c)} \begin{math}\ell_{\cC}({\bm\theta}_\cC^*)-\ell_{\cC}(\hat{\bm\theta}_{\cC})=\Theta(1).\end{math}
\item \label{lemma2:regularity:(d)} 
\begin{math}\underset{n \to \infty}{\operatorname{lim}}\ \left\{ p_{\btheta_\cC}^{(\cC)}(\yb^{(1)},\yb^{(2)},\Xb^{(1)},\Xb^{(2)} | \hat{\btheta}_{\cC})\hat \xi_\cC\right\}^{-1}p(\yb^{(1)},\yb^{(2)},\Xb^{(1)},\Xb^{(2)})=(2\pi)^{\frac {p_\cC} 2}\pi({\bm\theta}_\cC^*).\end{math}\\
\end{enumerate}
\end{lemma}

%%%%%%%%         Lemma 3        %%%%%%%%
Lemmas~\ref{lemma:ratio}--\ref{lemma:misspecified} concern the magnitudes of the joint marginal likelihood and marginal likelihood for each learner, under the case when the model linkage is well-specified or mis-specified.   Similar results were established in the  context of change point detection \citep{marginal}.   
Both lemmas will be used to prove that Algorithm~\ref{algo:greedy} will select the correct model linkage in each iteration of the algorithm.
\begin{lemma}
\label{lemma:ratio}
Assume that the regularity conditions in Appendix \ref{appendix:regularity} hold.  Let $\btheta_{\mathcal S_1,\mathcal S_2} \in \mathbb R^{p_s}$ be the shared parameter between $\cL_1$ and $ \cL_2$ as defined in Definition \ref{def:model linkage}. Let the interior value of $\btheta_{\mathcal S_1, \mathcal S_2}$ in $\cL_1$ and $\cL_2$ be $\btheta_{1,\mathcal S_1}^{*}$ and $\btheta_{2,\mathcal S_2}^{*}$, respectively. If $\btheta_{1,\mathcal S_1}^{*}= \btheta_{2,\mathcal S_2}^{*}$, 
and $n_1/n_2 = \Theta (1)$,
as $m=\min \left\{ n_1,n_2\right \} \to \infty$,  we have
\[
\frac{p(\yb^{(1)},\yb^{(2)}|\Xb^{(1)},\Xb^{(2)})}{p(\yb^{(1)}|\Xb^{(1)})p(\yb^{(2)}|\Xb^{(2)})}\overset{p}{\to} \Theta(m^{\frac{p_s}{2}}).
\]
\end{lemma}  
%%%%            Proof              %%%%

%  Lemma 3: When combine two datasets, but they didn't share parameter.
\begin{lemma}
	\label{lemma:misspecified}
Under the same conditions as in Lemma~\ref{lemma:ratio}, 
if $\btheta_{1,\mathcal S_1}^{*} \neq \btheta_{2,\mathcal S_2}^{*}$, as $\min \{n_1,n_2\} \rightarrow \infty$, we have 
\begin{align}
\frac{p(\yb^{(1)},\yb^{(2)}|\Xb^{(1)},\Xb^{(2)})}{p(\yb^{(1)}|\Xb^{(1)})p(\yb^{(2)}|\Xb^{(2)})}\overset{p}{\to} 0. \label{eq_new3}
\end{align}
\end{lemma}

 %%%%      Proof of Theorem 1     %%%%
 \subsection{Proof of Theorem~\ref{thm:consistency}}
 \begin{proof}
The main idea of the proof is to show that in each iteration, the proposed algorithm will integrate information from one additional learner based on the linkage set $ E^*$, and avoid incorporating information from learner that is not in the linkage set $E^*$.   Consequently, the final output of the algorithm $\hat{E} = E^*$. 
We start the proof at the $\ell$th iteration.

At the $\ell$th iteration, let $\zeta^{(\ell-1)}$ be the set of learners that are already included into the joint model built from the previous $\ell-1$ iterations. 
Let $\zeta_{\cC(G)}^{(\ell-1)} \subseteq \{2,\ldots,M\} \backslash \zeta^{(\ell-1)}$ be a non-empty set of learners with at least a path to $\cL_1$ as defined in the user-specified model linkage graph $G$.  Note that $\zeta_{\cC(G)}^{(\ell-1)}$ is non-empty, otherwise, the algorithm would have been terminated at the $(\ell-1)$th iteration.

There are two cases: (i)  there is at least a $j \in \zeta_{\cC(G)}^{(\ell-1)}$ and an $i\in \zeta^{(\ell-1)}$ such that the model linkage between $\cL_j$ and $\cL_i$ is well-specified; and (ii) there are no well-specified linkages between pairs of learners in  $\zeta_{\cC(G)}^{(\ell-1)}$ and $\zeta^{(\ell-1)}$.
 
 Case (i):
 Suppose that there is a well-specified model linkage between $\cL_j $ for a $j \in\zeta_{\cC(G)}^{(\ell-1)}$ and $\cL_i$ for an $i \in \zeta^{(\ell-1)}$. 
  By Lemma~\ref{lemma:ratio}, Step 2 of Algorithm~\ref{algo:greedy} will hold, and  $\cL_j$ will form a new set with $\zeta^{(\ell-1)}$, namely  $\zeta^{(\ell)} = \zeta^{(\ell-1)}\cup \{j\} $. 
More generally, if there are more than one learner in $\zeta_{\cC(G)}^{(\ell-1)}$ that have well-specified linkages with learners in $\zeta^{(\ell-1)}$, the algorithm will select $j_{\text{opt}}= \text{argmax}_{{j\in \zeta_{\cC(G)}^{(\ell-1)}}}
  p(\cup_{\kappa \in \zeta^{(\ell-1)}} \yb^{(\kappa)}|\yb^{(j)})$ and form a new set  $\zeta^{(\ell)} = \zeta^{(\ell-1)}\cup \{j_{\text{opt}}\}$.

Case (ii): On the other hand, if the model linkages between $\cL_j$ for  $j\in \zeta_{\cC(G)}^{(\ell-1)}$ and $\cL_i$ for  $i\in \zeta^{(\ell-1)}$ are misspecified for all $i$ and $j$, Lemma~\ref{lemma:misspecified} ensures that Algorithm~\ref{algo:greedy} will be terminated, and hence, the misspecified model linkages are not included into $\zeta^{(\ell-1)}$. 
  
As a result, the algorithm terminates when all well-specified linkages are included, and no misspecified linkages and misspecified models will be included, namely $\hat{E}= E^*$. This concludes the proof.
 \end{proof}
 
 %%%%%      Efficiency          %%%%%%%%
  \subsection{Proof of Theorem~\ref{thm:efficiency}}
 \begin{proof} 
 Recall that $\cC(G)$ is the set of indices recording the vertices that form a connected component with learner $\cL_1$ in the user-specified graph $G$, and $G^*$ is the true underlying graph after the statistical models for all learners and $G$ are specified.
 Let $\cG= \{\bar{G} = (V,\bar E):  \bar E \in E,~ \bar{E} \ne  E^*\}$ be a set of graphs that is a subgraph of $G$, but $\bar{E} \ne E^*$.
Let $\hat p_{\cC(G)}$ be the posterior predictive distribution constructed based on learners in $\cC(G)$ as defined in \eqref{eq:joint:predictive}, and let $G_i = (V,E_i)$ be a graph with edge set $E_i$. 
 We consider the prediction efficiency ratio defined as follows:
 \[
 \frac{ \mathbb E\{s(\hat{p}_{\cC(G^*)},\tilde {y},\tilde\xb) - s(p^*,\tilde {y},\tilde \xb)\}}{\Pr(\hat E =  E^*)\mathbb E\{s(\hat{p}_{\cC( G^*)},\tilde{y},\tilde\xb) - s(p^*,\tilde {y},\tilde\xb)\}+\sum_{G_i \in \cG}\Pr(\hat E = E_i)\mathbb E\{s(\hat{p}_{\cC(G_i)},\tilde{y},\tilde\xb) - s(p^*,\tilde {y},\tilde\xb)\}}.
 \]
To prove Theorem~\ref{thm:efficiency}, it suffices to show that
  \begin{equation}
 	\Pr(\hat E =  E^*)  \to 1
 	\label{eq:thm1:1}
 \end{equation}
and 
 \begin{equation}
 	\frac{\Pr(\hat E = E_i)\mathbb E\{s(\hat{p}_{\cC(G_i)},\tilde{y},\tilde\xb) - s(p^*,\tilde {y},\tilde\xb)\}}{\mathbb E\{s(\hat{p}_{\cC(G^*)},\tilde{y},\tilde\xb) - s(p^*,\tilde {y},\tilde\xb)\}}  \to 0
 	\label{eq:45}
 \end{equation}
 for all $G_i\in \cG$.
Equation~\eqref{eq:thm1:1} is a direct consequence of the result in Theorem 1.
In the remaining of the proof, we focus on establishing \eqref{eq:45}.

To show~\eqref{eq:45}, we consider two cases: (1) all learners in $\cC(G_i)$ are well-specified and that the model linkages that form a connected component with $\cL_1$ are well-specified; (2) there exist at least one learner with misspecified models or misspecified model linkages in  $\cC(G_i)$.

For case (1), Assumption (A2) in Section~\ref{app:assumption} indicates that $\mathbb E\{s(\hat{p}_{\cC(G_i)},\tilde{y},\tilde\xb) - s(p^*,\tilde {y},\tilde\xb)\} = \Theta(n_{\cC(G_i)}^{-1/2})$, and $\mathbb E\{s(\hat{p}_{\cC(G^*)},\tilde{y},\tilde\xb) - s(p^*,\tilde {y},\tilde\xb)\} = \Theta(n_{\cC(G^*)}^{-1/2})$, where $n_{\cC(G_i)}$ and $n_{\cC(G^*)}$ are the sample sizes of $\cC(G_i)$ and $\cC(G^*)$, respectively. 
 Combining the above and the result from Theorem~\ref{thm:consistency} that $\Pr(\hat E = E_i)\to 0$ leads to \eqref{eq:45}.
 
For case (2), consider a learner $\cL_w \in \cC(G_i)$ that has misspecified model linkages with some other learners in $\cC(G_i)$, denoted as $\cC(G_i)_{\mathrm{miss}}$.
By definition, $\cC(G_i)_{\mathrm{miss}}\subset\cC(G_i)$. 
If $\hat{E}=E_i$, then 
$G_i$ must  
include the path between  $\cL_w$ and the learners in $\cC( G_i)_{\mathrm{miss}}$, which further indicates that 
 \begin{align}
 \Pr(\hat E = E_i)\le 	\Pr\bigg\{\frac{p(\yb^{(w)},\cup_{\kappa\in \cC(G_i)_{\mathrm{miss}}}\yb^{(\kappa)}|\Xb^{(w)},\cup_{\kappa\in \cC(G_i)_{\mathrm{miss}}}\Xb^{(\kappa)})}{p(\yb^{(w)}|\Xb^{(w)})p(\cup_{\kappa\in \cC(G_i)_{\mathrm{miss}}}\yb^{(\kappa)}|\cup_{\kappa\in \cC(G_i)_{\mathrm{miss}}}\Xb^{(\kappa)})}>1\bigg\},
 	\label{eq:prob}
 \end{align}
 where the right side of the equation can be interpreted as the probability of including $\cL_w$ in Step 2 of Algorithm~\ref{algo:greedy} to build a joint model.
 
From the proof of Lemma~\ref{lemma:misspecified}, we have 
\begin{align}
	\frac{p(\yb^{(w)},\cup_{\kappa\in \cC(G_i)_{\mathrm{miss}}}\yb^{(\kappa)}|\Xb^{(w)},\cup_{\kappa\in \cC(G_i)_{\mathrm{miss}}}\Xb^{(\kappa)})}{p(\yb^{(w)}|\Xb^{(w)})p(\cup_{\kappa\in \cC(G_i)_{\mathrm{miss}}}\yb^{(\kappa)}|\cup_{\kappa\in \cC(G_i)_{\mathrm{miss}}}\Xb^{(\kappa)})} = \Theta\{n_w^{p^\prime/2}\exp(-n_w C_w)\}
	\label{eq:newnew10}
\end{align} 
for some positive finite constants $C_w$ and $p^\prime$, where $n_w$ is the number of samples in $\cL_w$. By an application of the Markov's inequality and \eqref{eq:newnew10}, 
we have 
\begin{align}
&\Pr\bigg\{\frac{p(\yb^{(w)},\cup_{\kappa\in \cC(G_i)_{\mathrm{miss}}}\yb^{(\kappa)}|\Xb^{(w)},\cup_{\kappa\in \cC(G_i)_{\mathrm{miss}}}\Xb^{(\kappa)})}{p(\yb^{(w)}|\Xb^{(w)})p(\cup_{\kappa\in \cC(G_i)_{\mathrm{miss}}}\yb^{(\kappa)}|\cup_{\kappa\in \cC(G_i)_{\mathrm{miss}}}\Xb^{(\kappa)})}>1\bigg\}	\\
&\le \mathbb E\bigg\{\frac{p(\yb^{(w)},\cup_{\kappa\in \cC(G_i)_{\mathrm{miss}}}\yb^{(\kappa)}|\Xb^{(w)},\cup_{\kappa\in \cC(G_i)_{\mathrm{miss}}}\Xb^{(\kappa)})}{p(\yb^{(w)}|\Xb^{(w)})p(\cup_{\kappa\in \cC(G_i)_{\mathrm{miss}}}\yb^{(\kappa)}|\cup_{\kappa\in \cC(G_i)_{\mathrm{miss}}}\Xb^{(\kappa)})}\bigg\} \\
&< n_w^{p^\prime/2}\exp(-\frac 1 2 n_w C_w),
\end{align}
implying $\Pr(\hat E = E_i)<n_w^{p^\prime/2}\exp(-0.5 n_w C_w) $.

Due to the existence of misspecified linkages in $\cC(G_i)$,
Assumption (A2) is no longer applicable to bound $\mathbb E\{s(\hat{p}_{\cC(G_i)},\tilde{y},\tilde\xb) - s(p^*,\tilde {y},\tilde\xb)\}$. We instead employ Assumption (A1) to bound $\mathbb E\{s(\hat{p}_{\cC(G_i)},\tilde{y},\tilde\xb) - s(p^*,\tilde {y},\tilde\xb)\}$ by a finite constant $C'$.  Combining the above, we have
\begin{equation}
\label{eq:emiss}
	{\Pr(\hat E = E_i)\mathbb E\{s(\hat{p}_{\cC(G_i)},\tilde{y},\tilde\xb) - s(p^*,\tilde {y},\tilde\xb)\}}<C^\prime{n_w^{p^\prime/2}\exp(-\frac 1 2 n_w C_w)}.
\end{equation}
By Assumption (A2) in Section~\ref{app:assumption} and \eqref{eq:emiss}, we conclude that 
\begin{align}
\frac{\Pr(\hat E = E_i)\mathbb E\{s(\hat{p}_{\cC(G_i)},\tilde{y},\tilde\xb) - s(p^*,\tilde {y},\tilde\xb)\}}{\mathbb E\{s(\hat{p}_{\cC(G^*)},\tilde{y},\tilde\xb) - s(p^*,\tilde {y},\tilde\xb)\}}\le 
	C^{\prime\prime}n_w^{(1+p^\prime)/2}{\exp(-\frac 1 2 n_w C_w)} \to 0
\end{align}
as $n_w\to \infty$, where $C^{\prime\prime}$ is some positive finite constant. 
This concludes the proof.

 \end{proof}

 %%%%%%%%%%%%%%%%%%%%%%%%%%%%%%%%%%%%%%%%%%%%%%%%%%%%%%%%%%%%%%%%
  %%%%%%%%%%%%%%%%%%%%%%%%%%%%%%%%%%%%%%%%%%%%%%%%%%%%%%%%%%%%%%%%
  % Proof of Lemmas
   %%%%%%%%%%%%%%%%%%%%%%%%%%%%%%%%%%%%%%%%%%%%%%%%%%%%%%%%%%%%%%%%
    %%%%%%%%%%%%%%%%%%%%%%%%%%%%%%%%%%%%%%%%%%%%%%%%%%%%%%%%%%%%%%%% 
\section{Proof of Lemmas~\ref{lemma:regularity}--\ref{lemma:misspecified}}
\subsection{Proof of Lemma~\ref{lemma:regularity}}
\label{app:lemma:regularity}
%%%%%%%%%%%%%%%%%%%%%%%%%%%%%%%%%%%%%
%%%%%%%%%%%%%%%%%%%%%%%%%%%%%%%%%%%%%
% Proof of Lemma 1
%%%%%%%%%%%%%%%%%%%%%%%%%%%%%%%%%%%%%
%%%%%%%%%%%%%%%%%%%%%%%%%%%%%%%%%%%%%
\begin{proof}
\textbf{Proof of Lemma~\ref{lemma:regularity}(\ref{lemma:regularity:(a)}}):
Let $\btheta \in \bTheta \setminus \btheta^*$ and let $Z_i=\log\left\{ p(y_i,\xb_{i}|\bm\theta)/p(y_i,\xb_{i}|\btheta^*)\right\}$ be the log ratio between two joint densities evaluated under $\btheta$ and $\btheta^*$. Let $\mathbb E(Z_i)$ be the expectation of $Z_i$ with respect to the density function $p(y_i,\xb_i\mid \btheta^*)$.  
We start with proving the following intermediate result that is helpful for the proof of Lemma~\ref{lemma:regularity}(\ref{lemma:regularity:(a)}):
\begin{equation}
\label{eq6}
\underset{n \to \infty}{\operatorname{lim}}\ \Pr\left [\frac 1 n \left\{ \ell(\bm\theta)-\ell(\btheta^*)\right\}<-c(\bm\theta)\right ]=1,
\end{equation}
where $c(\btheta)$ is a positive finite number that may depend on $\btheta$. 

We consider two cases when $\mathbb{E}( Z_i )$ is finite and infinite, respectively. 
When $\mathbb{E}( Z_i)$ is finite, it follows  from the Jensen's inequality that
\begin{equation}
	\mathbb{E}(Z_i)<\log \mathbb {E}\left\{ \exp (Z_i)\right\}=0.
	\label{eq:jensen}
\end{equation}
Thus, by the law of large number, we have $$n^{-1} \sum_{i=1}^n Z_i \overset{p}{\to}  \mathbb E(Z_i) <0 ,$$ implying $\Pr\{  n^{-1} \sum_{i=1}^n Z_i \geq 0.5 \mathbb E(Z_i)\} \to 0$. 
That is, $\Pr\{n^{-1} \sum_{i=1}^n Z_i \leq  0.5 \mathbb E(Z_i)\} \to 1.$
Pick $c(\bm\theta)=-0.5 \mathbb {E}(Z_i)$, and \eqref{eq6} is satisfied.
If $\mathbb {E}(Z_i)$ is not finite, then we have $\mathbb {E}(Z_i)= -\infty$. 
Let $Z_i^\ast=\max \{Z_i,k\}$, where $k<0$. Then $\mathbb E(|Z_i^\ast|)<\infty$. 
By the strong law of large number, we obtain
\begin{equation}
\frac 1 n \sum_{i=1}^n Z_i^\ast \overset{a.s.}{\to} \mathbb E(Z_i^\ast).
\label{eq:lln}
\end{equation}
 As $k\to -\infty,$ by the monotone convergence theorem, we obtain $\mathbb E(Z_i^\ast) \to \mathbb E(Z_i) = -\infty$. 
Moreover, \eqref{eq:lln} implies
 \begin{equation}
 	\underset{n\to \infty}{\mathrm{\lim\sup}}\ \frac 1 n \sum_{i=1}^n Z_i = \underset{n\to \infty}{\mathrm{\lim}}\underset{n\geq m}{\mathrm{\sup}}\ \frac 1 m \sum_{i=1}^m Z_i \leq \underset{n\to \infty}{\mathrm{\lim}} \frac 1 n \sum_{i=1}^n Z_i^\ast =  \mathbb E(Z_i^\ast) = -\infty
 	\label{eq:inftylln}
 \end{equation}
 almost surely . 
Thus, we obtain $\underset{n\to \infty}{\mathrm{\lim\sup}} ~n^{-1} \sum_{i=1}^n Z_i = -\infty$ almost surely. 
In other words,  $n^{-1} \sum_{i=1}^n Z_i \overset{a.s.}{\to} -\infty$. Any positive finite number $c(\btheta)$ guarantees that \eqref{eq6} will hold. 

We now apply~\eqref{eq6} to prove Lemma~\ref{lemma:regularity}(\ref{lemma:regularity:(a)}) holds in some open balls, where the union of these finite number of open balls covers $\bTheta \setminus N({\delta})$.
Consider $\btheta_j \in \bTheta$ and let $N_j(\delta_j) = \{\btheta:\|\btheta-\btheta_j\|_2<\delta_j\}$ be a ball of size $\delta_j$ centered at $\btheta_j$.
By the regularity condition (\uppercase\expandafter{\romannumeral 4}) in Appendix \ref{appendix:regularity}, we have 
\begin{equation}
\label{eqnew2}
\begin{split}
	\underset{\bm\theta \in N_j(\delta_j)}{\operatorname{sup}}\ \frac 1 n \left\{ \ell(\btheta)-\ell(\btheta^*) \right\} &= 	\underset{\bm\theta \in N_j(\delta_j)}{\operatorname{sup}}\ \left[\frac 1 n \left\{ \ell(\btheta)-\ell(\btheta_j) \right\}+\frac 1 n \left\{ \ell(\btheta_j)-\ell(\btheta^*) \right\}
\right]\\
&<\frac 1 n\sum_{i=1}^n H_\delta(y_i,\xb_{i},\btheta_j)+\frac 1 n \left\{ \ell(\btheta_j)-\ell(\btheta^*) \right\}.
\end{split}
\end{equation} 
By the weak law of large number, as $n \rightarrow \infty$, we have 
\begin{equation}
\underset{\delta \to 0}{\operatorname{lim}}\ \frac 1 n\sum_{i=1}^n H_\delta(y_i,\xb_{i},\bm\theta_j) \overset{p}{\to}   \mathbb E\{H_\delta(y_i,\xb_{i},\bm\theta_j)\}= 0
 \label{eq:new6}
\end{equation}
Applying \eqref{eq6} with $\btheta=\btheta_j$, we obtain
\begin{equation}
\label{eq:new5}
\underset{n \to \infty}{\operatorname{lim}}\ \Pr\left [\frac 1 n \left\{ \ell(\bm\theta_j)-\ell(\btheta^*)\right\}<-c_j \right ]=1,
\end{equation}
where $c_j$ is a positive constant that depends on $\btheta_j$.
Applying \eqref{eq:new6} and \eqref{eq:new5}, we get the upper bound in \eqref{eqnew2}, which is shown below:
\[
\underset{n \to \infty}{\operatorname{lim}}\ \Pr\left\{ \underset{\bm\theta \in N_j(\delta_j)}{\operatorname{sup}}\frac 1 n \{\ell(\btheta)-\ell(\btheta^*)\}< \frac 1 n\{\ell(\btheta_j)-\ell(\btheta^*)\}+\frac 1 n\sum_{i=1}^n H_\delta(y_i,\xb_{i},\bm\theta_j)
<-c_j \right \} =1.
\]
Then we get
\begin{equation}
\label{eq19}
\underset{n \to \infty}{\operatorname{lim}}\ \Pr\left\{ \underset{\bm\theta \in N_j(\delta_j)}{\operatorname{sup}}\frac 1 n \{\ell(\btheta)-\ell(\btheta^*)\}< -c_j \right \} =1.
\end{equation}
thereafter. 

If $\bm\Theta$ is bounded, the compact set $\bm\Theta \setminus N(\delta)$ can be covered by a finite number of balls, namely $N_1(\delta_1),N_2(\delta_2),\ldots,N_m(\delta_m)$,  centered at $\btheta_1,\btheta_2,\ldots,\btheta_m$, respectively. 
%\textcolor{orange}{(JZ: Jie, I don't get the point why it should be moved before (11).)}
Then Lemma~\ref{lemma:regularity}(\ref{lemma:regularity:(a)}) holds by \eqref{eq19} with
\[
k(\delta)=\min\{c_1,c_2,\ldots,c_m\}.
\]

If $\bm\Theta$ is unbounded, we apply the same argument to the bounded compact set $\bm\Theta \setminus \left\{N(\delta) \cup S(\Delta)\right\}$, where $S(\Delta)=\left\{ \bm\theta: \|\bm\theta\|_2 >\Delta \right \} $ for sufficiently large $\Delta,$ from (\uppercase\expandafter{\romannumeral 5}) in Appendix \ref{appendix:regularity}, we have
\begin{equation} 
\label{eq7}
\underset{\bm\theta \in S(\Delta)}{\operatorname{sup}}\frac 1 n \{\ell(\btheta)-\ell(\btheta^*)\}< \frac 1 n \sum_{i=1}^{n}K_\Delta(y_i,\xb_i,\btheta^*).
  \end{equation}
If $\mathbb E\{K_\Delta(y_i,\xb_i,\btheta^*)\}>-\infty$, by the weak law of large number, we get
$n^{-1} \sum_{i=1}^{n}K_\Delta(y_i,\xb_i,\btheta^*) \overset{p}\to \mathbb E\{K_\Delta(y_i,\xb_i,\btheta^*)\} < 0 $, thus we have
\begin{equation}
	\label{eq:lemma1}
	\underset{n \to \infty}{\operatorname{lim}}\ \Pr\left[ \underset{\bm\theta \in S(\Delta)}{\operatorname{sup}}\frac 1 n \left\{\ell(\btheta)-\ell(\btheta^*)\right\}< \mathbb E\left\{K_\Delta(y_i,\xb_i,\btheta^*)\right\}
  \right] =1.
\end{equation}
Under this case, Lemma~\ref{lemma:regularity}(\ref{lemma:regularity:(a)})
holds with
 $k(\delta)=\min\left[c_1,\ldots,c_m,-\mathbb E\{K_\Delta(y_i,\xb_i,\btheta^*)\}\right].$ 
 
If $\mathbb E\{K_\Delta(y_i,\xb_i,\btheta^*)\}= -\infty$, 
using the similar argument in \eqref{eq:lln} and \eqref{eq:inftylln}, we can derive the conclusion that
\[
\frac 1 n \sum_{i=1}^n K_\Delta(y_i,\xb_i,\btheta^*) \overset{a.s.}{\to} \mathbb E\{K_\Delta(y_i,\xb_i,\btheta^*)\} = -\infty.
\]
Lemma~\ref{lemma:regularity}(\ref{lemma:regularity:(a)})
still holds with
 $k(\delta)=\min\left\{c_1,\ldots,c_m \right\}.$  \\

%%%%%%%%%%%%%%%%%%%%%%%%%%%%%%%%%%%%%%%%%%%%%%%
%%%%%%%%%%%%%%%%%%%%%%%%%%%%%%%%%%%%%%%%%%%%%%%
% Proof of Lemma 1(ii)
%%%%%%%%%%%%%%%%%%%%%%%%%%%%%%%%%%%%%%%%%%%%%%%
%%%%%%%%%%%%%%%%%%%%%%%%%%%%%%%%%%%%%%%%%%%%%%%
\textbf{Proof of Lemma~\ref{lemma:regularity}(\ref{lemma:regularity:(b)}}):
Since $\delta$ can be sufficiently small and $\underset{n \to \infty}{\operatorname{lim}}\ \Pr\left\{ \|\hat{\bm\theta}-\btheta^*\|_2<\delta\right \} =1$, 
we have $n^{-1}|\{ (\bm I_n(\hat{\bm\theta})-\bm I_n(\btheta^*)\} _{i,j}|<n^{-1} \sum_{i=1}^n M_\delta(y_i,\xb_{i},\btheta^*)$ from (\uppercase\expandafter{\romannumeral 9}) in Appendix \ref{appendix:regularity}, the limiting property
 \[
 \underset{n \to \infty}{\operatorname{lim}}\ \frac 1 n \sum_{i=1}^n M_\delta (y_i,\xb_{i},\btheta^*)=\mathbb{E} \left\{ M_\delta (y_i,\xb_{i},\btheta^*)\right \}  \to 0
 \]
 and the weak law of large number imply that 
\begin{align}
	\underset{n \to \infty}{\operatorname{lim}}\ \frac 1 n \left\{ \bm I_n(\hat\btheta)\right \} _{i,j}=\underset{n \to \infty}{\operatorname{lim}}\ \frac 1 n \left\{ \bm I_n(\btheta^*)\right \} _{i,j}\overset{p}{\to} \left\{ \bm I(\btheta^*)\right \} _{i,j},
	\label{eq7.5}
\end{align}
Finally, by continuous mapping theorem, we obtain $n^{-p} { \mathrm{det}}|\bm I_n(\hat{\bm\theta})|\overset{p}{\to} { \mathrm{det}}|\bm I(\btheta^*)|$.\\

%%%%%%%%%%%%%%%%%%%%%%%%%%%%%%%%%%%%%%%%%%%%%%%
%%%%%%%%%%%%%%%%%%%%%%%%%%%%%%%%%%%%%%%%%%%%%%%
% Proof of Lemma 1(iii)
%%%%%%%%%%%%%%%%%%%%%%%%%%%%%%%%%%%%%%%%%%%%%%%
%%%%%%%%%%%%%%%%%%%%%%%%%%%%%%%%%%%%%%%%%%%%%%%
\textbf{Proof of Lemma~\ref{lemma:regularity}(\ref{lemma:regularity:(c)}}): Recall that $\hat{\btheta}$ is the maximum likelihood estimator of $\btheta$.  Thus, $\nabla \ell(\hat{\btheta})=\bm 0$. By a second-order Taylor expansion, for $\btheta^*\in \bTheta$, there exists a $t\in [0,1]$ such that 
\begin{equation}
\label{eq8}
\ell(\btheta^*)=\ell(\hat{\bm\theta})-\frac 1 2(\btheta^*-\hat{\bm\theta})^\T \left[ \bm I_n\{ \hat{\btheta}+ t({\btheta}^*-\hat{\btheta})\}\right ] (\btheta^*-\hat{\bm\theta}).
\end{equation}
It suffices to show $ (\btheta^*-\hat{\bm\theta})^\T \left[ \bm I_n\{ \hat{\btheta}+ t({\btheta}^*-\hat{\btheta})\}\right ] (\btheta^*-\hat{\bm\theta}) = \Theta(1)$.
Since $\hat{\bm\theta}\stackrel{p}{\rightarrow} \btheta^*$, 
by~\eqref{eq7.5} in the proof of Lemma~\ref{lemma:regularity}(\ref{lemma:regularity:(b)}), we have $ n^{-1} [ \bm I_n\{ \hat{\btheta}+ t({\btheta}^*-\hat{\btheta})\}] _{i,j} \overset{p}{\to} \left\{ \bm I(\btheta^*)\right \} _{i,j}$.
Also, we have $\hat{\bm\theta}=\btheta^*+\Theta(n^{-1/2})$.  
Consequently, we obtain
\begin{equation}
	\begin{split}
\label{eq9}
(\btheta^*-\hat{\bm\theta})^\T\left[ \bm I_n\{ \hat{\btheta}+ t({\btheta}^*-\hat{\btheta})\}\right ] (\btheta^*-\hat{\bm\theta})
&\overset{p}{\to}\left\{ n^{\frac 1 2}(\hat{\bm\theta}-\btheta^*)\right \} ^\T\bm I(\btheta^*) \left\{ n^{\frac 1 2}(\hat{\bm\theta}-\btheta^*)\right \}\\
& = \Theta(1)\bm I(\btheta^*)\Theta(1) \\
& =\Theta(1).
\end{split}
\end{equation}

%%%%%%%%%%%%%%%%%%%%%%%%%%%%%%%%%%%%%%%%%%%%%%%
%%%%%%%%%%%%%%%%%%%%%%%%%%%%%%%%%%%%%%%%%%%%%%%
% Proof of Lemma 1(iv)
%%%%%%%%%%%%%%%%%%%%%%%%%%%%%%%%%%%%%%%%%%%%%%%
%%%%%%%%%%%%%%%%%%%%%%%%%%%%%%%%%%%%%%%%%%%%%%%
\textbf{\textbf{Proof of Lemma~\ref{lemma:regularity}(\ref{lemma:regularity:(d)}}):} Recall that $p(\yb,\Xb|\btheta)= \exp \{\ell(\btheta)\}$.  Thus, the marginal likelihood can be written as
\begin{equation}
\small
\begin{split}
p(\mathbf{y,X}) &= \int_{\btheta\in \bTheta} \pi(\btheta) \exp \{\ell(\btheta)\}d\btheta\\
&= p(\yb,\Xb|\hat{\btheta}) \int_{\btheta\in \bTheta} \pi(\btheta)  \exp \{\ell(\btheta)-\ell(\hat{\btheta})\}d\btheta\\
&=p(\yb,\Xb|\hat{\btheta}) \int_{\btheta\in \bTheta\setminus N(\delta)} \pi(\btheta)  \exp \{\ell(\btheta)-\ell(\hat{\btheta})\}d\btheta+p(\yb,\Xb|\hat{\btheta}) \int_{\btheta\in  N(\delta)} \pi(\btheta)  \exp \{\ell(\btheta)-\ell(\hat{\btheta})\}d\btheta\\
&:= I_1+I_2.
\end{split}
\end{equation}
It suffices to show that $\left\{ p(\yb,\Xb | \hat{\bm\theta}) \hat\xi \right\} ^{-1} I_1 \overset{p}{\to} 0$ and $\left\{p(\yb,\Xb | \hat{\bm\theta})\hat\xi\right \} ^{-1}I_2 \overset{p}{\to}  (2\pi)^{p/2}\pi(\btheta^*)$, respectively. 

We first show $\left\{ p(\yb,\Xb | \hat{\bm\theta}) \hat\xi \right\} ^{-1} I_1 \overset{p}{\to} 0$.
Note that 
\[
I_1 = p(\yb,\Xb | \hat{\bm\theta}) \exp\left\{ \ell(\btheta^*)-\ell(\hat{\bm\theta})\right \} \int_{\btheta\in \bm\Theta \setminus N(\delta)}\pi({\bm\theta})\exp\left\{ \ell({\bm\theta})-\ell(\btheta^*)\right \} d{\bm\theta},
\]
We start the proof by conditioning on the event $ \exp \left\{ \ell({\bm\theta})-\ell(\btheta^*)\right \} \le \exp \{-nk(\delta)\}$.
 Thus, 
\begin{equation}
	\begin{split}
	\int_{\btheta\in\bm\Theta \setminus N(\delta)} \pi({\bm\theta}) \exp \left\{ \ell({\bm\theta})-\ell(\btheta^*)\right \} d\btheta&  \leq \exp\left\{ -nk(\delta)\right \} \int_{\btheta \in \bm\Theta \setminus N(\delta)}\pi({\bm\theta})d{\bm\theta}\leq  \exp\left\{ -nk(\delta)\right \}.
\end{split}
\label{eq:I_1 ineq}
\end{equation}
Multiplying $I_1$ with $\left\{ p(\yb,\Xb | \hat{\bm\theta}) \hat\xi \right \} ^{-1}$ and by \eqref{eq:I_1 ineq}, we obtain $\left\{ p(\yb,\Xb | \hat{\bm\theta}) \hat\xi \right \} ^{-1}I_1 \leq  \exp\{\ell(\btheta^*)-\ell(\hat{\bm\theta})\}\hat\xi^{-1}\exp\left\{ -nk(\delta)\right\}$. 
By Lemma~\ref{lemma:regularity}(\ref{lemma:regularity:(c)}), we have $\ell(\btheta^*)-\ell(\hat{\bm\theta})=\Theta(1)$. Moreover, by Lemma~\ref{lemma:regularity}(\ref{lemma:regularity:(b)}) and Slutsky's Theorem, we obtain
\begin{equation}\label{eq11}
\begin{split}
\underset{n \to \infty}{\operatorname{lim}}\ \hat\xi^{-1}\exp\left\{ -nk(\delta)\right \} &=\underset{n \to \infty}{\operatorname{lim}}\ n^{-p/2}(\hat\xi^{-2})^{1/2}n^{p/2}\exp\left\{ -nk(\delta)\right \} \\
&= ( { \mathrm{det}}|\bm I(\btheta^*)|)^{1/2}\underset{n \to \infty}{\operatorname{lim}}\ [\exp\left\{ -nk(\delta) + p \log (n)/2\right \}]\\
&=0.
\end{split}
\end{equation}
Finally, by Lemma~\ref{lemma:regularity}(\ref{lemma:regularity:(a)}), the event $ \exp \left\{ \ell({\bm\theta})-\ell(\btheta^*)\right \} \le \exp \{-nk(\delta)\}$ holds with probability one  as $n\rightarrow \infty$ for any $\btheta \in \bTheta\setminus N(\delta)$.
Combining the above, we have
\[
 \left\{ p(\yb,\Xb | \hat{\bm\theta}) \hat\xi \right \} ^{-1}I_1 \overset{p}{\to} 0.
 \]

%%%%%%%%%%%%%%%%%%%%%%%%%%%%%%%%%%%
%%%%%%%%%%%%%%%%%%%%%%%%%%%%%%%%%%%
% We now prove the second case on I2
%%%%%%%%%%%%%%%%%%%%%%%%%%%%%%%%%%%
%%%%%%%%%%%%%%%%%%%%%%%%%%%%%%%%%%%
Next, we show that $ \left\{ p(\yb,\Xb | \hat{\bm\theta})\hat\xi\right \} ^{-1}I_2 \overset{p}{\to}  (2\pi)^{p/2}\pi(\btheta^*)$. By a second-order Taylor expansion, we have
\begin{equation} 
\label{eq10}
\begin{split}
 \ell({\bm\theta})&=\ell(\hat{\bm\theta})-\frac 1 2 ({\bm\theta}-\hat{\btheta})^\T\bm I_n\{\hat{\btheta}+ t({\btheta}-\hat{\btheta})\}({\bm\theta}-\hat{\btheta})\\
 &=\ell(\hat{\bm\theta})-\frac 1 2({\bm\theta}-\hat{\btheta})^\T\bm I_n(\hat{\bm\theta})({\bm\theta}-\hat{\btheta})-\frac 1 2 ({\bm\theta}-\hat{\btheta})^\T\left[\bm I_n\{\hat{\btheta} + t({\btheta}-\hat{\btheta})\}-\bm I_n(\hat{\bm\theta})\right]({\bm\theta}-\hat{\btheta}).
 \end{split}
\end{equation}
For notational simplicity, let $R_n = 0.5({\bm\theta}-\hat{\btheta})^\T \left[\bm I_n\{\hat{\btheta} + t({\btheta}-\hat{\btheta})\}-\bm I_n(\hat{\bm\theta})\right]({\bm\theta}-\hat{\btheta})$.
By \eqref{eq10}, $I_2$ can be rewritten as
\begin{equation}\label{eq13}
\begin{split}
\emph{$I_2$}=&p(\yb,\Xb| \hat{\bm\theta}) \int_{\btheta\in N(\delta)}\pi({\bm\theta})\exp\left\{ \ell({\bm\theta})-\ell(\hat{\bm\theta})\right \} d{\bm\theta}\\
=&p(\yb,\Xb| \hat{\bm\theta}) \int_{\btheta\in N(\delta)}\pi({\bm\theta})\exp\left [ -\frac 1 2({\bm\theta}-\hat{\btheta})^\T \bm I_n(\hat{\bm\theta})({\bm\theta}-\hat{\btheta})-R_n \right ]d{\bm\theta}.
\end{split}
\end{equation}
Under (\uppercase\expandafter{\romannumeral 10}) in Appendix \ref{appendix:regularity}, given any $\epsilon^\prime >0$, let $\epsilon = 2\epsilon^\prime/\pi(\btheta^*)$, since the prior function $\pi(\btheta)$ is continuous around $\btheta^*$, thus, 
we can choose  a $\delta$ such that 
\begin{equation}
	|\pi({\bm\theta})-\pi(\btheta^*)|< \epsilon^\prime <\epsilon \pi(\btheta^*)  \quad \text{if}\quad {\bm\theta} \in N(\delta).
\label{eq:new4}
\end{equation}
Let 
\begin{equation}\label{eq15}
\emph{$I_3$}=\int_{\btheta\in N(\delta)}\exp\left \{ -\frac 1 2({\bm\theta}-\hat{\btheta})^\T\bm I_n(\hat{\bm\theta}) ({\bm\theta}-\hat{\btheta})-R_n \right \}d{\bm\theta}.
\end{equation}
By~\eqref{eq:new4}, we obtain
\begin{equation}
\label{eq14}
(1-\epsilon)\pi(\btheta^*)\emph{$I_3$}< \left\{ p(\yb,\Xb | \hat{\bm\theta})\right \} ^{-1}\emph{$I_2$} < (1+\epsilon)\pi(\btheta^*)\emph{$I_3$}.
\end{equation}
It suffices to obtain lower and upper bounds for $\hat\xi^{-1}I_3$.

We divide the derivation of upper and lower bounds for $\hat\xi^{-1}I_3$ into two parts, we first show that $\hat\xi^{-1} \int_{\btheta\in N(\delta)}\exp\left \{ -\frac 1 2({\bm\theta}-\hat{\btheta})^\T\bm I_n(\hat{\bm\theta}) ({\bm\theta}-\hat{\btheta})\right \}d{\bm\theta} \to (2\pi)^{p/2}$, and prove later that $|R_n| < \epsilon$ for some proper $\delta$.

Recall that $\hat\xi^{-1} = \det|\bm I_n(\hat{\btheta})|^{1/2}$, and by the regularity condition (\uppercase\expandafter{\romannumeral 6}) in Appendix \ref{appendix:regularity}, $\{\bm I_n(\hat{\bm\theta})\}^{-1}$ exists since $\bm I_n(\hat{\bm\theta})$ is non-singular. 
We have 
\begin{equation}
\begin{split}
&\hat\xi^{-1} \int_{\btheta \in N(\delta)}\exp\left \{-\frac 1 2({\bm\theta}-\hat{\bm\theta})^\T\bm I_n(\hat{\bm\theta}) ({\bm\theta}-\hat{\bm\theta})\right \}d{\bm\theta}\\
&= \sqrt{(2\pi)^p }\int_{\btheta \in N(\delta)}\frac{\exp\left\{ -\frac 1 2({\bm\theta}-\hat{\bm\theta})^\T \bm I_n(\hat{\bm\theta}) ({\bm\theta}-\hat{\bm\theta})\right \} }{\sqrt{(2\pi)^p { \mathrm{det}}|{ \{\bm I_n(\hat{\bm\theta})\}^{-1}}|}}d{\bm\theta}.
\end{split}
\label{eq:cg note}
\end{equation} 
Because the part inside the integral is the density function of multivariate normal distribution $\btheta \sim N(\hat\btheta, \{\bm I_n(\hat{\bm\theta})\}^{-1})$, we have that \eqref{eq:cg note} will be less than
\begin{equation}
\begin{split}
&\sqrt{(2\pi)^p }\int \frac{\exp\left\{ -\frac 1 2({\bm\theta}-\hat{\bm\theta})^\T \bm I_n(\hat{\bm\theta}) ({\bm\theta}-\hat{\bm\theta})\right \} }{\sqrt{(2\pi)^p { \mathrm{det}}|{ \{\bm I_n(\hat{\bm\theta})\}^{-1}}|}}d{\bm\theta}
= (2\pi)^{\frac p 2}.
\end{split}
\label{eq16}
\end{equation}
Moreover, the symmetry property of $\bm I_n(\hat{\bm\theta})$ indicates that there exists a $p\times p$ matrix $\bm V$ such that $\bm I_n(\hat{\bm\theta}) = \bm V^\T \bm V$. Change variable ${\bm\theta}^\prime=\bm V{\bm\theta}$, \eqref{eq:cg note} will be greater than
\begin{equation}
\begin{split}
&\sqrt{(2\pi)^p }\int_{\btheta^\prime \in N^\prime(\delta^\prime)} \frac{\exp\left\{ -\frac 1 2({\bm\theta}^\prime-\bm V\hat{\bm\theta})^\T({\bm\theta}^\prime-\bm V\hat{\bm\theta})\right \} }{\sqrt{(2\pi)^p { \mathrm{det}}|{ \{\bm I_n(\hat{\bm\theta})\}^{-1}}|}} \mathrm{det}|\bm V^{-1}|d{\bm\theta}^\prime \\
&=\sqrt{(2\pi)^p }\int_{\btheta^\prime \in N^\prime(\delta^\prime)}\frac{\exp\left\{ -\frac 1 2({\bm\theta}^\prime-\bm V\hat{\bm\theta})^\T({\bm\theta}^\prime-\bm V\hat{\bm\theta})\right \} }{\sqrt{(2\pi)^p}}d{\bm\theta}^\prime,
\label{eqnew1}
\end{split}
\end{equation}
where $N^\prime(\delta^\prime)=\left\{\bm\theta^\prime: \|\bm\theta^\prime-\bm V\btheta^*\|_2 < \delta^\prime \right\}$, and $\delta^\prime$ is determined by $\delta^\prime=\underset{\bm\theta: \|\bm\theta-\btheta^*\|_2 = \delta}\min\|\bm V\bm\theta-\bm V\btheta^*\|_2$.
We show that $\delta^\prime = \underset{\bm\theta: \|\bm\theta-\btheta^*\|_2 = \delta}\min\|\bm V\bm\theta-\bm V\btheta^*\|_2 \to \infty$.
Let $\delta(\btheta) = \|\bm V\bm\theta-\bm V\btheta^*\|_2$ under the condition that 
$\|\bm\theta-\btheta^*\|_2 = \delta$, if $\delta(\btheta) < \infty$, then $\bm V\bm\theta-\bm V\btheta^*$ and $\btheta-\btheta^*$ are both elementwise finite, and their lengths are both finite number $p$, we conclude from above that $\det|(\bm V\bm\theta-\bm V\btheta^*)(\bm\theta-\btheta^*)^\T|<\infty$. 
However, $\det|\bm V|=(\det|\bm I_n(\hat\btheta)|)^{\frac 1 2}\to \infty$ from Lemma \ref{lemma:regularity}(\ref{lemma:regularity:(b)}),
 thus $\det|(\bm V\bm\theta-\bm V\btheta^*)(\bm\theta-\btheta^*)^\T|=\det|\bm V|\cdot \det|(\bm\theta-\btheta^*)(\bm\theta-\btheta^*)^\T|\to \infty$, which is a contradiction.
  All the above indicates that $\delta (\btheta) \to \infty$ for any $\|\bm\theta-\btheta^*\|_2=\delta$, which concludes $\delta^\prime = \underset{\bm\theta: \|\bm\theta-\btheta^*\|_2 = \delta}\min\|\bm V\bm\theta-\bm V\btheta^*\|_2 \to \infty$.
  Therefore, we have \eqref{eqnew1} converging to $(2\pi)^{p/2}$ in probability. The conclusion $\hat\xi^{-1} \int_{\btheta\in N(\delta)}\exp\left \{ -\frac 1 2({\bm\theta}-\hat{\btheta})^\T\bm I_n(\hat{\bm\theta}) ({\bm\theta}-\hat{\btheta})\right \}d{\bm\theta} \to (2\pi)^{p/2}$ is then derived.

We next derive the upper bound of $|R_n|$.
By the triangle inequality, we have 
\begin{equation}
\begin{split}
\frac 1 n |R_n|&=\frac 1 n\left|\frac 1 2 ({\bm\theta}-\hat{\btheta})^\T[\bm I_n\{\hat{\btheta} + t({\btheta}-\hat{\btheta})\}-\bm I_n(\hat{\bm\theta})]({\bm\theta}-\hat{\btheta})\right| \\
&\leq \frac {1}{2n}\left| ({\bm\theta}-\hat{\btheta})^\T[\bm I_n\{\hat{\btheta} + t({\btheta}-\hat{\btheta})\}-\bm I_n({\btheta}^*)]({\bm\theta}-\hat{\btheta})\right|\\
&\qquad+\frac {1}{2n}\left|({\bm\theta}-\hat{\bm\theta})^\T \left\{ \bm I_n(\btheta^*)-\bm I_n(\hat{\bm\theta})\right \} ({\bm\theta}-\hat{\btheta})\right|.
\end{split}	
\label{ineq:eq1}
\end{equation}

To further derive the upper bound of \eqref{ineq:eq1}, consider the length $p$ vector $\bb=(b_1,b_2,\ldots,b_p)^\T$ and $p\times p$ matrix $\Ab$ with $ \{\Ab\}_{i,j}= a_{i,j},\,{i,j\in\{1,2,\ldots,p\}}$. Let $g(\Ab,\bb) = \tr(\bb^{\T}\Ab \bb)$, this function can be formalized as $g(\Ab,\bb)  = \tr (\bb \bb^\T \Ab) = \sum_{j=1}^{p}\sum_{i=1}^{p} b_ib_j a_{i,j}$. Let $\bb = \btheta - \hat{\btheta}$ and $\Ab = \bm I_n\{\hat{\btheta} + t({\btheta}-\hat{\btheta})\}-\bm I_n({\btheta}^*)$, we have $\|\bb\|_2 = \|(\btheta-\btheta^*)+(\btheta^*-\hat\btheta)\|_2\leq \|\btheta-\btheta^*\|_2+\|\btheta^*-\hat\btheta\|_2\leq \delta$ in probability, because $\btheta \in N(\delta)$ and $\hat{\btheta} - \btheta^* = \Theta(n^{-1/2})$. Thus, $|b_i|\leq \delta$ for $i=1,2,\ldots,p$. 
 We can get the inequality that $|g(\Ab,\bb)|\leq \delta^2 \sum_{j=1}^{p}\sum_{i=1}^{p}|a_{i,j}|$.
From triangle inequality we have $\|\hat{\btheta} + t({\btheta}-\hat{\btheta})-\btheta^*\|_2 \leq t\|\btheta-\btheta^*\|_2 + (1-t)\|\hat\btheta - \btheta^*\|_2 
\overset{p}{\to}t\delta \leq \delta$, thus $\hat{\btheta} + t({\btheta}-\hat{\btheta}) \in N(\delta)$ in probability. From (\uppercase\expandafter{\romannumeral 9}) in Appendix \ref{appendix:regularity}, we have $|a_{i,j}| = |[\bm I_n\{\hat{\btheta} + t({\btheta}-\hat{\btheta})\}-\bm I_n({\btheta}^*)]_{i,j}|
\leq \sum_{i=1}^n M_\delta(y_i,\xb_i,\btheta^*)$. Thus we have $\bb^\T \Ab \bb \leq \delta^2 \sum_{j=1}^{p}\sum_{k=1}^{p}\sum_{i=1}^n M_\delta(y_i,\xb_i,\btheta^*)$. Note that this inequality also holds when $\Ab = \bm I_n(\btheta^*)-\bm I_n(\hat\btheta)$, since $\hat\btheta-\btheta^* = \Theta(n^{-1/2})$, which indicates $\hat\btheta \in N(\delta)$ almost surely, thus (\uppercase\expandafter{\romannumeral 9}) in Appendix \ref{appendix:regularity} can be applied and get the same conclusion as well.
 Based on \eqref{ineq:eq1} and the weak law of large number, $n^{-1}|R_n|$ is less than
\begin{equation}
\frac {1} {n} \delta^2\sum_{j=1}^p\sum_{k=1}^p\sum_{i=1}^n M_\delta(y_i,\xb_{i},\btheta^*)\overset{p}{\to} p^2\delta^2 \mathbb{E} \left\{ M_\delta(y_i,\xb_{i},\btheta^*)\right \}  \quad \text{when}\quad n \to \infty.
\end{equation}
Under (\uppercase\expandafter{\romannumeral 9}) in Appendix \ref{appendix:regularity}, $\underset{\delta \to 0}\lim \ \mathbb{E} \left\{ M_\delta(y_i,\xb_{i},\btheta^*)\right\} = 0$, given the condition that $\delta$ is chosen to make \eqref{eq:new4} hold, then for any $\epsilon>0$, if $\delta$ is also chosen such that
\[\mathbb{E} \left\{ M_\delta(y_i,\xb_{i},\btheta^*)\right \} <\frac{\epsilon}{2 np^2\delta^2}.\]
Therefore
\begin{equation}
\label{neq1}
\underset{n \to \infty}{\operatorname{lim}}\ \Pr\left\{ \underset{{\bm\theta} \in N(\delta)}{\operatorname{sup}}|R_n|<\epsilon\right \} =1.
\end{equation}
Hence, we get the conclusion
\begin{equation}
\label{eq:I_3}
\underset{n \to \infty}{\operatorname{\lim}}\Pr\left\{(2\pi)^{\frac p 2}\exp(-\epsilon) < \hat\xi^{-1}I_3 <(2\pi)^{\frac p 2}\exp(\epsilon)\right\} =1.
\end{equation}
Since $\delta$ can be chosen so that \eqref{eq:I_3} and \eqref{eq14} both hold for arbitrary small $\epsilon$, we deduce the result \[
\underset{n \to \infty}{\operatorname{lim}}\ \Pr\left [(2\pi)^{\frac p 2}\pi(\btheta^*)(1-\epsilon)\exp(-\epsilon)<\left\{ p(\yb,\Xb | \hat{\bm\theta})\hat\xi\right \} ^{-1}\emph{$I_2$}<(2\pi)^{\frac p 2}\pi(\btheta^*)(1+\epsilon)\exp(\epsilon)\right ]=1,
\]
which leads to $ \left\{ p(\yb,\Xb | \hat{\bm\theta})\hat\xi\right \} ^{-1}I_2 \overset{p}{\to}   (2\pi)^{p/2}\pi(\btheta^*)$ when $n\to \infty$.
Then we get the conclusion
\begin{equation}\label{eq17}
\underset{n \to \infty}{\operatorname{lim}}\ \left\{ (\yb,\Xb | \hat{\btheta})\hat\xi\right \} ^{-1}p(\yb,\Xb)=(2\pi)^{\frac p 2}\pi(\btheta^*).
\end{equation}
\end{proof}

\subsection{Proof of Lemma~\ref{lemma:multiple}}
\begin{proof}	
	We start with the proof of Lemma~\ref{lemma:multiple}(\ref{lemma2:regularity:(a)}). 
	Recall that $\bTheta_\cC$ is the parameter space of $\btheta_\cC$, and $\tilde \btheta_1 = (\btheta_{1,-\cS_1}^\T,\btheta_{\cS_1,\cS_2}^\T)^\T$ and $\tilde \btheta_2 = (\btheta_{2,-\cS_2}^\T,\btheta_{\cS_1,\cS_2}^\T)^\T$ as the parameters for $\cL_1$ and $\cL_2$ after incorporating information from the model linkage between the two learners.
	 Let $N_{\kappa}(\delta_\kappa)=\left\{ {\bm\theta}:\|{\bm\theta}-\btheta_{\kappa}^*\|_2<\delta_\kappa\right \}$ be the neighborhood of $\btheta_\kappa^*$, and let $\bTheta_{1,\delta_1} = \{\btheta_{\cC}: \tilde{\btheta}_1 \in N_1(\delta_1),\,\btheta_{2,-\cS_2} = \btheta_{2,-\cS_2}^*\}$ and $\bTheta_{2,\delta_2} = \{\btheta_{\cC}: \tilde{\btheta}_2 \in N_2(\delta_2),\,\btheta_{1,-\cS_1} = \btheta_{1,-\cS_1}^*\}$. The parameter space $\bTheta_\cC$ can be divided into $\bTheta_{1,\delta_1}$, $\bTheta_{2,\delta_2}$, and $\bTheta_{\cC} \setminus (\bTheta_{1,\delta_1}\cup \bTheta_{2,\delta_2})$. 
For a fixed $\delta$, there exist $\delta_1$ and $\delta_2$, such that $\bTheta_{\kappa,\delta_\kappa}\in N_\cC(\delta)$, 
	$\kappa=1,2$. 
	Lemma~\ref{lemma:regularity}(\ref{lemma:regularity:(a)}) indicates that $\underset{{\tilde\btheta_\kappa} \in \bm\Theta_\kappa \setminus N_{\kappa}({\delta_\kappa})}{\operatorname{sup}}n_\kappa^{-1}\left\{ \ell_\kappa({\tilde\btheta_\kappa})-\ell_\kappa(\btheta_{\kappa}^*)\right\}< -k_\kappa(\delta_\kappa)$ for some positive functions $k_\kappa(\delta_\kappa),\,\kappa=1,2.$
	Let $k(\delta) = c_1\cdot k_1(\delta_1)+c_2 \cdot k_2(\delta_2)$. Then, 
	 	\begin{align}
		&\underset{{\bm\theta_\cC} \in \bm\Theta_\cC \setminus N_{\cC}({\delta})}{\operatorname{sup}}n^{-1}\left\{ \ell_{\cC}({\bm\theta_\cC})-\ell_{\cC}(\btheta_{\cC}^*)\right\}\\
		&\leq \underset{{\bm\theta_\cC} \in \bm\Theta_\cC \setminus (\bTheta_{1,\delta_1}\cup \bTheta_{2,\delta_2})}{\operatorname{sup}}n^{-1}\left\{ \ell_{\cC}({\bm\theta_\cC})-\ell_{\cC}(\btheta_{\cC}^*)\right\} \label{eq:lemma2:proof:48}\\
%		&\leq \underset{{\bm\theta_\cC} \in \bm\Theta_\cC \setminus (\bTheta_{1,\delta_1}\cup \bTheta_{2,\delta_2})}{\operatorname{sup}}n^{-1}\left\{ \ell_1({\tilde\btheta_1})-\ell_1(\btheta_{1}^*)\right\} + \underset{{\bm\theta_\cC} \in \bm\Theta_\cC \setminus (\bTheta_{1,\delta_1}\cup \bTheta_{2,\delta_2})}{\operatorname{sup}}n^{-1}\left\{ \ell_2({\tilde\btheta_2})-\ell_2(\btheta_{2}^*)\right\}\\
%		&\leq \underset{{\tilde\btheta_1} \in \bm\Theta_\cC \setminus \bTheta_{1,\delta_1}}{\operatorname{sup}}n^{-1}\left\{ \ell_1({\tilde\btheta_1})-\ell_1(\btheta_{1}^*)\right\} + \underset{{\tilde\btheta_2} \in \bm\Theta_\cC \setminus \bTheta_{2,\delta_2}}{\operatorname{sup}}n^{-1}\left\{ \ell_2({\tilde\btheta_2})-\ell_2(\btheta_{2}^*)\right\}\\
		&\leq \underset{{\tilde\btheta_1} \in \bm\Theta_1 \setminus N_{1}({\delta_1})}{\operatorname{sup}}n^{-1}\left\{ \ell_1({\tilde\btheta_1})-\ell_1(\btheta_{1}^*)\right\} + \underset{{\tilde\btheta_2} \in \bm\Theta_2 \setminus N_{2}({\delta_2})}{\operatorname{sup}}n^{-1}\left\{ \ell_2({\tilde\btheta_2})-\ell_2(\btheta_{2}^*)\right\}\label{eq:lemma2:proof:49}\\
		&=c_1\underset{{\tilde\btheta_1} \in \bm\Theta_1 \setminus N_{1}({\delta_1})}{\operatorname{sup}}n_1^{-1}\left\{ \ell_1({\tilde\btheta_1})-\ell_1(\btheta_{1}^*)\right\} + c_2\underset{{\tilde\btheta_2} \in \bm\Theta_2 \setminus N_{2}({\delta_2})}{\operatorname{sup}}n_2^{-1}\left\{ \ell_2({\tilde\btheta_2})-\ell_2(\btheta_{2}^*)\right\}\label{eq:lemma2:proof:50}\\
		& \leq c_1\times \{-k_1(\delta_1)\}+c_2\times \{-k_2(\delta_2)\}\label{eq:lemma2:proof:51}\\
		& = -k(\delta) ,
\end{align}
where \eqref{eq:lemma2:proof:49} holds by the fact that  $\ell_\cC(\btheta_\cC)-\ell_\cC(\btheta_\cC^*) = \{\ell_1(\tilde\btheta_1)-\ell_1(\btheta_1^*)\}+ \{\ell_2(\tilde\btheta_2)-\ell_2(\btheta_2^*)\}$ and \eqref{eq:lemma2:proof:51} holds by applications of Lemma~\ref{lemma:regularity}(\ref{lemma:regularity:(a)}). Therefore, Lemma~\ref{lemma:multiple}(\ref{lemma2:regularity:(a)}) is proved.

For Lemma~\ref{lemma:multiple}(\ref{lemma2:regularity:(b)}),   
we prove that $n^{-1}\bI_n(\hat\btheta_\cC)$ is positive definite by showing that $n^{-1}\bI_n(\hat\btheta_\cC)$ can be rewritten as a sum of two matrices, namely $ n^{-1}\bI_n(\hat\btheta_\cC)= \bM_1 + \bM_2$, where  $\bM_1$, $\bM_2$ are positive definite matrices. 
The proof can then be concluded by the fact that all elements in $\bM_1$ and $\bM_2$ are bounded by some finite constants.
 
 Recall that $\btheta_{\cC} = (\btheta_{1,-\cS_1}^\T, \btheta_{\cS_1,\cS_2}^\T,\btheta_{2,-\cS_2}^\T)^\T \in \mathbb R^{p_\cC}$.
Let $\hat\btheta_{\cC,\kappa}$ be the MLE of $\tilde\btheta_\kappa$ obtained by maximizing $\ell_\cC$, $\kappa = 1,2$. 
Let $\lambda_\kappa$ be the smallest eigenvalue in $n_\kappa^{-1}\bI_{n_\kappa}^{(\kappa)}(\hat\btheta_{\cC,\kappa})$.  
By Condition VI in Appendix \ref{appendix:regularity}  (\uppercase\expandafter{\romannumeral 6}, 
the positive definite property of $\bI_{n_\kappa}^{(\kappa)}(\hat\btheta_{\cC,\kappa})$ indicates that $\lambda_\kappa>0$, $\kappa=1,2$.
In the following proof, we will focus on constructing $\bM_1$.  Construction of $\bM_2$ is similar as $\bM_1$ and is omitted. 

We now define the elements in $\bM_1$. For $i,j\leq p_1$, let 
\begin{align}
	\{\bM_1\}_{i,j} = c_1 \frac{1}{n_1}\{\bI_{n_1}^{(1)}(\hat\btheta_{\cC,1})\}_{i,j}-\frac {c_1} 2 \lambda_1  
	\label{eq:lm2:1}
\end{align}
for $i=j$ and $i\leq p_1-p_s$, and let
\begin{align}
	\{\bM_1\}_{i,j} = c_1 \frac{1}{n_1}\{\bI_{n_1}^{(1)}(\hat\btheta_{\cC,1})\}_{i,j}  
	\end{align}
for the other elements.

When $p_1+1\leq i \leq p_1+p_2-p_s$ or $p_1+1\leq j \leq p_1+p_2-p_s$, all the elements are zeros except when $i=j$, we set 
\begin{align}
	\{\bM_1\}_{i,j} = \frac {c_2} 2 \lambda_2.
	\label{eq:lm2:2}
\end{align}
By construction, $\bM_1$ is a $2\times 2 $ block diagonal matrix, where each block matrix is positive definite. 
The upper left diagonal block is the difference between $c_1n_1^{-1}\{\bI_{n_1}^{(1)}(\hat \btheta_{\cC,1})\}$ and a  diagonal matrix, with the first $p_1-p_s$ diagonal entries are set to equal $0.5c_1\lambda_1$.  The bottom right block matrix is a diagonal matrix with all diagonal entries equaling $0.5c_2\lambda_2$. Thus $\bM_1$ is positive definite.
The matrix $\bM_2$ is constructed in a similar fashion and can be shown to be positive definite.
  
We now proceed to prove the limiting property of $n_1^{-1}\bI_{n_1}^{(1)}(\hat\btheta_{\cC,1})$. 
We have 
 \begin{equation}
   \frac {1}{n_1}|\{ \bm I_{n_1}^{(1)}(\hat{\bm\theta}_{\cC,1})-\bm I_{n_1}^{(1)}({\bm\theta}_1^*)\} _{i,j}| 
  <\frac {1} {n_1} \sum_{i=1}^{n_1} M_{\delta_1}(y_i^{(1)},\xb_{i}^{(1)},{\bm\theta}_1^*)	
  \label{eq:lm2:3}
\end{equation}
  from (\uppercase\expandafter{\romannumeral 9}) in Appendix \ref{appendix:regularity}. 
  Moreover,  the limiting property
 \begin{align}
 	\underset{n_1 \to \infty}{\operatorname{lim}}\ \frac 1 {n_1} \sum_{i=1}^{n_1} M_{\delta_1}(y_i^{(1)},\xb_{i}^{(1)},{\bm\theta}_1^*)=\mathbb{E}\left\{ M_{\delta_1}(y_i^{(1)},\xb_{i}^{(1)},{\bm\theta}_1^*)\right \}  \to 0
 \end{align}
implies that
\begin{align}
\underset{n_1 \to \infty}{\operatorname{lim}}\ \frac 1 {n_1} \left\{ \bm I_{n_1}^{(1)}(\hat\btheta_{\cC,1})\right \} _{i,j}=\underset{n_1 \to \infty}{\operatorname{lim}}\ \frac 1 {n_1} \left\{ \bm I_{n_1}^{(1)}({\bm\theta}_1^*)\right \} _{i,j}=\left\{ \bm I^{(1)}({\bm\theta}_1^*)\right \} _{i,j}.
\label{eq:lm2:4}
\end{align}
Equations \eqref{eq:lm2:1}--\eqref{eq:lm2:4} imply that $\bM_1$ and $\bM_2$ are positive definite and elementwise finite. 
Combining this with the continuous mapping theorem, we obtain the conclusion that $\det|\bM_1+\bM_2| = \det|n^{-1}\bI_n(\hat \btheta_\cC)|= \Theta(1)$, which concludes Lemma~\ref{lemma:multiple}(\ref{lemma2:regularity:(b)}).

Lemma~\ref{lemma:multiple}(\ref{lemma2:regularity:(c)}) is implied by Lemma~\ref{lemma:regularity}(\ref{lemma2:regularity:(c)}).  The proof of Lemma~\ref{lemma:multiple}(\ref{lemma2:regularity:(d)}) is similar to the proof of Lemma~\ref{lemma:regularity}(\ref{lemma:regularity:(d)}), and is omitted.

%, it is known that $\ell_{\cC}({\bm\theta}_\cC^*)-\ell_{\cC}(\hat{\bm\theta}_{\cC}) = \sum_{\kappa=1}^2 \{\ell_\kappa({\bm\theta}_{\kappa}^*)-\ell_\kappa(\hat{\bm\theta}_{\cC,\kappa})\}$. Combining this and Lemma~\ref{lemma:regularity}(\ref{lemma2:regularity:(c)}) concludes the proof.

%For Lemma~\ref{lemma:multiple}(\ref{lemma2:regularity:(d)}), the argument is similar to the proof of Lemma~\ref{lemma:regularity}(\ref{lemma:regularity:(d)}).
\end{proof}

\subsection{Proof of Lemma~\ref{lemma:ratio}}
\begin{proof}
Recall from Lemma \ref{lemma:regularity}(\ref{lemma:regularity:(d)}) and Lemma \ref{lemma:multiple}(\ref{lemma2:regularity:(d)}) that 
  $(\hat \xi_1^2)^{-1}= { \mathrm{det}} |\bm I_{n_1}^{(1)}(\hat{\bm\theta}_{1}) |$, $(\hat \xi_2^2)^{-1}= { \mathrm{det}} |\bm I_{n_2}^{(2)}(\hat{\bm\theta}_{2}) |$, and $(\hat \xi_\cC^2)^{-1}= { \mathrm{det}} |\bm I_n(\hat{\bm\theta}_{\cC}) |$ with $n = n_1+n_2$.
Let $h_1(\Xb^{(1)})$ and $h_2(\Xb^{(2)})$ be the  density function of $\Xb^{(1)}$ and $\Xb^{(2)}$, respectively.
Since the covariates between two learners are independent, as $m\to \infty$, we have 
\begin{align}
%\begin{split}
 \frac{p(\yb^{(1)},\yb^{(2)}|\Xb^{(1)},\Xb^{(2)})}{p(\yb^{(1)}|\Xb^{(1)})p(\yb^{(2)}|\Xb^{(2)})}
 &=\frac{p(\yb^{(1)},\yb^{(2)},\Xb^{(1)},\Xb^{(2)})/h_1(\Xb^{(1)})h_2(\Xb^{(2)})}{p(\yb^{(1)},\Xb^{(1)})/h_1(\Xb^{(1)})\times p(\yb^{(2)},\Xb^{(2)})/h_2(\Xb^{(2)})} \nonumber \\
  &=\frac{p(\yb^{(1)},\yb^{(2)},\Xb^{(1)},\Xb^{(2)})}{p(\yb^{(1)},\Xb^{(1)})p(\yb^{(2)},\Xb^{(2)})} \nonumber \\
 &=\Theta\left\{ \frac{p_{\btheta_\cC}^{(\cC)}(\yb^{(1)},\yb^{(2)},\Xb^{(1)},\Xb^{(2)})| \hat{\bm\theta}_{\cC})}{p_{\tilde\btheta_1}^{(1)}(\yb^{(1)},\Xb^{(1)}| \hat{\bm\theta}_{1})p_{\tilde\btheta_2}^{(2)}(\yb^{(2)}, \Xb^{(2)}|\hat{\bm\theta}_{2})}\cdot \frac{\hat\xi_\cC}{\hat\xi_1 \hat\xi_2}\right \}, \label{eq18}
%\end{split}
\end{align}
where the third equality holds by an application of Lemma \ref{lemma:regularity}(\ref{lemma:regularity:(d)}) and Lemma \ref{lemma:multiple}(\ref{lemma2:regularity:(d)}).

Since the model linkages are well-specified, the joint density can be factored as 
 \[p_{\btheta_\cC}^{(\cC)}(\yb^{(1)},\yb^{(2)},\Xb^{(1)},\Xb^{(2)}| \bm\theta_\cC^\ast) = p_{\tilde\btheta_1}^{(1)}(\yb^{(1)},\Xb^{(1)}| \bm\theta_1^\ast) p_{\tilde\btheta_2}^{(2)}(\yb^{(2)},\Xb^{(2)}| \bm\theta_2^\ast).
 \] 
 %where
% $\btheta_\cC^\ast,\btheta_1^\ast,\btheta_2^\ast$ are the interior parameters coming from likelihood functions $p_{\btheta_\cC}^{(\cC)}(\yb^{(1)},\yb^{(2)},\Xb^{(1)},\Xb^{(2)}| \bm\theta_\cC)$,
 % $ p_{\btheta_1}^{(1)}(\yb^{(1)},\Xb^{(1)}| \bm\theta_1)$,
 % $ p_{\btheta_2}^{(2)}(\yb^{(2)},\Xb^{(2)}| \bm\theta_2)$, respectively.
Thus, we have
\begin{align}
 &\frac{p_{\btheta_\cC}^{(\cC)}(\yb^{(1)},\yb^{(2)},\Xb^{(1)},\Xb^{(2)}| \hat{\bm\theta}_{\cC})}{p_{\tilde\btheta_1}^{(1)}(\yb^{(1)},\Xb^{(1)}|\hat{\bm\theta}_{1})p_{\tilde\btheta_2}^{(2)}(\yb^{(2)},\Xb^{(2)}|\hat{\bm\theta}_{2})}\nonumber \\
 &=\frac{p_{\btheta_\cC}^{(\cC)}(\yb^{(1)},\yb^{(2)},\Xb^{(1)},\Xb^{(2)}| \hat{\bm\theta}_{\cC})\left\{ p_{\btheta_\cC}^{(\cC)}(\yb^{(1)},\yb^{(2)},\Xb^{(1)},\Xb^{(2)}| \bm\theta_\cC^\ast)\right \} ^{-1}}{p_{\tilde\btheta_1}^{(1)}(\yb^{(1)},\Xb^{(1)}| \hat{\bm\theta}_{1})\left\{ p_{\tilde\btheta_1}^{(1)}(\yb^{(1)},\Xb^{(1)} | {\bm\theta}_1^\ast)\right \} ^{-1}p_{\tilde\btheta_2}^{(2)}(\yb^{(2)},\Xb^{(2)}| \hat{\bm\theta}_{2})\left\{ p_{\tilde\btheta_2}^{(2)}(\yb^{(2)},\Xb^{(2)}| {\bm\theta}_2^\ast)\right \} ^{-1}} \nonumber \\
 &=\Theta(1), \label{eq_new2}
\end{align}
where the second equality holds by Lemma~\ref{lemma:regularity}(\ref{lemma:regularity:(c)}). 

Also, Lemma \ref{lemma:regularity}(\ref{lemma:regularity:(b)}) and Lemma \ref{lemma:multiple}(\ref{lemma2:regularity:(b)}) imply that
\begin{align}
\frac{\hat\xi_\cC}{\hat\xi_1 \hat\xi_2}=\frac{\hat\xi_\cC (n_1+n_2)^{\frac {p_1+p_2-p_s}{2}}}{\hat\xi_1 n_1^{\frac {p_1}{2}}\cdot \hat\xi_2 n_2^{\frac{p_2}{2}}}\cdot{\frac{n_1^{\frac{p_1}{2}}n_2^{\frac {p_2}{2}}}{(n_1+n_2)^{\frac {p_1+p_2-p_s}{2}}}}\overset{p}{\to} \Theta(1)\cdot {\frac{n_1^{\frac{p_1}{2}}n_2^{\frac {p_2}{2}}}{(n_1+n_2)^{\frac {p_1+p_2-p_s}{2}}}}=\Theta(m^{\frac {p_s} 2}).\label{eq_new1}
\end{align}
Substituting (\ref{eq_new2}) and (\ref{eq_new1}) into (\ref{eq18}) concludes the proof.
\end{proof}

\subsection{Proof of Lemma~\ref{lemma:misspecified}}

\begin{proof}
From the proof of Lemma~\ref{lemma:ratio}, we have 
\[
	\frac{p(\yb^{(1)},\yb^{(2)}|\Xb^{(1)},\Xb^{(2)})}{p(\yb^{(1)}|\Xb^{(1)})p(\yb^{(2)}|\Xb^{(2)})}	= \frac{p(\yb^{(1)},\yb^{(2)},\Xb^{(1)},\Xb^{(2)})}{p(\yb^{(1)},\Xb^{(1)})p(\yb^{(2)},\Xb^{(2)})},
\]
and it remains to show that 
\begin{equation}
\frac{p(\yb^{(1)},\yb^{(2)},\Xb^{(1)},\Xb^{(2)})}{p(\yb^{(1)},\Xb^{(1)})p(\yb^{(2)},\Xb^{(2)})}\overset{p}{\to} 0.
\end{equation}
By definition, $p(\yb^{(1)},\yb^{(2)},\Xb^{(1)},\Xb^{(2)}) = \int_{\bTheta_{\cC}} p_{\tilde\btheta_1}^{(1)}(\yb^{(1)},\Xb^{(1)}|\tilde\btheta_1)p_{\tilde\btheta_2}^{(2)}(\yb^{(2)},\Xb^{(2)}|\tilde\btheta_2)\pi(\btheta_\cC)d\btheta_\cC$.
Choose a $\delta>0$ such that $N_1(\delta)$ and $ N_2(\delta)$ are non-overlapping neighborhoods of $\btheta_1^*$ and $\btheta_2^*$, respectively. We split $p(\yb^{(1)},\yb^{(2)},\Xb^{(1)},\Xb^{(2)})$ into three integrals, $I_1, \ I_2$, and $I_3$, taken on sets
$\bTheta_{1,\delta}$, $\bTheta_{2,\delta}$, and $\bTheta_{\cC} \setminus (\bTheta_{1,\delta}\cup \bTheta_{2,\delta})$, where $\bTheta_{1,\delta} = \{\btheta_{\cC}: \tilde{\btheta}_1 \in N_1(\delta)\}$ and $\bTheta_{2,\delta} = \{\btheta_{\cC}: \tilde{\btheta}_2 \in N_2(\delta)\}$.  Note that $\tilde{\btheta}_1 \subseteq \btheta_{\cC}$ and $\tilde{\btheta}_2 \subseteq \btheta_{\cC}$

For the first integral, we have
\begin{equation}
	\begin{split}
		I_1 &= \int_{\bTheta_{1,\delta}} p_{\tilde\btheta_1}^{(1)}(\yb^{(1)},\Xb^{(1)}|\tilde\btheta_1)p_{\tilde\btheta_2}^{(2)}(\yb^{(2)},\Xb^{(2)}|\tilde\btheta_2)\pi(\btheta_\cC)d \btheta_\cC \\
		&=p_{\tilde\btheta_2}^{(2)}(\yb^{(2)},\Xb^{(2)} | \hat{\bm\theta}_2)\hat\xi_2 \exp\left\{ \ell_2({\bm\theta}_2^*)-\ell_2(\hat{\bm\theta}_2)\right \}\\
		&\times \int_{\bTheta_{1,\delta}} \hat\xi_2^{-1}\exp\left\{ \ell_2({\tilde\btheta_2})-\ell_2({\bm\theta}_2^*)\right \}p_{\tilde\btheta_1}^{(1)}(\yb^{(1)},\Xb^{(1)}|\tilde\btheta_1)\pi({\bm\theta_\cC}) d{\bm\theta_\cC}.\\
		\end{split}
		\label{eq:44}
\end{equation}
Since $\tilde{\btheta}_2 \notin N_1(\delta)$ in \eqref{eq:44}, 
according to Lemma~\ref{lemma:regularity}(\ref{lemma:regularity:(a)}), the integral on the right hand side in \eqref{eq:44} is less than 
\begin{equation}
\begin{split}
& \int_{\bTheta_{1,\delta}} \hat\xi_2^{-1}\exp\left\{ \ell_2({\tilde\btheta_2})-\ell_2({\bm\theta}_2^*)\right \}p_{\tilde\btheta_1}^{(1)}(\yb^{(1)},\Xb^{(1)}|\tilde\btheta_1)\pi({\bm\theta_\cC}) d{\bm\theta_\cC}\\	 &\le\hat\xi_2^{-1}\exp\{ -n_2k_2(\delta)\}\int_{\bTheta_{1,\delta}} p_{\tilde\btheta_1}^{(1)}(\yb^{(1)},\Xb^{(1)}|\tilde\btheta_1)\pi({\bm\theta_\cC}) d{\bm\theta_\cC}\\
	 & \leq\hat\xi_2^{-1}\exp\{ -n_2k_2(\delta)\}\int_{\bTheta_{1}} p_{\tilde\btheta_1}^{(1)}(\yb^{(1)},\Xb^{(1)}|\tilde\btheta_1)\pi({\tilde{\btheta}_1}) d{\tilde{\btheta}_{1}}\\
	 &= \{n_2^{p_2} \hat\xi_2^2\}^{-\frac 1 2}n_2^{\frac {p_2} 2}\exp\{ -n_2k_2(\delta)\} p(\yb^{(1)},\Xb^{(1)})\\
\end{split}
\label{eq1:lemma3}
\end{equation}
with probability tending to 1 as $n_2\rightarrow \infty$. 
From Lemmas~\ref{lemma:regularity}(\ref{lemma:regularity:(b)})--(\ref{lemma:regularity:(d)}), as $n_2\to \infty$, we have 
\begin{equation}
\begin{split}
	&\{n_2^{p_2} \hat\xi_2^2\}^{-\frac 1 2} \overset{p}{\to}(\det|\bI^{(2)}(\btheta_2^*)|)^{\frac 1 2};\\
	&\exp\left\{ \ell_2({\bm\theta}_2^*)-\ell_2(\hat{\bm\theta}_2)\right \} \to \Theta(1);\\
	&\{p_{\tilde\btheta_2}^{(2)}(\yb^{(2)},\Xb^{(2)} | \hat{\bm\theta}_2)\hat\xi_2\}^{-1}p(\yb^{(2)},\Xb^{(2)}) \overset{p}{\to} (2\pi)^{\frac {p_2}{2}}\pi(\btheta_2^*).
	\end{split}
	\label{eq2:lemma3}
\end{equation}
It follows that 
\begin{equation}
	\frac{I_1}{p(\yb^{(1)},\Xb^{(1)})p(\yb^{(2)},\Xb^{(2)})} =  \Theta(n_2^{\frac {p_2} 2}\exp\{-n_2k_2(\delta)\})\overset{p}{\to}0.
\end{equation}
Using a similar argument for $I_2$, as $n_1 \to \infty$, we have
\begin{equation}
	\frac{I_2}{p(\yb^{(1)},\Xb^{(1)})p(\yb^{(2)},\Xb^{(2)})} =  \Theta(n_1^{\frac {p_1} 2}\exp\{-n_1k_1(\delta)\})\overset{p}{\to}0.
\end{equation}
For the integral $I_3$, we apply a similar argument as in the proof of $I_1$ and $I_2$.  Specifically, 
\[
\begin{split}
		I_3 &= \int_{\bTheta_{\cC}\setminus \{\bTheta_{1,\delta}\cup\bTheta_{2,\delta}\}} p_{\tilde\btheta_1}^{(1)}(\yb^{(1)},\Xb^{(1)}|\tilde\btheta_1)p_{\tilde\btheta_2}^{(2)}(\yb^{(2)},\Xb^{(2)}|\tilde\btheta_2)\pi(\btheta_\cC)d\btheta_\cC \\
		& = p_{\tilde\btheta_1}^{(1)}(\yb^{(1)},\Xb^{(1)} | \hat{\bm\theta}_1)\hat\xi_1 p_{\tilde\btheta_2}^{(2)}(\yb^{(2)},\Xb^{(2)} | \hat{\bm\theta}_2)\hat\xi_2 \exp\left\{ \ell_1({\bm\theta}_1^*)-\ell_1(\hat{\bm\theta}_1)\right \}\exp\left\{ \ell_2({\bm\theta}_2^*)-\ell_2(\hat{\bm\theta}_2)\right \}\\
		&\times \int_{\bTheta_{\cC}\setminus \{\bTheta_{1,\delta}\cup\bTheta_{2,\delta}\}} \hat\xi_1^{-1}\hat\xi_2^{-1}\exp\left\{ \ell_1({\tilde\btheta_1})-\ell_1({\btheta}_1^*)\right \}\exp\left\{ \ell_2({\tilde\btheta_2})-\ell_2({\bm\theta}_2^*)\right \}\pi({\bm\theta_\cC}) d{\bm\theta_\cC},\\
		&\leq p_{\tilde\btheta_1}^{(1)}(\yb^{(1)},\Xb^{(1)} | \hat{\bm\theta}_1)\hat\xi_1 p_{\tilde\btheta_2}^{(2)}(\yb^{(2)},\Xb^{(2)} | \hat{\bm\theta}_2)\hat\xi_2 \exp\left\{ \ell_1({\bm\theta}_1^*)-\ell_1(\hat{\bm\theta}_1)\right \}\exp\left\{ \ell_2({\bm\theta}_2^*)-\ell_2(\hat{\bm\theta}_2)\right \}\\
		&\times \hat\xi_1^{-1}\hat\xi_2^{-1}\int_{\bTheta_\cC}\exp\left\{ \ell_1({\tilde\btheta_1})-\ell_1({\bm\theta}_1^*)\right \}\exp\left\{ \ell_2({\tilde\btheta_2})-\ell_2({\bm\theta}_2^*)\right \}\pi(\btheta_\cC)d\btheta_\cC.
		\end{split}
\]
Since the region $\bTheta_{\cC}\setminus \{\bTheta_{1,\delta}\cup\bTheta_{2,\delta}\}$ contains neither the neighborhood of $\btheta_1^*$ nor the neighborhood of $\btheta_2^*$, by an application of Lemma~\ref{lemma:regularity}(\ref{lemma:regularity:(a)}),  we have 
\[
\begin{split}
I_3 & \leq p_{\tilde\btheta_1}^{(1)}(\yb^{(1)},\Xb^{(1)} | \hat{\bm\theta}_1)\hat\xi_1 p_{\tilde\btheta_2}^{(2)}(\yb^{(2)},\Xb^{(2)} | \hat{\bm\theta}_2)\hat\xi_2 \exp\left\{ \ell_1({\bm\theta}_1^*)-\ell_1(\hat{\bm\theta}_1)\right \}\exp\left\{ \ell_2({\bm\theta}_2^*)-\ell_2(\hat{\bm\theta}_2)\right \}\\
		&\times \hat\xi_1^{-1}\hat\xi_2^{-1}\exp\left\{ -n_1k_1(\delta)\right \}\exp\left\{-n_2k_2(\delta)\right\}.
\end{split}
\]
Using an argument similar to that of $I_1$ and Lemmas~\ref{lemma:multiple}(\ref{lemma2:regularity:(b)})--(\ref{lemma2:regularity:(d)}), we have 
\begin{equation}
	\frac{I_3}{p(\yb^{(1)},\Xb^{(1)})p(\yb^{(2)},\Xb^{(2)})} =  \Theta(n_1^{\frac {p_1} 2}n_2^{\frac {p_2} 2}\exp\{-n_1k_1(\delta)-n_2k_2(\delta)\})\overset{p}{\to}0.
\end{equation}
Combining the above, we have
\begin{equation}
	\frac{p(\yb^{(1)},\yb^{(2)},\Xb^{(1)},\Xb^{(2)})}{p(\yb^{(1)},\Xb^{(1)})p(\yb^{(2)},\Xb^{(2)})} = \frac{I_1 +I_2 +I_3}{p(\yb^{(1)},\Xb^{(1)})p(\yb^{(2)},\Xb^{(2)})}\overset{p}{\to} 0.
\end{equation}
This concludes the proof of Lemma \ref{lemma:misspecified}.
\end{proof}

\bibliography{reference}

\end{document}